\documentclass[aps, showpacs, floatfix, twocolumn, reprint, twoside, prc, preprintnumbers, superscriptaddress]{revtex4-2}
\usepackage{array}  
\usepackage{graphicx}  
\usepackage[export]{adjustbox}  
\usepackage{float}
\usepackage{epstopdf}
\usepackage{amsmath}
\usepackage{amssymb}
\usepackage{bm}
\usepackage{xcolor,xspace}
\usepackage{multirow}
\usepackage{hyperref}
\hypersetup{colorlinks=True,urlcolor=blue,linkcolor=blue,
citecolor=blue,filecolor=black}
\usepackage{braket}
\usepackage{wasysym}  
\usepackage{scrextend}  
\usepackage[percent]{overpic}  
\usepackage[capitalise]{cleveref}  
\newcommand{\be}[4]{B(E2;#1^+_{#2} \!\to\! #3^+_{#4}) }

\begin{document}
\title{Zr Isotopes as a region of intertwined quantum phase transitions}
\author{N.~Gavrielov}\email{noam.gavrielov@yale.edu}
\affiliation{Racah Institute of Physics, The Hebrew University, 
Jerusalem 91904, Israel}
\affiliation{Center for Theoretical Physics, Sloane Physics Laboratory, 
Yale University, New Haven, Connecticut 06520-8120, USA}
\author{A.~Leviatan}\email{ami@phys.huji.ac.il}
\affiliation{Racah Institute of Physics, The Hebrew University, 
Jerusalem 91904, Israel}
\author{F.~Iachello}\email{francesco.iachello@yale.edu}
\affiliation{Center for Theoretical Physics, Sloane Physics Laboratory, 
Yale University, New Haven, Connecticut 06520-8120, USA}

\date{\today}  
\begin{abstract}
The zirconium isotopes with $A=$ 92--110 have one of the most complicated evolution of structure in the nuclear chart. In order to understand the structural evolution of these isotopes, we carry a detailed calculation in a definite symmetry-based framework, the interacting boson model with configuration mixing (IBM-CM). We compare our calculation to a large range of experimental data, such as energy levels, two neutron separation energies, $E2$ and $E0$ transition rates, isotope shifts and magnetic moments. 
The structural evolution of the low lying spectra of these isotopes is explained using the notion of intertwined quantum phase transitions (IQPTs), for which a QPT involving a crossing of two configurations (Type II) is accompanied by a QPT involving a shape evolution of each configuration separately (Type I). 
In our study, we find the occurrence of Type I QPT within the intruder configuration, changing from weakly deformed to prolate deformed and finally to $\gamma$-unstable, associated with the U(5), SU(3) and SO(6) dynamical symmetry limits of the IBM, respectively. Alongside the Type I QPT, we also find the occurrence of Type II QPT between the normal and intruder configurations, where both Types I and II have a critical-point near $A\approx100$. The good agreement of our calculation with the vast empirical data along the chain of isotopes demonstrates the relevance of IQPTs to the zirconium isotopes, and can serve as a case study to set path for new investigations of IQPTs in other nuclei and other physical systems.
\end{abstract}

\maketitle

\section{Introduction}
\subsection{Intertwined quantum phase transitions (IQPTs)}
\begin{figure}[t]
\centering
\includegraphics[width=1\linewidth]{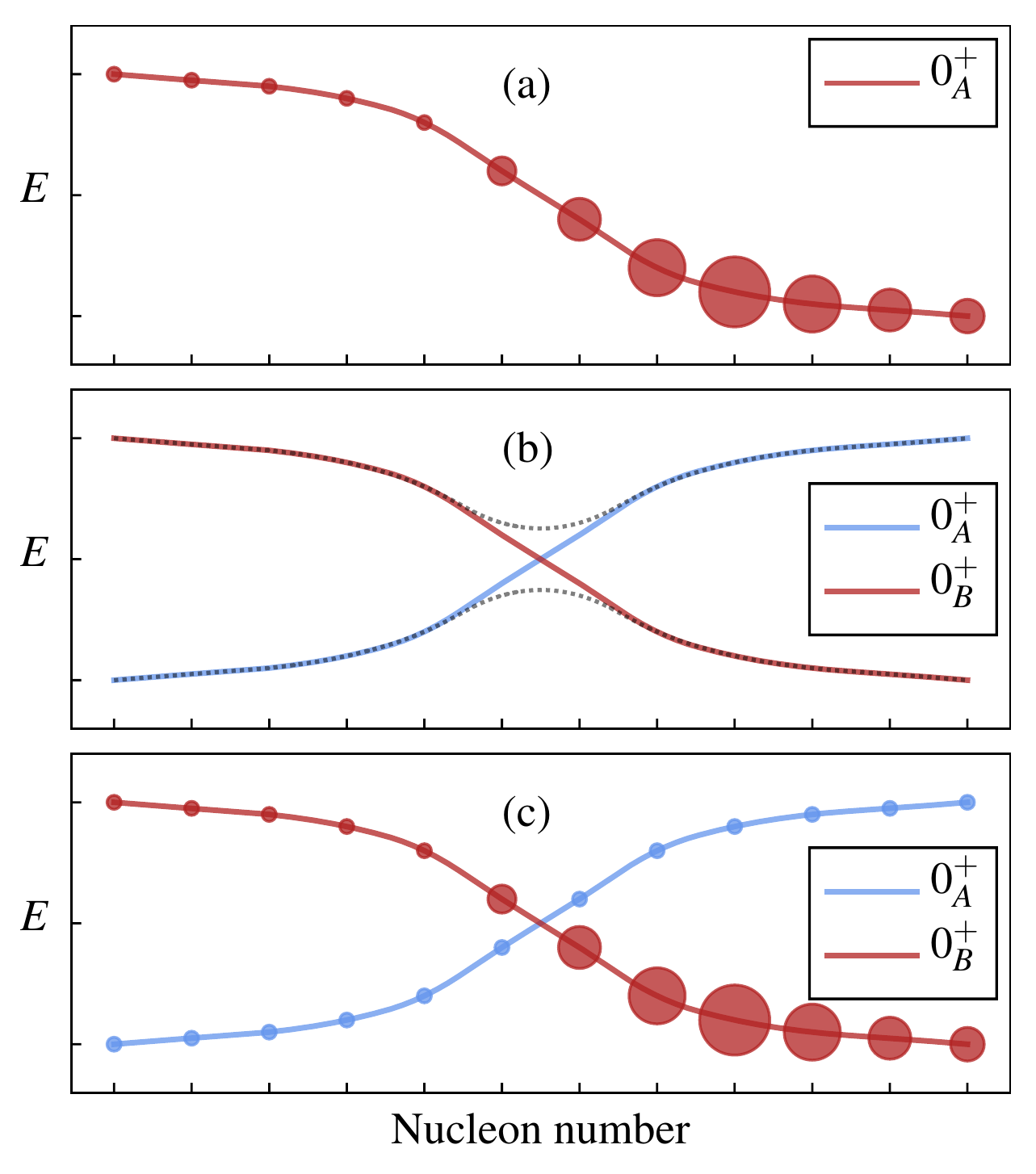} %
\caption{Schematic illustration for the evolution with nucleon number of energies (in arbitrary units) of the lowest $0^{+}$ states of one or two configurations, $A$ and $B$. (a)~Type~I QPT: shape changes within a single configuration (small and large circles denote weak and strong deformation, respectively). {(b)~Type~II QPT:} coexisting and possibly crossing of two configurations, usually, with strong mixing. The dashed lines depict the mixing, as in a two states mixing scenario.
(c)~IQPTs: abrupt crossing of two configurations, with weak mixing, accompanied by a pronounced gradual shape evolution within each configuration.\label{fig:iqpt-schem}}
\end{figure}

Quantum phase transitions (QPTs) \cite{Gilmore1978b, Gilmore1979} have been the subject of great interest for many years in atomic nuclei \cite{Cejnar2010} and in other fields \cite{carr2010QPT}. These are structural changes in a system induced by variations of coupling constants in its quantum Hamiltonian. In atomic nuclei, two types of QPTs are mainly encountered. The first describes shape phase transitions in a single configuration as the number of nucleons is varied. We denote this QPT as Type I. One common approach for investigating Type I QPTs is by using Hamiltonians composed of two different parts
\begin{equation}\label{eq:type-i}
\hat H = (1-\xi)\hat H_1 + \xi\hat H_2~.
\end{equation}
As the control parameter $\xi$ varies from 0 to 1, the equilibrium shape and symmetry of the Hamiltonian vary from those of $\hat H_1$ to those of $\hat H_2$. 
Type I QPT has been established in the neutron number 90 region for Nd-Sm-Gd-Dy isotopes, where the shape of the nuclei evolves from spherical to deformed. 
Such an evolution in deformation is portrayed schematically in \cref{fig:iqpt-schem}(a), where the size of the circles depicts the amount of deformation. 
From a shell-model perspective, when few nucleons interact within a single configuration, low-lying levels of nuclei exhibit characteristics of single-particle excitations, with a seniority-like structure and weak collectivity. This is denoted by small circles in \cref{fig:iqpt-schem}(a). As nucleons are added, they drive collective modes of excitations and onset of deformation in the ground state, which lowers its energy. This is denoted by large circles in \cref{fig:iqpt-schem}(a). 

A different type of phase transitions occurs when two (or more) configurations coexist \cite{Heyde2011}. We denote this QPT as Type II. In this case, the quantum Hamiltonian has a matrix form \cite{Frank2006}
\begin{equation}\label{eq:type-ii}
\hat H =
\left [
\begin{array}{cc}
\hat H_A(\xi_A) & \hat W(\omega) \\ 
\hat W(\omega)  & \hat H_B(\xi_B)
\end{array}
\right]~,
\end{equation}
given here for two configurations, where the indices $A$ and $B$ denote the two configurations and $\hat W$ denotes their coupling.
In such cases, the wave function of the ground state is composed of mixed configurations and evolves from having a dominant component of one configuration to another. 
Type II QPT has been established in nuclei near shell closure, e.g., in the light Pb-Hg isotopes, with strong mixing between the configurations. Such QPT is depicted schematically for two configurations in Fig.~\ref{fig:iqpt-schem}(b). The ground state starts with having a single dominant configuration in its wave function. As nucleons are added, it becomes more mixed with an excited configuration and at some point the latter dominates the ground state. 

Such a scenario follows when protons and neutrons, occupying spin-orbit partner orbitals, interact via the residual isoscalar proton-neutron interaction, $V_{pn}$~\cite{Federman1979}, where the resulting gain in $n$-$p$ interaction energy can compensate the loss in single-particle and pairing energy. Consequently, a mutual polarization effect is enabled and single-particle orbitals at higher configurations are lowered near the ground state configuration, effectively reversing their order. 

Although the two types of QPTs are usually discussed separately, we note that as the control parameters ($\xi_A, \xi_B, \omega$) in \cref{eq:type-ii} are varied, each of the Hamiltonians $\hat H_A$ and $\hat H_B$ can undergo a separate shape-phase transition of Type I, and the combined Hamiltonian can experience a Type II QPT in which there is a crossing of configurations $A$ and $B$. 

In most cases encountered in nuclei, the separate Type I QPTs are masked by the strong mixing between the two configurations. 
In this paper, we present a situation where the Type I QPTs are distinguished.
This is achieved in a situation where within the Type II QPT the mixing between the configurations is weak and as a consequence one can identify the Type I QPT within each configuration separately. This results in an intricate interplay of \textit{intertwined quantum phase transitions} (IQPTs)~\cite{Gavrielov2019, Gavrielov2020}, depicted schematically in Fig.~\ref{fig:iqpt-schem}(c). One can see the energies of the lowest $0^+$ state in each configuration cross while also their individual shape evolve.

\subsection{The zirconium isotopes}
There are several regions in the nuclear chart that are considered to accommodate mixed configurations. One of them is the $Z\!\approx\!40$, $A\!\approx\!100$ region, with coexisting spherical and deformed configurations.
The spherical configuration seems to dominate the ground state wave function for neutron number 50--58 and the deformed configuration dominates for neutron number larger than 58 \cite{Heyde1985, Heyde1987, Cheifetz1970, Federman1979, Sambataro1982, Mach1989} due to a sudden onset of deformation at neutron number 60.
The sudden onset of deformation has been ascribed in the shell-model to $V_{pn}$ between nucleons that occupy the $\pi(1g_{9/2})$--$\nu(1g_{7/2})$ spin-orbit partners~\cite{Federman1977, Federman1979, Heyde1985, Heyde1987}, which induces the normal and intruder configurations to cross. The crossing arises from promotion of protons across the $Z\!=\!40$ sub-shell gap, which creates $2p\text{--}2h$ intruder excitations \cite{Federman1979, Federman1979b}.

These dramatic structural changes have attracted considerable theoretical and experimental interest in the Zr chain.
Different theoretical approaches have studied them, including mean-field based methods, both non-relativistic~\cite{Delaroche2010, Nomura2016c} and relativistic~\cite{Mei2012}, large-scale shell-model calculations~\cite{Sieja2009, Petrovici2012}, the Monte-Carlo shell-model (MCSM)~\cite{Togashi2016} and algebraic models \cite{Gavrielov2019,Gavrielov2020,GarciaRamos2019,GarciaRamos2020}. Recently, several experimental investigations have also come to light \cite{Chakraborty2013, Browne2015a, Browne2015b, Kremer2016, Ansari2017, Paul2017, Witt2018, Singh2018, Karayonchev2020}, opening the door for understanding the properties of both yrast and non-yrast states.

In the present paper, we expand our work from~\cite{Gavrielov2019, Gavrielov2020} and explain how the indication for changes in the content of configuration and the amount of deformation suggests the occurrence of IQPTs in the zirconium isotopes.
This is done by presenting a detailed comparison between our calculation and the empirical data for many observables. This comparison is further supported by analyzing the chain's configuration and symmetry content of the wave functions and the shape evolution.

\subsection{Layout}
The paper is divided into the following sections. In \cref{sec:symm-base} we introduce the theoretical framework, which includes the IBM, its geometric interpretation and Type I QPTs (\cref{sec:ibm-single}), and the IBM with configuration mixing, its geometric interpretation and Type II QPTs (\cref{sec:ibm-cm}). 
In \cref{sec:qpts-zr} we discuss QPTs in the zirconium chain, where we present the model space (\cref{sec:model-space}), the Hamiltonian and its energy surface (\cref{sec:ham-pes}) and the configuration and symmetry assignment for the wave functions (\cref{sec:conf-sym}).

Our results are divided into three main sections.
In \cref{sec:results-indi} we present our results for the individual isotopes, which include spectrum analysis and decomposition of wave functions. This section is further partitioned into the $^{92-96}$Zr region (\cref{sec:92-96zr-region}), the $^{98-102}$Zr region (\cref{sec:98-102zr-region}) and the $^{104-110}$Zr region (\cref{sec:104-110zr-region}).
In \cref{sec:results-evo} we present our results for the configuration (\cref{sec:evo-conf}) and symmetry (\cref{sec:evo-sym}) evolution of wave functions and the evolution of order parameters (\cref{sec:evo-order}).
In \cref{sec:pes} we present a classical analysis for each isotope.
In \cref{sec:evo-obs} we present our results for the evolution of more observables. This includes energy levels (\cref{sec:evo-energy}), two-neutron separation energies (\cref{sec:evo-s2n}), $E2$ transition rates (\cref{sec:evo-e2}), isotope shifts and $E0$ transitions (\cref{sec:evo-iso-e0}) and magnetic moments (\cref{sec:evo-mag-m1}).

We compare our work with other works in \cref{sec:compar-works}. This includes a comparison for $^{98}$Zr (\cref{sec:comp-98zr}), for $^{100}$Zr (\cref{sec:comp-100zr}), for the heavier isotopes (\cref{sec:comp-heavier}) and some general remarks (\cref{sec:general-remarks}).
The conclusions and outlook are given in \cref{sec:conclu}. 
The fitting procedure for determining the Hamiltonian parameters is discussed in \cref{app:fitting}.
\section{Theoretical framework}\label{sec:symm-base}
We employ an algebraic approach to study QPTs in the zirconium isotopes. In order to do so, we use the interacting boson model (IBM)~\cite{IachelloArimaBook}, which describes low lying quadrupole states in even-even nuclei in terms of
a system of monopole ($s$) and quadrupole ($d$) bosons representing valence nucleon pairs. The IBM provides a simple and tractable shell-model-inspired framework, where global trends of structure and symmetries can be clearly identified and a diversity of observables calculated. Below we present a brief introduction to the model.
\subsection{The IBM for a single configuration}\label{sec:ibm-single}
For a single configuration, the IBM Hamiltonian consists of Hermitian and  rotational-scalar interactions that conserve the total number of $s$ and $d$ bosons,
\begin{equation}\label{eq:boson-number}
\hat N=\hat n_s+\hat n_d\!=\!s^\dagger s+\sum_\mu d^\dagger_\mu d_\mu~.
\end{equation}
The latter is fixed by the microscopic interpretation of the IBM \cite{IachelloTalmi1987} to be $N\!=\!N_{\pi}+N_{\nu}$, where $N_{\pi}$ ($N_{\nu}$) is the number of proton (neutron) particle or hole pairs counted from the nearest closed shell. 
\paragraph{Basis states and dynamical symmetries.}
In its simplest version, the IBM has U(6) as a spectrum generating algebra and exhibits three dynamical symmetry (DS) limits
\begin{align}\label{eq:ds-chains}
\text{U(6)} \supset \begin{cases}
	  &\text{U(5)} \supset \text{SO(5)} \supset \text{SO(3)},\\
      &\text{SU(3)} \supset \text{SO(3)}, \\
	  &\text{SO(6)} \supset \text{SO(5)} \supset \text{SO(3)}.
\end{cases}
\end{align}
In a DS, the Hamiltonian is written in terms of Casimir operators of the algebras of a given chain. In such a case, the spectrum is completely solvable and resembles known paradigms of collective motion: spherical vibrator [U(5)], axially symmetric [SU(3)] and $\gamma$-soft deformed rotor [SO(6)]. 
In each case, the energies and eigenstates are labeled by quantum numbers that are the labels of irreducible representations (irreps) of the algebras in the chain. The corresponding basis states for each of the chains \eqref{eq:ds-chains} are
\begin{subequations}
\begin{align}
\text{U(5)}: & \quad\ket{N,n_d,\tau,n_\Delta,L},\\
\text{SU(3)}: & \quad\ket{N,(\lambda,\mu),K,L},\\
\text{SO(6)}: & \quad\ket{N,\sigma,\tau,n_\Delta,L},
\end{align}
\end{subequations}
where $N,n_d,(\lambda,\mu),\sigma,\tau,L$ label the irreps of U(6), U(5), SU(3), SO(6), SO(5) and SO(3), respectively, and $n_\Delta,K$ are multiplicity labels. For a general Hamiltonian, the wave functions with a given boson number, $N$, and angular momentum, $L$, can be expanded in terms of the DS bases in the following manner
\begin{subequations}\label{eq:wf-ds}
\begin{align}
\ket{\Psi; [N],L} & = \sum_{n_d,\tau,n_\Delta} C^{(N,L)}_{n_d,\tau,n_\Delta}\ket{N,n_d,\tau,n_\Delta,L},\\
\ket{\Psi; [N],L} & = \sum_{(\lambda,\mu),K} C^{(N,L)}_{(\lambda,\mu),K}\ket{N,(\lambda,\mu),K,L},\\
\ket{\Psi; [N],L} & = \sum_{\sigma,\tau,n_\Delta} C^{(N,L)}_{\sigma,\tau,n_\Delta}\ket{N,\sigma,\tau,n_\Delta,L},
\end{align}
\end{subequations}
where the coefficients $C^{(N,L)}_\alpha$, with quantum numbers $\alpha$, give the weight of each component in the wave function.
\paragraph{Geometry.}
A geometric visualization of the IBM is obtained by a coherent (intrinsic) state~\cite{Ginocchio1980a,Dieperink1980},
\begin{eqnarray}\label{eq:coherent}
&&  \ket{\beta,\gamma;N} = (N!)^{-1/2}(b^\dagger_c)^N\ket{0}~,\nonumber\\
&&b^\dagger_c = (1+\beta^2)^{-1/2}[\beta\cos\gamma~d^\dagger_0\\
&&  \qquad\qquad
  + \beta\sin\gamma (d^\dagger_2 + d^\dagger_{-2})/\sqrt{2} + s^\dagger]~.\nonumber
\end{eqnarray}
and taking the expectation value of the Hamiltonian to form an energy surface
\begin{eqnarray}\label{eq:surface}
E_N(\beta,\gamma) &=&
\bra{\beta,\gamma; N} \hat H \ket{\beta,\gamma;N}~.
\end{eqnarray} 
Here $(\beta,\gamma)$ are quadrupole shape parameters whose values, $(\beta_{\rm eq},\gamma_{\rm eq})$, at the global minimum of $E_{N}(\beta,\gamma)$ define the equilibrium shape for a given Hamiltonian. The values are $(\beta_{\rm eq}\!=\!0)$, $(\beta_{\rm eq}\!=\!\sqrt{2},\gamma_{\rm eq}\!=\!0)$ and $(\beta_{\rm eq}\!=\!1,\gamma_{\rm eq}\text{~arbitrary})$ for the U(5), SU(3) and SO(6) DS limits, respectively. Furthermore, for these values the ground-band intrinsic state, $\ket{\beta_{\rm eq},\gamma_{\rm eq};N}$, becomes a lowest (or highest) weight state in the irrep of the leading subalgebra of the DS chain, with quantum numbers $(n_d\!=\!0)$, $(\lambda,\mu)\!=\!(2N,0)$ and $(\sigma\!=\!N)$ for the U(5), SU(3) and SO(6) DS limits, respectively.
\paragraph{QPTs: Type I.}
The energy surface $E_{N}(\beta,\gamma;\xi)$, which depends also on the Hamiltonian parameters [e.g. $\xi$ of \cref{eq:type-i}], serves as the Landau potential, whose topology determines the type of phase transition (Ehrenfest classification). 
The correspondence between the DS limits and shapes, identifies the DSs as possible phases of the system. QPTs involving a single configuration (Type I) can be studied in the IBM using a Hamiltonian $\hat{H}(\xi)$, as in \cref{eq:type-i}, that interpolates between different DS limits (phases) by varying its control parameters $\xi$. Such QPTs have been studied extensively in the IBM framework~\cite{Dieperink1980,Cejnar2009,Cejnar2010,Iachello2011}. 

In Type I QPTs, the order parameter is taken to be the expectation value of the $d$-boson number operator, $\hat n_d$, in the ground state, with the following values for the DS limits 
\begin{subequations}\label{eq:order-param-ds}
\begin{align}
\text{U(5)}:\quad\braket{\hat n_d}_{0^+_1} & = 0,\\
\text{SU(3)}:\quad\braket{\hat n_d}_{0^+_1} & = \frac{4N(N-1)}{3(2N-1)},\\
\text{SO(6)}:\quad\braket{\hat n_d}_{0^+_1} & = \frac{N(N-1)}{2(N+1)}.
\end{align}
\end{subequations}
The expressions of \cref{eq:order-param-ds} converge in the large-$N$ limit to the geometric form of the order parameter in terms of the corresponding equilibrium deformation, $\beta_{\rm eq}$,
\begin{equation}\label{eq:order-param}
\frac{\braket{\hat n_d}_{0^+_1}}{N}\approx
\frac{\beta_{\rm eq}^2}{1+\beta_{\rm eq}^2}~.
\end{equation}
\paragraph{Example: Hamiltonian for Type I QPTs.}
A typical Hamiltonian frequently used for Type I QPTs, has the form~\cite{Warner1983,Lipas1985}
\begin{equation}\label{eq:ham-q}
\hat H(\epsilon_d,\kappa,\chi) =
\epsilon_d\, \hat n_d + \kappa\, \hat Q_\chi \cdot \hat Q_\chi~,
\end{equation}
where the quadrupole operator is given by
\begin{equation}\label{eq:q-op}
\hat Q_\chi =
d^\dag s+s^\dag \tilde d \!+\!\chi (d^\dag \times \tilde d)^{(2)} ~.
\end{equation}
Here $\tilde d_m = (-1)^m d_{-m}$ and standard notation of angular
momentum coupling is used. The control parameters, $(\epsilon_d,\kappa,\chi)$, in Eq.~\eqref{eq:ham-q} with values $(\kappa\!=\!0)$, $(\epsilon_d\!=\!0,\chi\!=\!-\sqrt{7}/2)$ and $(\epsilon_d\!=\!0,\chi\!=\!0)$, interpolate between the respective U(5), SU(3) and SO(6) DS~limits. The U(5)-SU(3) transition is found to be first-order, the U(5)-SO(6) transition is second order and the SU(3)-SO(6) transition is a crossover.
For the Hamiltonian~\eqref{eq:ham-q}, the associated Landau potential \eqref{eq:surface} reads
\begin{eqnarray} \label{eq:surface-single}
&&E_{N}(\beta,\gamma;\epsilon_d,\kappa,\chi) =
\nonumber\\
&&\quad 5\kappa\, N + \frac{N\beta^2}{1+\beta^2} 
\left[\epsilon_d + \kappa (\chi^2-4)\right]
\nonumber\\
&&\quad + \frac{N(N-1)\beta^2}{(1+\beta^2)^2}\kappa
\left[4 - 4\bar{\chi}\beta\,\Gamma + \bar\chi^2\beta^2\right] ,\qquad
\end{eqnarray} 
where $\bar\chi\!=\!\sqrt{\frac{2}{7}}\chi$ and $\Gamma\!=\!\cos3\gamma$.

\subsection{The IBM for configuration mixing}\label{sec:ibm-cm}
An extension of the IBM to include intruder excitations is based on associating the different shell-model spaces of 0p-0h, 2p-2h, 4p-4h, $\dots$ particle-hole excitations, with the corresponding boson spaces comprising $N,\, N\!+\!2,\, N\!+\!4,\ldots$ bosons, respectively, which are subsequently mixed.
In this case, the resulting interacting boson model with configuration mixing (IBM-CM)~\cite{Duval1981,Duval1982} Hamiltonian has the form as in \cref{eq:type-ii}.
In the present work, we write it not in matrix form, but rather in the equivalent form
\begin{equation}\label{eq:ham-cm}
\hat H = \hat H_A^{(N)} + \hat H_B^{(N+2)} + \hat W^{(N,N+2)}~.
\end{equation}
Here, $\hat{\cal O}^{(N)}=\hat{P}_{N}^{\dag }\hat{\cal O}\hat{P}_{N}$ and
$\hat{\cal O}^{(N,N^{\prime })}=
\hat{P}_{N}^{\dag }\hat{\cal O}\hat{P}_{N^{\prime }}$, 
for an operator $\hat{\cal O}$, with $\hat{P}_{N}$, a projection operator 
onto the $N$ boson space. The Hamiltonian $\hat{H}_{A}^{(N)}$ represents
the normal $A$ configuration ($N$ boson space) and $\hat{H}_{B}^{(N+2)}$
represents the intruder $B$ configuration ($N\!+\!2$ boson space).

The $E2$ operator for the two configurations is expanded accordingly
\begin{equation}\label{eq:te2}
\hat{T}(E2) = e^{(A)}\hat Q^{(N)}_{\chi} + e^{(B)}\hat Q^{(N+2)}_{\chi},
\end{equation} 
with~$\hat Q_\chi^{(N)}\!=\!\hat P_N^\dag\hat Q_\chi\hat P_N$
and $\hat Q_\chi$, defined in Eq.~\eqref{eq:q-op}, is the same quadrupole operator appearing in the Hamiltonian~\eqref{eq:ham-q}.
In \cref{eq:te2}, $e^{(A)}$ and $e^{(B)}$ are the boson effective charges
for the configurations $A$ and $B$, respectively. No mixing term appears in \cref{eq:te2}, since we assume that the $E2$ operator is a one-body operator and therefore cannot change the boson number by 2.
\paragraph{Wave functions.}
The resulting eigenstates $\ket{\Psi;L}$ of the Hamiltonian \eqref{eq:ham-cm} with angular momentum $L$, are linear combinations of the wave functions, $\Psi_A$ and $\Psi_B$, in the two spaces $[N]$ and $[N+2]$,
\begin{equation}\label{eq:wf}
\ket{\Psi; L} = a\ket{\Psi_A; [N], L} + b\ket{\Psi_B; [N\!+\!2], L}\, ,\;
\end{equation}
with $a^{2} + b^{2} = 1$. We note that each of the components in \cref{eq:wf}, $\ket{\Psi_A; [N], L}$ and $\ket{\Psi_B; [N\!+\!2], L}$, can be expanded in terms of the different DS limits of \cref{eq:wf-ds} with its corresponding boson number.
\paragraph{Geometry.}
A geometric interpretation~\cite{Frank2004} is obtained by means of the matrix 
$E(\beta,\gamma)$,
\begin{align}\label{eq:surface-mat}
E(\beta,\gamma) =
\left [
\begin{array}{cc}
E_A(\beta,\gamma;\xi_A) & \Omega(\beta,\gamma;\omega) \\ 
\Omega(\beta,\gamma;\omega) & E_B(\beta,\gamma;\xi_B)
\end{array}
\right ] ,
\end{align} 
whose entries are the matrix elements of the corresponding terms in the Hamiltonian~\eqref{eq:type-ii}, between  the intrinsic states~\eqref{eq:coherent} of the two configurations, with appropriate boson numbers,
\begin{subequations}\label{eq:surface-elem}
\begin{align}
E_A(\beta,\gamma) & =
\braket{\beta,\gamma;N|\hat H_A|\beta,\gamma;N},\\
E_B(\beta,\gamma) & =
\braket{\beta,\gamma;N+2|\hat H_B|\beta,\gamma;N+2},\\
\Omega(\beta,\gamma) & =
\braket{\beta,\gamma;N|\hat{W}|\beta,\gamma;N+2}.
\end{align}
\end{subequations}
Diagonalization of this two-by-two matrix produces the so-called eigen-potentials, $E_{\pm}(\beta,\gamma)$.
\paragraph{QPTs: Type II.}
$E(\beta,\gamma)$ of \cref{eq:surface-mat}, which depends also on the Hamiltonian parameters, serves as the Landau potential matrix \cite{Frank2006}. 
QPTs involving multiple configurations (Type II) can be studied in the IBM-CM using a Hamiltonian $\hat{H}(\xi_A,\xi_B,\omega)$ as in \cref{eq:type-ii}, that interpolates between the different configurations by varying its control parameters $\xi_A,\xi_B,\omega$. 
Configuration-mixed QPTs and coexistence phenomena in nuclei have been studied extensively in the  IBM-CM framework~\cite{Duval1981, Duval1982, Sambataro1982, Duval1983, PadillaRodal2006, Fossion2003, Frank2006, GarciaRamos2011, GarciaRamos2014a, GarciaRamos2014b, GarciaRamos2015a, Nomura2016c, Leviatan2018a}.

In Type II QPTs, the order parameters are taken to be the expectation value of $\hat n_d$ in the ground state wave function, $\ket{\Psi;L=0^+_1}$, and in its $\Psi_A$ and $\Psi_B$ components, \cref{eq:wf}, denoted by $\braket{\hat n_d}_{0^+_1}$, $\braket{\hat n_d}_A$ and $\braket{\hat n_d}_B$, respectively. As can be inferred from \cref{eq:order-param}, the shape-evolution in each of the configurations~$A$ and $B$ , is depicted by $\braket{\hat n_d}_A$ and $\braket{\hat n_d}_B$, respectively. Their sum weighted by the probabilities of the $\Psi_A$ and and $\Psi_B$ components
\begin{equation}\label{eq:order-param-cm}
\braket{\hat n_d}_{0^+_1} =
a^2\braket{\hat n_d}_A + b^2\braket{\hat n_d}_B ~,
\end{equation}
portrays the evolution of the normal-intruder mixing.
\section{QPTs in the Zr chain}\label{sec:qpts-zr}
\subsection{Model space}\label{sec:model-space}
\begin{figure}[t]
\centering
\includegraphics[width=1\linewidth]{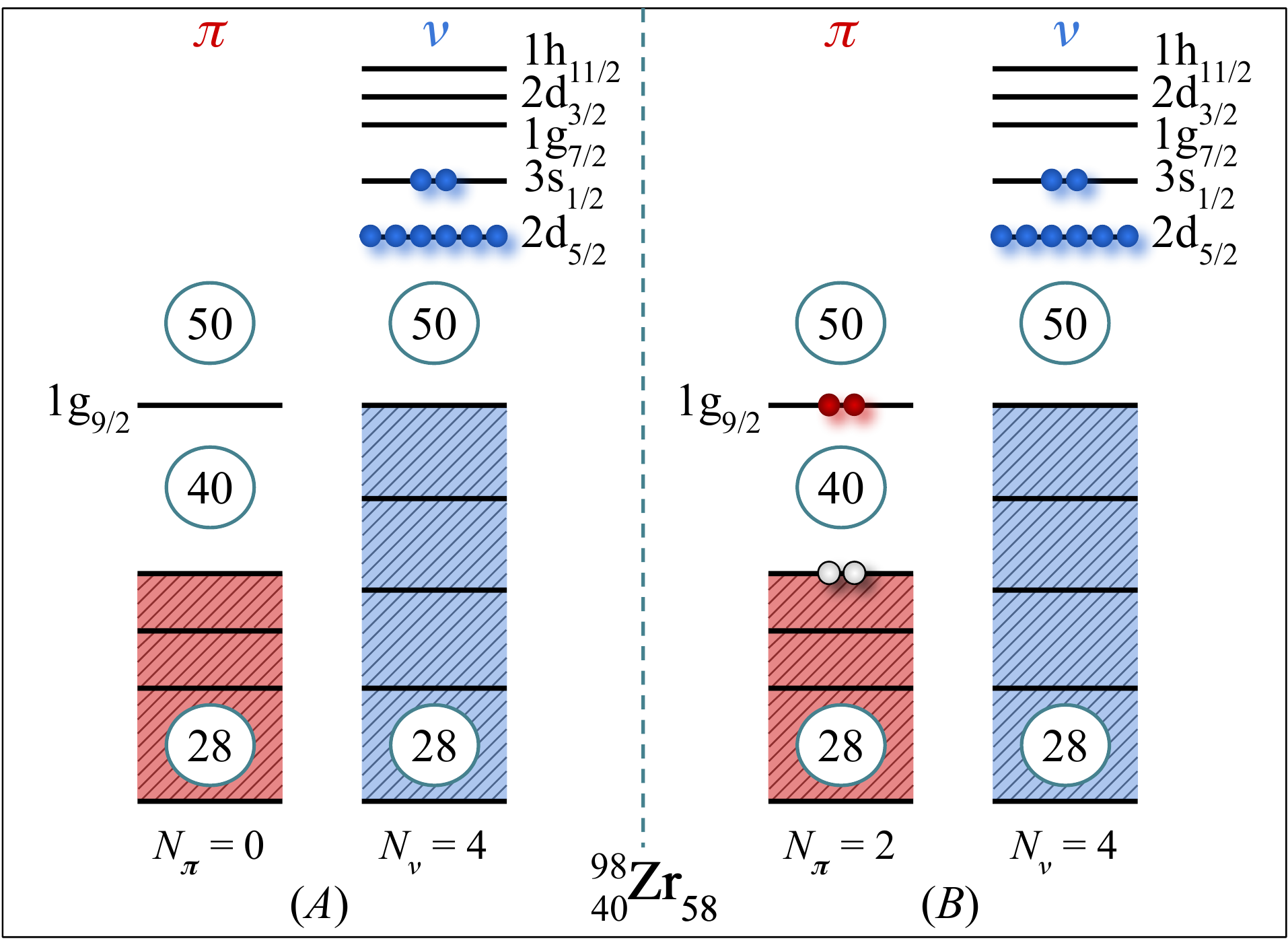}
\caption{Schematic representation of the two coexisting shell-model
configurations ($A$ and $B$) for $^{98}_{40}$Zr$_{58}$. The corresponding numbers of proton bosons ($N_{\pi}$) and neutron bosons ($N_{\nu}$), relevant to the IBM-CM, are listed for each configuration.\label{fig:zr-98-shell}}
\end{figure}
To describe the $_{40}$Zr isotopes in the IBM-CM framework, we consider $_{40}^{90}$Zr as a core and valence neutrons in the 50--82 major
shell, similar to a calculation done for the $_{42}$Mo isotopes
in~\cite{Sambataro1982}.
The normal $A$ configuration corresponds to having no active protons above $Z\!=\!40$ sub-shell gap, and the intruder $B$ configuration corresponds to two-proton excitation from below to above this gap, creating 2p-2h states.
According to the usual boson-counting, the corresponding bosonic configurations have proton bosons $N_{\pi}\!=\!0$ for configuration~$A$ and $N_{\pi}\!=\!2$ for configuration~$B$ . Both configurations have neutron bosons $N_{\nu}\!=\!1,~2, \ldots,~8$ for neutron number 52--66, and $\bar N_\nu\!=\!\bar 7,\,\bar 6$ for neutron number 68--70, where the bar over a number indicates that these are hole bosons. These two configurations are shown schematically in \cref{fig:zr-98-shell} for $^{98}$Zr. Altogether, the IBM-CM model space employed in the current study, consists of $[N]\oplus[N+2]$ boson spaces with total boson number $N=1,2,\ldots8$ for $^{92-106}$Zr and $\bar N=\bar 7,\bar 6$ for $^{108,110}$Zr, respectively.

\subsection{Hamiltonian and energy surface}\label{sec:ham-pes}
For two configurations,  the Hamiltonian has a form as in \cref{eq:type-ii} or \cref{eq:ham-cm}, with a typical choice 
\begin{subequations}
\label{eq:ham_ab}
\begin{align}
\hat H_A = & \hat H(\epsilon^{(A)}_d,\kappa^{(A)},\chi)~,
\label{eq:ham_a}
\\
\hat H_B = & \hat H(\epsilon^{(B)}_d,\kappa^{(B)},\chi)
+ \kappa^{\prime(B)} \hat L \cdot \hat L + \Delta_p~,
\label{eq:ham_b}
\end{align}
\end{subequations}
where $\hat H$ is given in \cref{eq:ham-q}. The Hamiltonian $\hat{H}_B$ of Eq.~(\ref{eq:ham_b}), contains an additional rotational term, ${\hat L \cdot \hat L}$ with parameter $\kappa^{\prime(B)}$, where $\hat L = \sqrt{10}(d^\dagger\tilde d)^{(1)}$ is the angular momentum operator. $\Delta_p$ is the off-set energy between configurations~$A$ and $B$ , and the index $p$ denotes the fact that this is a proton excitation. The mixing term in \cref{eq:ham-cm} between configurations $(A)$ and $(B)$ has the
form~\cite{IachelloArimaBook, Duval1981, Duval1982}
\begin{equation} \label{eq:mixing-s-d}
\hat W = \,[\omega_d\,(d^\dag\times d^{\prime\dag})^{(0)} + \omega_s\,(s^\dag s^{\prime\dag})\,] + {\rm H.c.}~,
\end{equation}
where primed (unprimed) bosons denote hole (particle) bosons with respect to the proton shell gap at $Z\!=\!40$. We take in this paper $\omega_s\!=\!\omega_d\!=\!\omega$ and $s^\prime,d^\prime=s,d$ in order to avoid proliferation of parameters and as done in all previous IBM-CM calculations~\cite{Duval1981, Duval1982}. The mixing term, has then the form
\begin{equation} \label{eq:mixing}
\hat W = \omega\,[\,(d^\dag\times d^\dag)^{(0)} + (s^\dag)^2\,] + {\rm H.c.} ~,
\end{equation}
where H.c. stands for Hermitian conjugate.

For the energy surface matrix \eqref{eq:surface-mat}, we calculate the expectation values of the Hamiltonians
$\hat H_A$~\eqref{eq:ham_a} and $\hat H_B$~\eqref{eq:ham_b}, in the intrinsic state~\eqref{eq:coherent}, with $N$ and $N\!+\!2$ bosons respectively, and a non-diagonal matrix element of the mixing term $\hat W$~\eqref{eq:mixing} between them.
The explicit expressions are found to be
\begin{subequations}\label{eq:surface-elements}
\begin{align}
E_A(\beta,\gamma) & = 
E_N(\beta,\gamma;\epsilon^{(A)}_d,\kappa^{(A)},\chi)~,
\label{eq:e_a}
\\
E_B(\beta,\gamma)  & = 
E_{N+2}(\beta,\gamma;\epsilon^{(B)}_d,\kappa^{(B)},\chi)
\nonumber\\
& \quad + 6\kappa^{\prime(B)}\frac{(N+2)\beta^2}{1+\beta^2}
+ \Delta_p~,
\label{eq:e_b}
\\
\Omega(\beta,\gamma) & = 
\frac{\sqrt{(N+2)(N+1)}}{1+\beta^2}
\omega\Bigg(1 + \frac{1}{\sqrt{5}}\beta^2\Bigg)~,
\label{eq:e_w}
\end{align}
\end{subequations}
where the surfaces on the right-hand-side of \cref{eq:e_a,eq:e_b} are obtained from Eq.~\eqref{eq:surface-single}.
\subsection{Configuration and symmetry assignment}\label{sec:conf-sym}
Given an eigenstate of the form as in \cref{eq:wf}, one can calculate for either the $A$  or $B$ part its decomposition in the DS bases, \cref{eq:wf-ds}. This defines the probability of having definite quantum numbers of a given symmetry,
\begin{subequations}\label{eq:decomp-ds}
\begin{align}
\text{U(5)}: \quad & P^{(N_i,L)}_{n_d} = \sum_{\tau,n_\Delta}[C^{(N_i,L)}_{n_d,\tau,n_\Delta}]^2, \label{eq:decomp-u5}\\
\text{SU(3)}: \quad & P^{(N_i,L)}_{(\lambda,\mu)} = \sum_{K}[C^{(N_i,L)}_{(\lambda,\mu),K}]^2,\label{eq:decomp-su3}\\
\text{SO(6)}: \quad & P^{(N_i,L)}_{\sigma} = \sum_{\tau,n_\Delta}[C^{(N_i,L)}_{\sigma,\tau,n_\Delta}]^2,\label{eq:decomp-so6}\\
\text{SO(5)}: \quad & P^{(N_i,L)}_{\tau} = \sum_{n_d,n_\Delta}[C^{(N_i,L)}_{n_d,\tau,n_\Delta}]^2\label{eq:decomp-so5}.
\end{align}
\end{subequations}
Here the subscripts $i\!=\!A,B$ denote the different configurations, i.e. $N_A\!=\!N$ and $N_B\!=\!N+2$.
Furthermore, for each eigenstate \cref{eq:wf}, one can also examine its coefficients $a$ and $b$, which portray the probability of the normal-intruder mixing,
\begin{align}
P^{(N_A,L)}_a & = a^2,\nonumber\\
P^{(N_B,L)}_b & = b^2.
\end{align}
$a^2$ and $b^2$ can be evaluated from the sum of the squared coefficients of an IBM basis [U(5), SU(3) or SO(6)] in their respective boson-space, $N$ and $N+2$. For the U(5) basis, we have
\begin{subequations}\label{eq:decomp-cm}
\begin{align}
a^2 & = \sum_{n_d,\tau,n_\Delta} |C^{(N_A,L)}_{n_d,\tau,n_\Delta}|^2,\\
b^2 & = \sum_{n_d,\tau,n_\Delta} |C^{(N_B,L)}_{n_d,\tau,n_\Delta}|^2,
\end{align}
\end{subequations}
where the sum goes over all possible values of $(n_d,\tau,n_\Delta)$ in the $(N_i,L)$ space, $i=A,B$, and $a^2 + b^2 = 1$.
\section{Results: Detailed Quantum Analysis of individual isotopes}\label{sec:results-indi} 
\begin{figure}[t]
\centering
\includegraphics[width=1\linewidth]{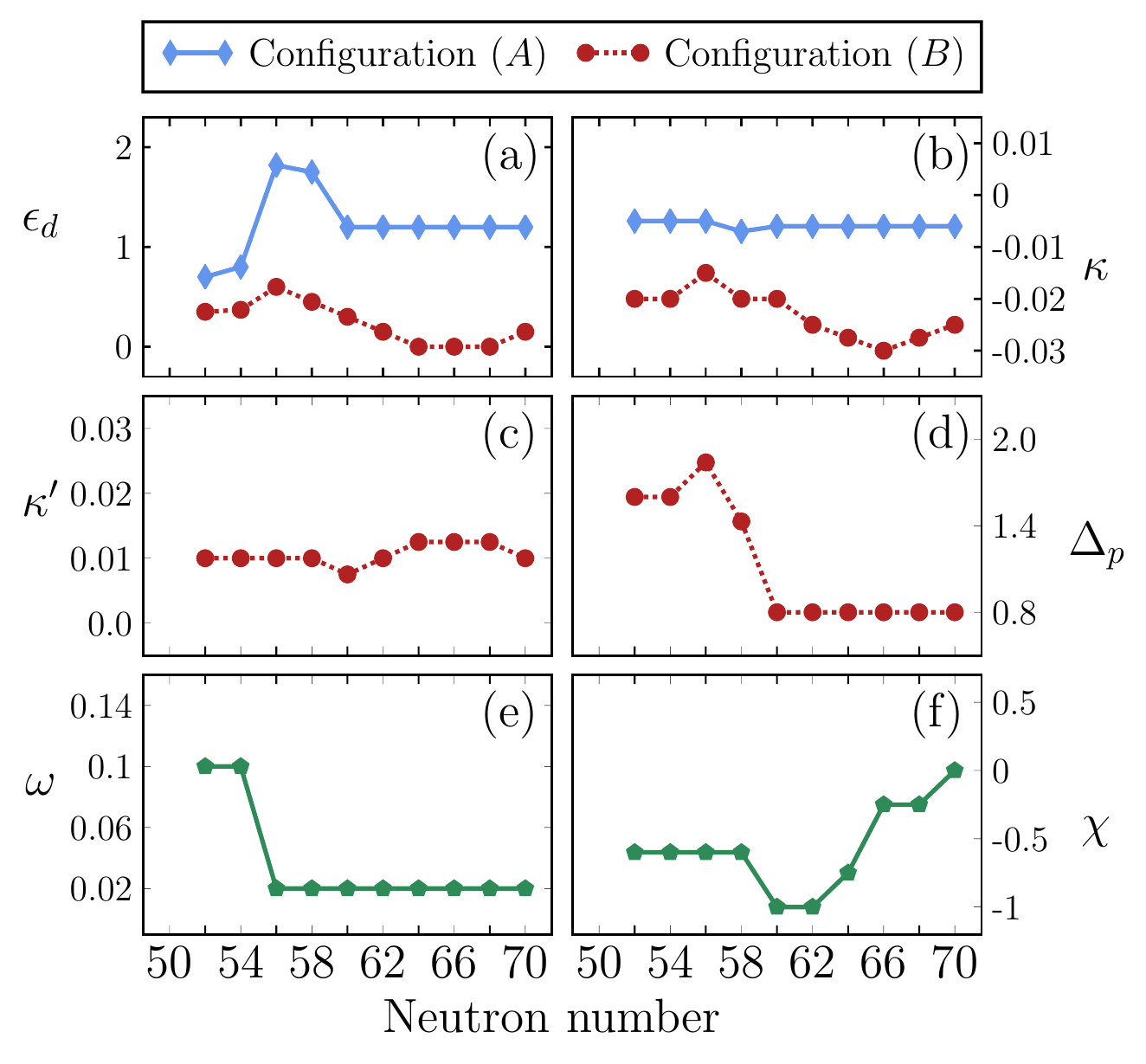}
\caption{Parameters of the IBM-CM Hamiltonians, Eqs.~\eqref{eq:ham_a},
\eqref{eq:ham_b}, \eqref{eq:mixing}, are in MeV and the parameter $\chi$
of Eq.~\eqref{eq:q-op}, is dimensionless.\label{fig:params}}
\end{figure}
The quantum analysis for $^{92-110}$Zr entails a detailed comparison of the experimental energies and $E2$ transition rates with the results of our calculation. The Hamiltonian parameters used are shown in \cref{fig:params} and \cref{tab:parameters}. The fitting procedure employed to obtain them and their trends, are discussed in \cref{app:fitting}.

We now discuss our calculation for individual isotopes, dividing them into different regions. Each region is defined by the symmetry properties of the intruder $B$ configuration. The first region $^{92-96}$Zr, with coexistence of two U(5)-configurations, the second region $^{98-102}$Zr, with Type I [U(5)-SU(3)] and Type II QPTs, and the third region $^{104-110}$Zr, with SU(3)-SO(6) crossover. For each region, we also discuss the configuration and symmetry content of selected eigenstates. Information on the symmetry structure within each configuration is obtained by examining the decomposition of the wave functions defined in \cref{eq:decomp-ds}. Information on the configuration content of each eigenstate is obtained from \cref{eq:decomp-cm}.
\subsection{The $^{92-96}$Zr region: U(5)-coexistence}\label{sec:92-96zr-region}
\begin{figure*}[t]
\begin{overpic}[width=0.49\linewidth]{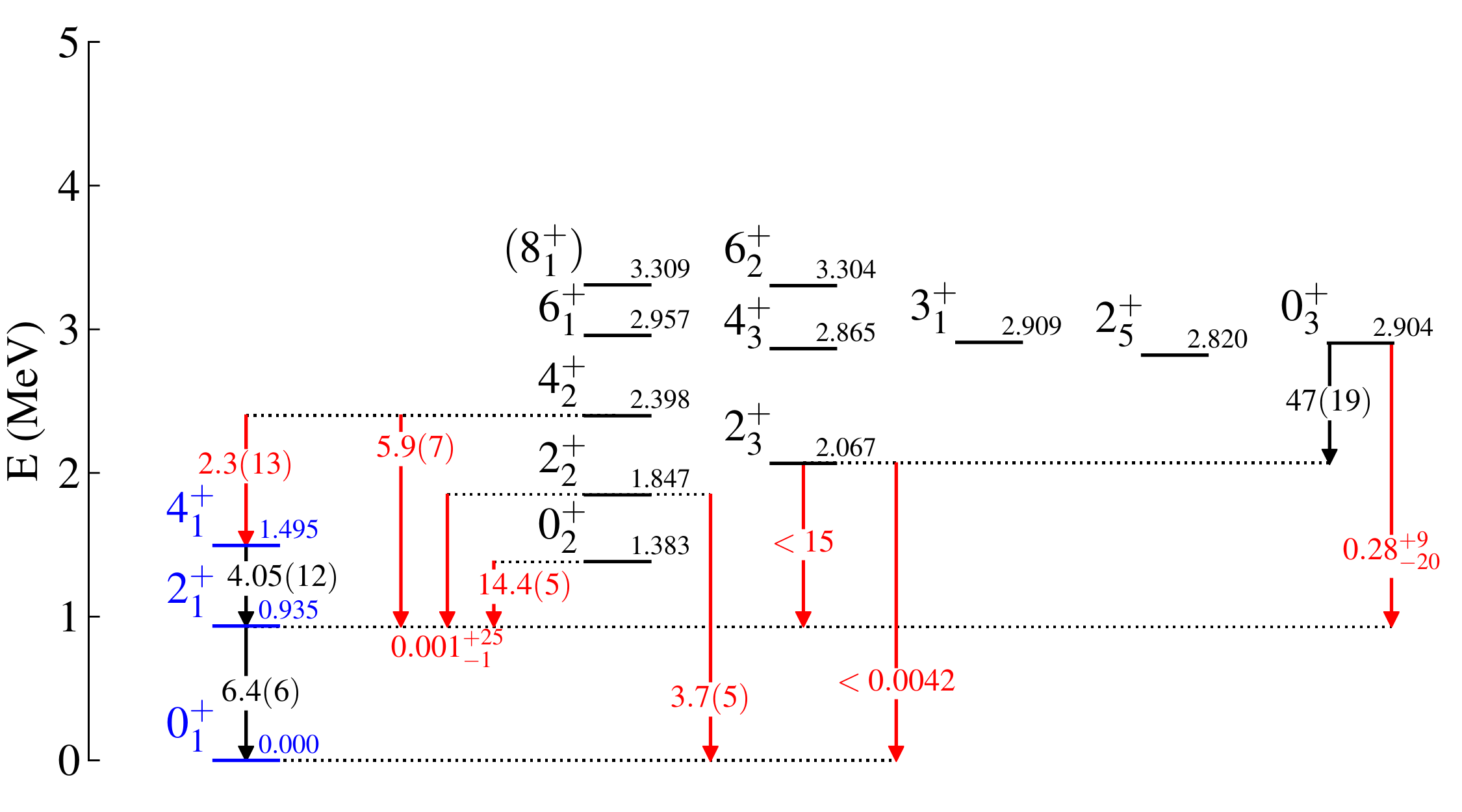}
\put (10,50) {\large (a)}
\put (10,42.5) {\large {\bf $^{92}$Zr exp}}
\end{overpic}
\begin{overpic}[width=0.49\linewidth]{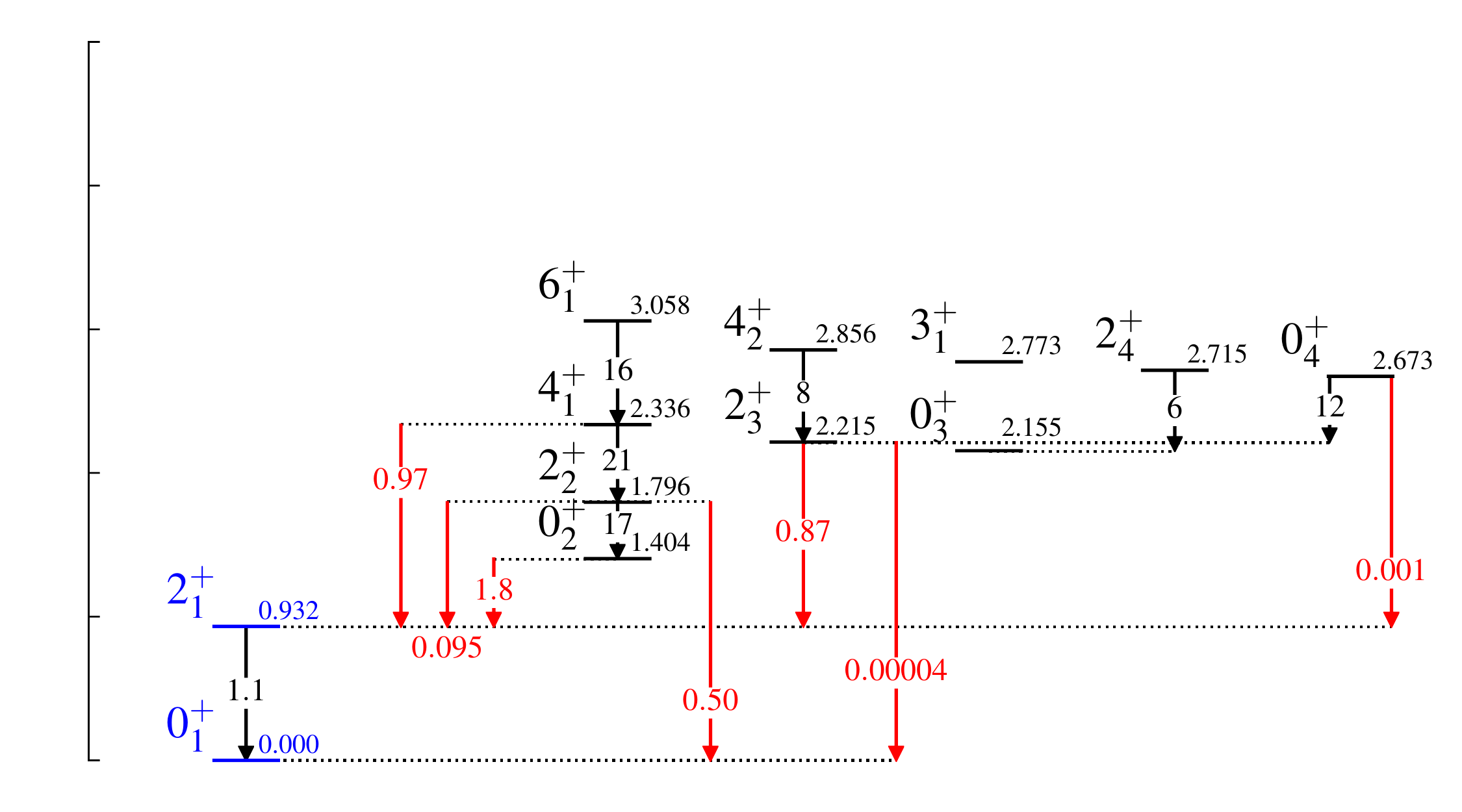}
\put (10,50) {\large (b)}
\put (10,42.5) {\large {\bf $^{92}$Zr calc}}
\end{overpic}\\
\begin{overpic}[width=0.49\linewidth]{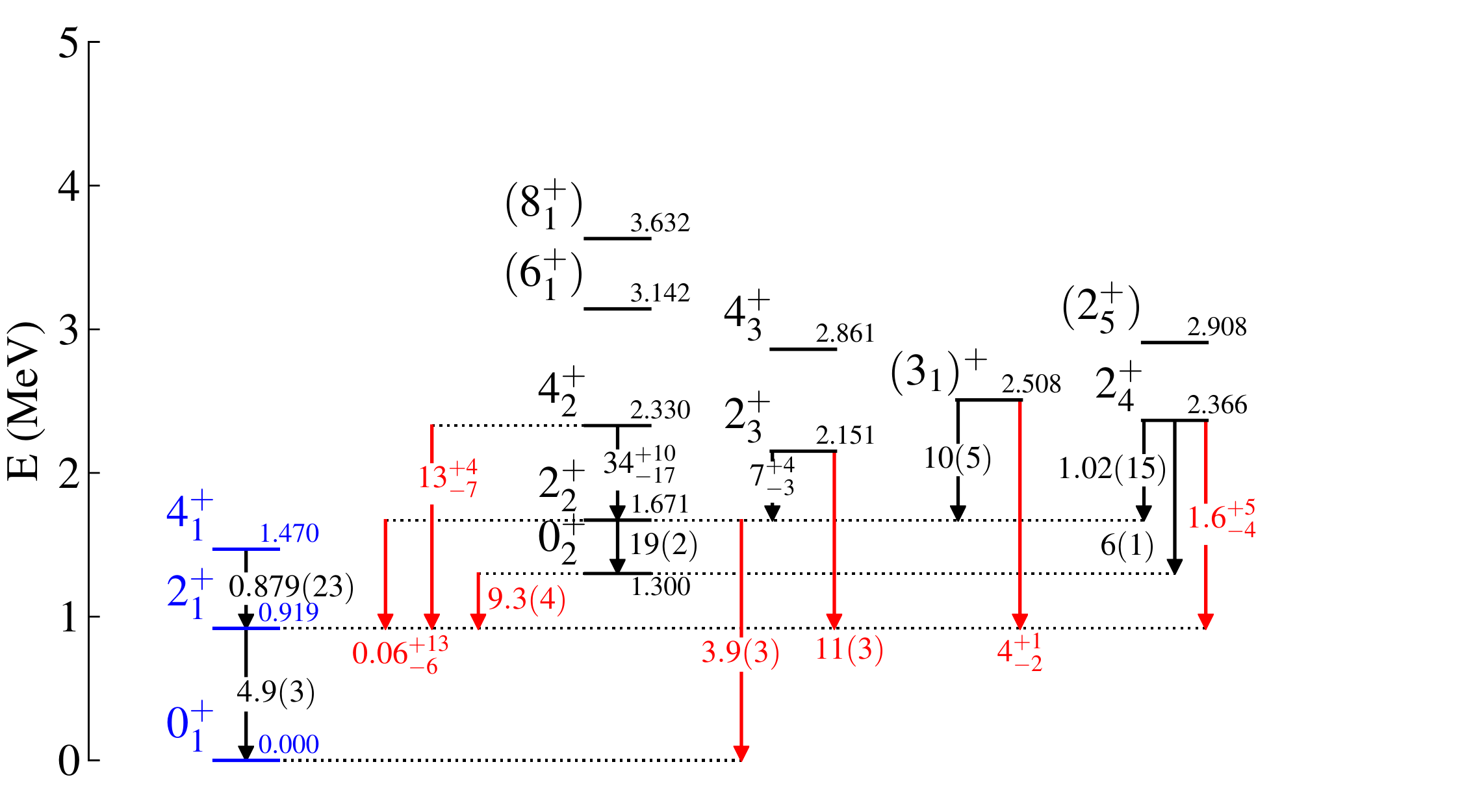}
\put (10,50) {\large (c)}
\put (10,42.5) {\large {\bf $^{94}$Zr exp}}
\end{overpic}
\begin{overpic}[width=0.49\linewidth]{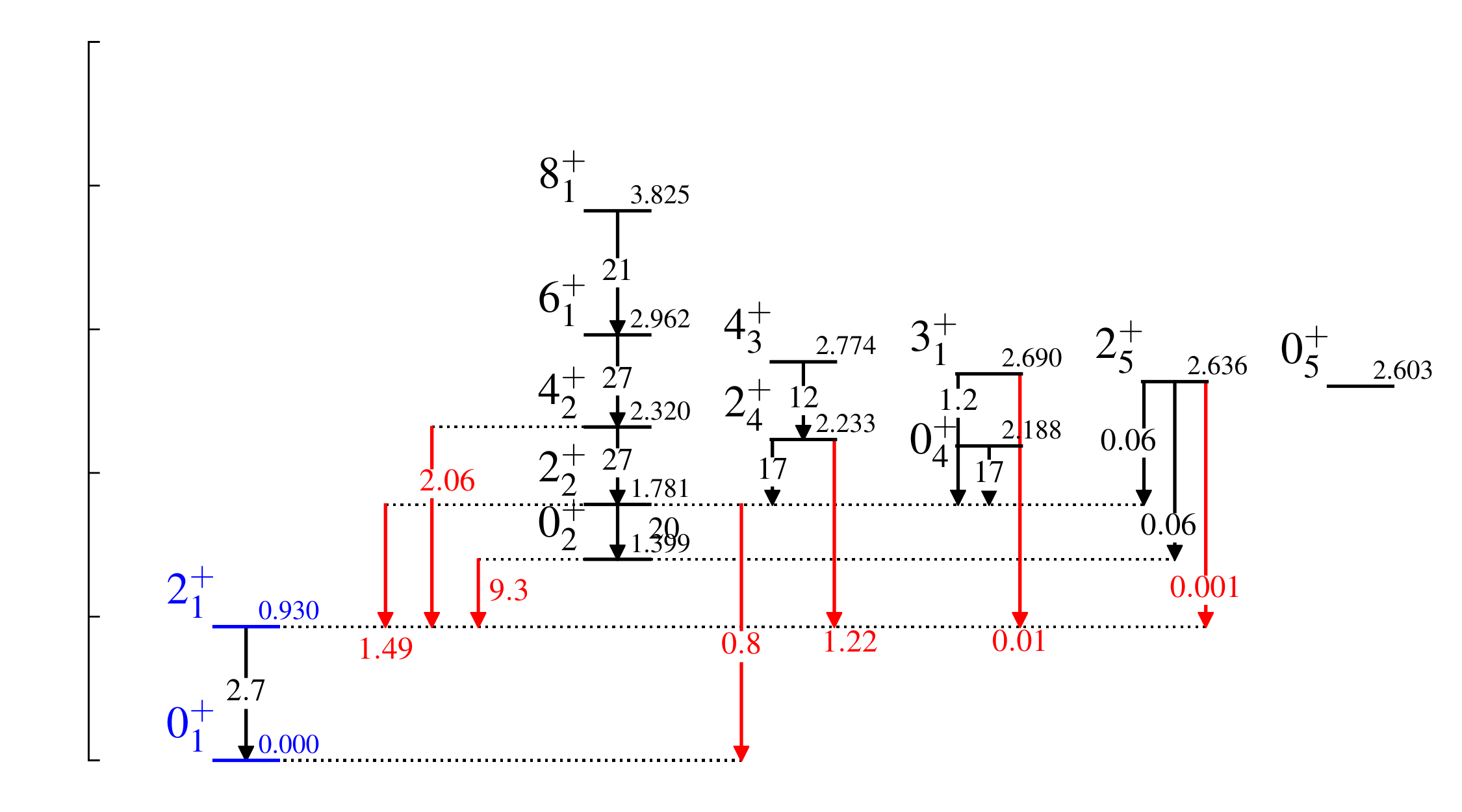}
\put (10,50) {\large (d)}
\put (10,42.5) {\large {\bf $^{94}$Zr calc}}
\end{overpic}\\
\begin{overpic}[width=0.49\linewidth]{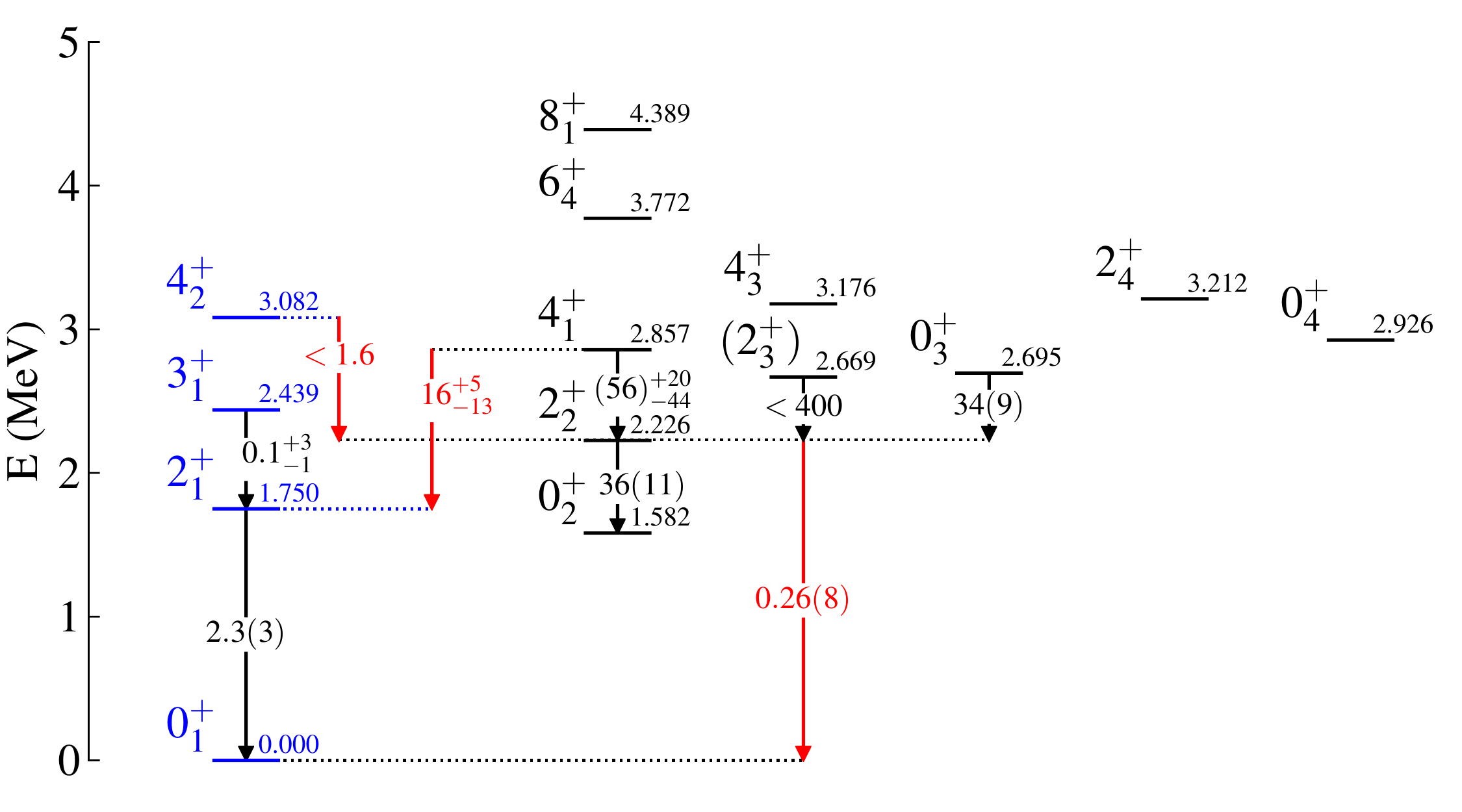}
\put (10,50) {\large (e)}
\put (10,42.5) {\large {\bf $^{96}$Zr exp}}
\end{overpic}
\begin{overpic}[width=0.49\linewidth]{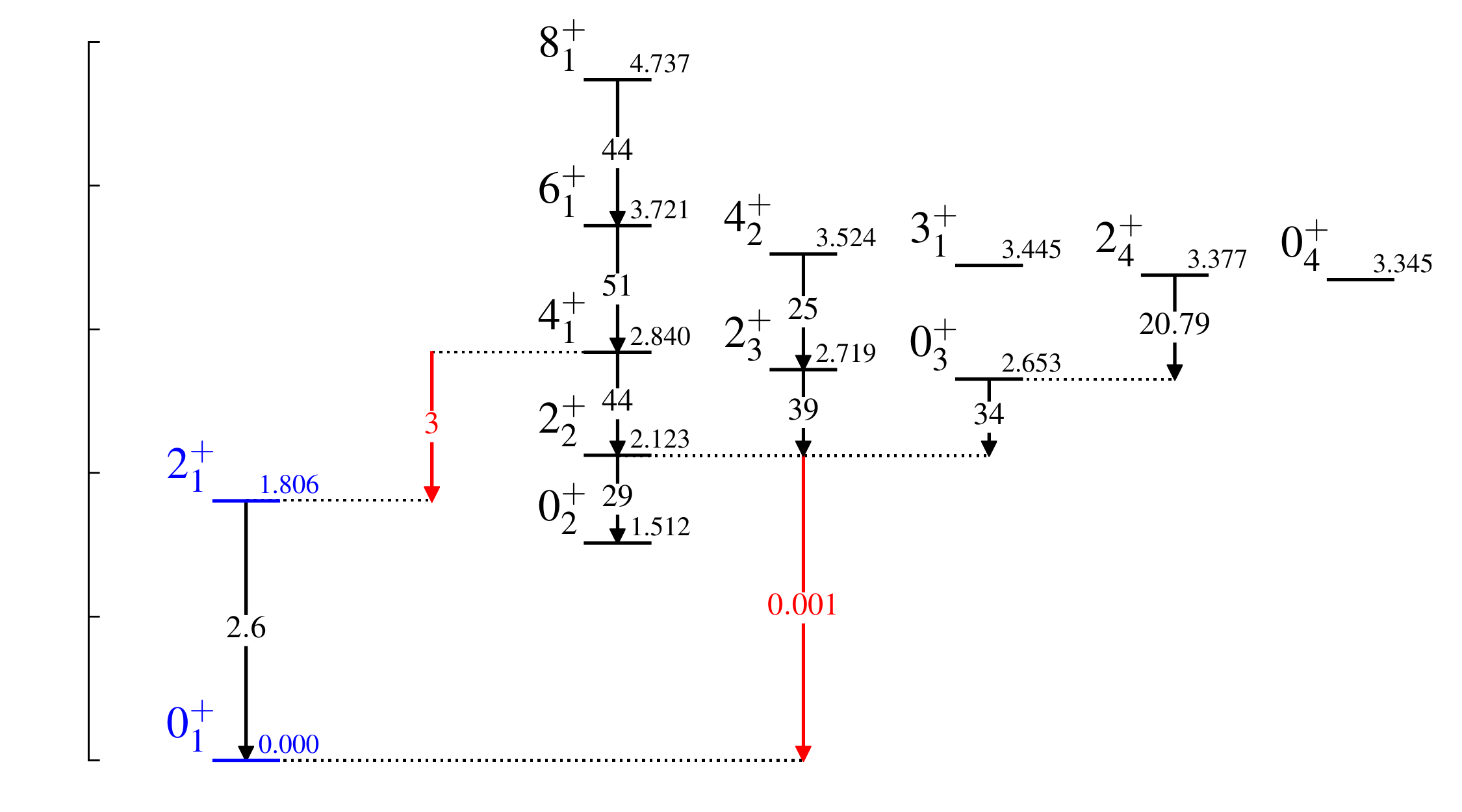}
\put (10,50) {\large (f)}
\put (10,42.5) {\large {\bf $^{96}$Zr calc}}
\end{overpic}
\caption{Experimental and calculated energy levels in MeV and $E2$ transition rates in W.u. Levels in blue (black) belong to the $A$ ($B$)  configuration. Transitions between different configurations are denoted in red.
For the configuration~$A$ experimental levels that have no corresponding calculated levels; see the Appendix.
Data are taken from \cite{NDS.113.2187.2012} for $^{92}$Zr, \cite{NDS.107.2423.2006,Chakraborty2013} for $^{94}$Zr, and \cite{NDS.109.2501.2008,Kremer2016} for $^{96}$Zr.
\label{fig:92-96Zr-scheme}}
\end{figure*}
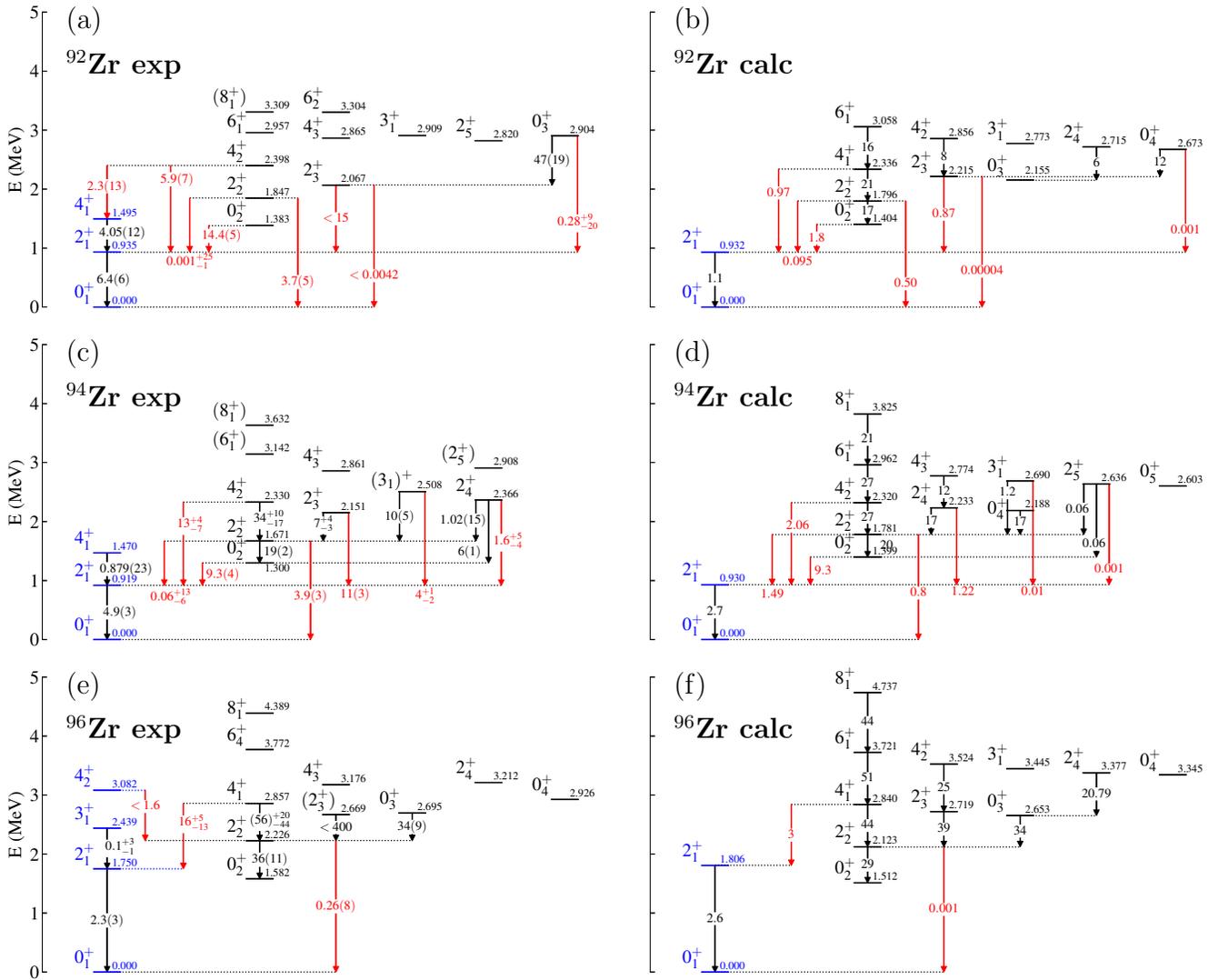
We begin by comparing our calculation to the experimental values for the region of $^{92-96}$Zr, shown in \cref{fig:92-96Zr-scheme}. For each of these isotopes, the spectrum exhibits coexistence of two spherical configurations with weak mixing between them and is divided into sectors of configuration~$A$ normal states (in blue, left) and configuration~$B$  intruder states (in black, right).

For $^{92,94}$Zr the experimental energies are reproduced well, while the $E2$ transition rates are reproduced more qualitatively than quantitatively, due to the small boson number ($N=1,2$, respectively). We note that in configuration~$B$ some of the proposed U(5) multiplets are incomplete (see \cref{tab:92-98Zr-levels} in the Appendix for more details). For both $^{92,94}$Zr, there is no experimental $0^+$ state to correspond to the calculated $0^+$ ($n_d\!\approx\!2$) state. For $^{94}$Zr, there is no experimental $0^+$ state to correspond to the calculated $0^+$ ($n_d\!\approx\!3$) state and for $^{96}$Zr there is no experimental $3^+$ state to correspond to the calculated $3^+$ ($n_d\!\approx\!3$) state. 

For~$^{96}$Zr, the boson number is increased ($N\!=\!3$) and configuration~$B$ becomes more collective. The $\be{2}{1}{0}{1}\!=\!2.3(3)$ W.u. is reproduced well, suggesting single-particle characteristics for the $0^+_1$ and $2^+_1$ states. The transitions within configuration~$B$ states, $\be{2}{2}{0}{2}\!=\!36(11)$, $\be{4}{1}{2}{2}\!=\!56^{+20}_{-44}$, $\be{2}{3}{2}{2}\!<\!400$ and $\be{0}{3}{2}{2}\!=\!34(9)$~W.u. are all reproduced well by the calculation and conform with the IBM-CM interpretation of quasi-phonon structure for configuration~$B$. The experimental transitions between the configurations, $\be{4}{1}{2}{1}\!=\!16^{+5}_{-13}$ and $\be{2}{2}{0}{1}\!=\!0.26(8)$~W.u. do not conform well to the calculated values of 3 (which is within the error rage) and 0.001 W.u. This suggests that perhaps a larger value for the $\omega$-mixing term in \cref{eq:mixing} could be used. Such an increase in $\omega$ (from 0.02 to 0.04 MeV), with only a minute variation to $\Delta_p$, results in a significant increase of the calculated values, placing them within the experimental error range, while keeping the rest of the calculated transitions approximately the same. Nevertheless, in such a scenario, the mixing between configuration~$A$ and $B$ $0^+_1$ and $0^+_2$ states, respectively, is still very weak.
Above the energy of the experimental states that correspond to the $n_d\!\approx\!2$ multiplet, it is more difficult to assign states to a certain phonon-multiplet due to the lack of data. Specifically, the experimental $8^+_1$ has a dominant branch to the $6^+_4$, which in turn has a dominant branch to the $4^+_1$ \cite{Witt2019} and therefore are assigned to configuration~$B$. Accordingly, they correspond to the calculated states with dominant $n_d\!\approx\!3~(6^+_1)$ and $n_d\!\approx\!4~(8^+_1)$ components. \paragraph*{Wave functions.}
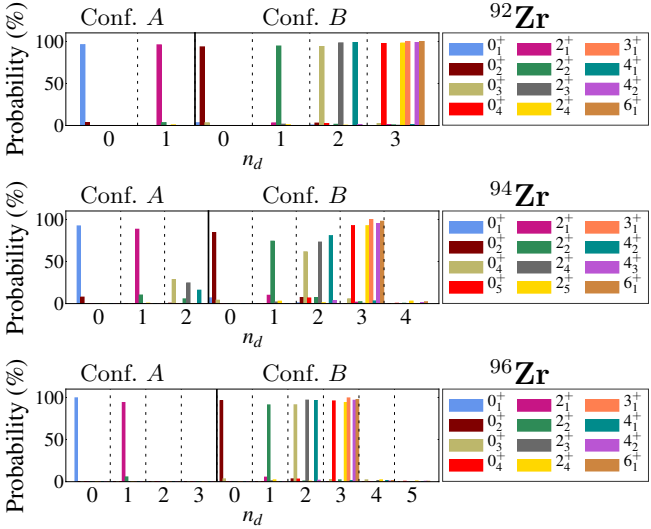
\begin{figure}[t]
\begin{overpic}[width=1\linewidth]{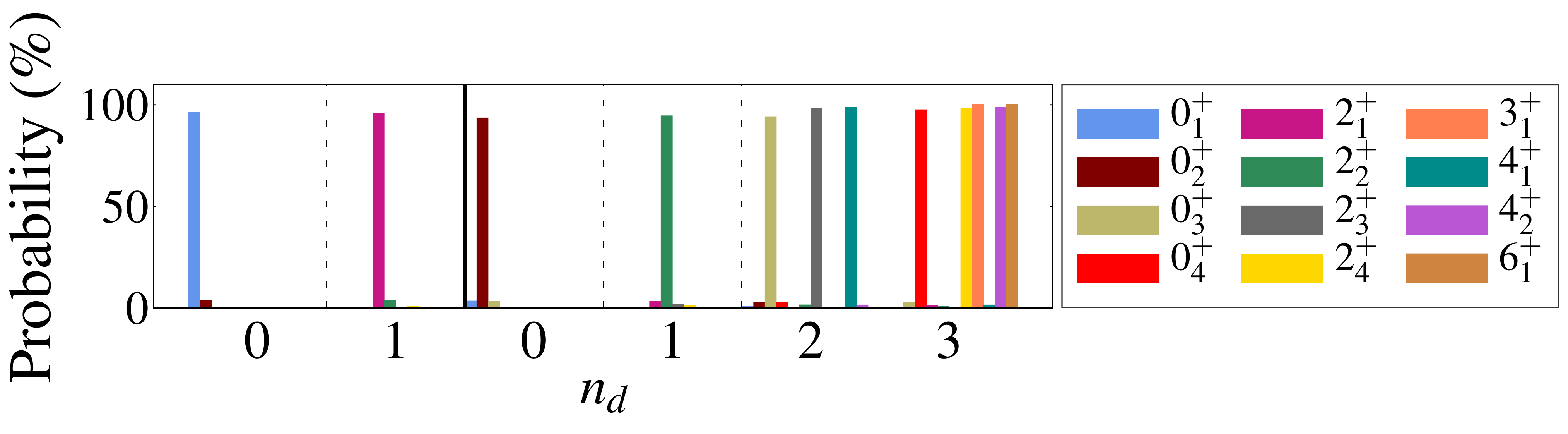}
\put (75,23) {\large {\bf $^{92}$Zr}}
\put (12,23.5) {\small Conf.~$A$}
\put (40,23.5) {\small Conf.~$B$}
\end{overpic}\\
\begin{overpic}[width=1\linewidth]{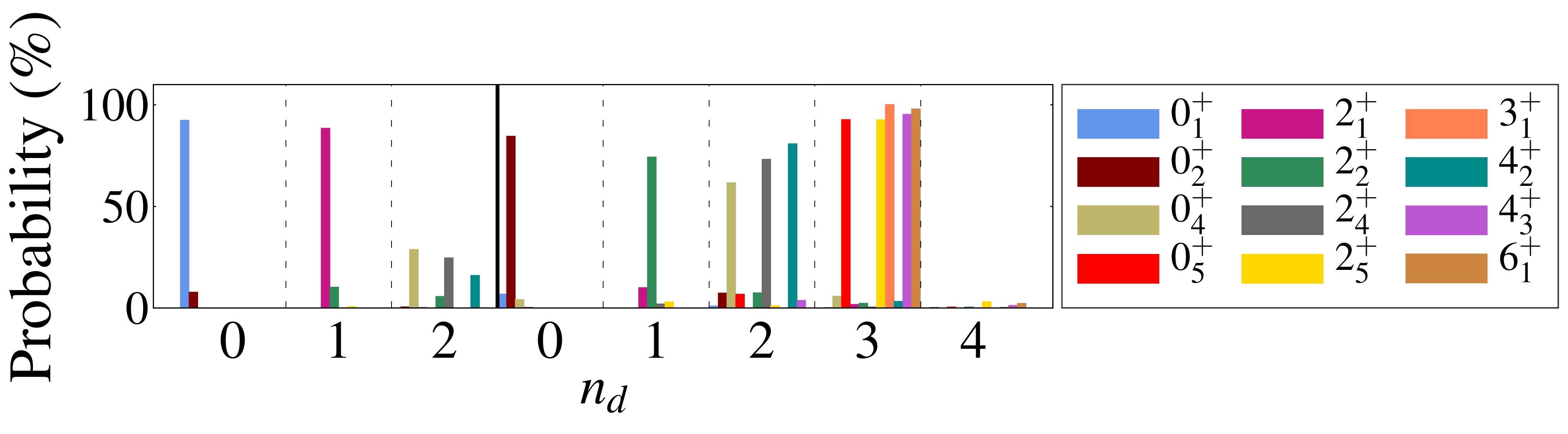}
\put (75,23) {\large {\bf $^{94}$Zr}}
\put (12,23.5) {\small Conf.~$A$}
\put (40,23.5) {\small Conf.~$B$}
\end{overpic}\\
\begin{overpic}[width=1\linewidth]{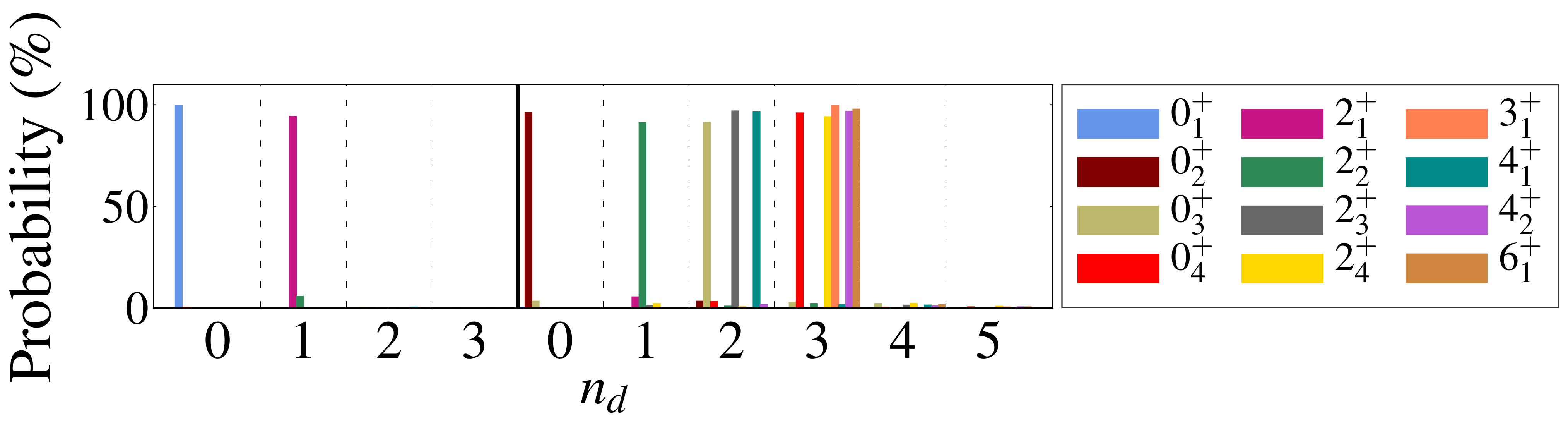}
\put (75,23) {\large {\bf $^{96}$Zr}}
\put (12,23.5) {\small Conf.~$A$}
\put (40,23.5) {\small Conf.~$B$}
\end{overpic}\\
\caption{U(5) $n_d$-decomposition of eigenstates of the Hamiltonian~\eqref{eq:ham-cm} for $^{92-96}$Zr. 
Each panel represents a single isotope and is divided into two parts, the decomposition within configuration~$A$ (left) and within configuration~$B$ (right). For each isotope, the histograms shown from left-to-right correspond to the calculated $L_i$ states listed in the right legend in the order top-to-bottom left-to-right.
\label{fig:92-96Zr-decomp}}
\end{figure}
For $^{92-96}$Zr, the calculated ground state ($0^+_1$) has $b^2\!=\!3.9\%,7.7\%$ and 0.4\% and the $2^+_1$ state has $b^2\!=\!4.2\%,11.6\%,6.8\%$, respectively, hence they are assigned to the $(A)$ configuration.
The $0^+_2$ state is almost purely configuration $B$ lowest state with $b^2\!=\!96.3\%,91.8\%$ and $99.6\%$, respectively.
\cref{fig:92-96Zr-decomp} depicts the $n_d$-decompositions for the $A$ and $B$ configuration of each eigenstate. We observe a clear single dominant $n_d$-component for each of the states, with weak mixing between the different configurations, suggesting a spherical structure for both of them. The $0^+_1,2^+_1$ states belong to configuration~$A$.
For configuration $B$, which has collective attributes, the calculation suggests that the states are almost purely spherical, as is clearly seen in \cref{fig:92-96Zr-decomp}, with large $n_d\approx0,1,2,3$ components for the states $(0^+_2),(2^+_2),(4^+_1,2^+_3,0^+_3)$ and $(6^+_1,4^+_2,3^+_1,2^+_4,0^+_4)$, respectively. For $^{94}$Zr, it is the calculated $2^+_4$ and $4^+_3$ instead of the $2^+_3$ and $4^+_2$ state.
As seen in the middle panel of \cref{fig:92-96Zr-decomp}, larger mixing is observed in $^{94}$Zr for the calculated $n_d\!\approx\!2$ triplet, $4^+_2,2^+_4,0^+_4$ with $b^2\!=\!84\%,~75\%,~71\%$, respectively. The reason is that these states have a smaller energy difference from the normal $4^+_1,2^+_3,0^+_3$ states ($\approx\!0.4$ MeV) and thus mix with them more strongly, compared to $^{96}$Zr ($\approx1$ MeV). The stronger mixing scenario could be reduced by adding an $\hat n_d(\hat n_d-1)$ interaction to the normal Hamiltonian~\eqref{eq:ham_a} (see the Appendix for more details). For $^{92}$Zr, such normal states are not generated due to the small boson number ($N\!=\!1$) of configuration~$A$.

\subsection{The $^{98-102}$Zr region: IQPT}\label{sec:98-102zr-region}
\begin{figure*}[t]
\begin{overpic}[width=0.49\linewidth]{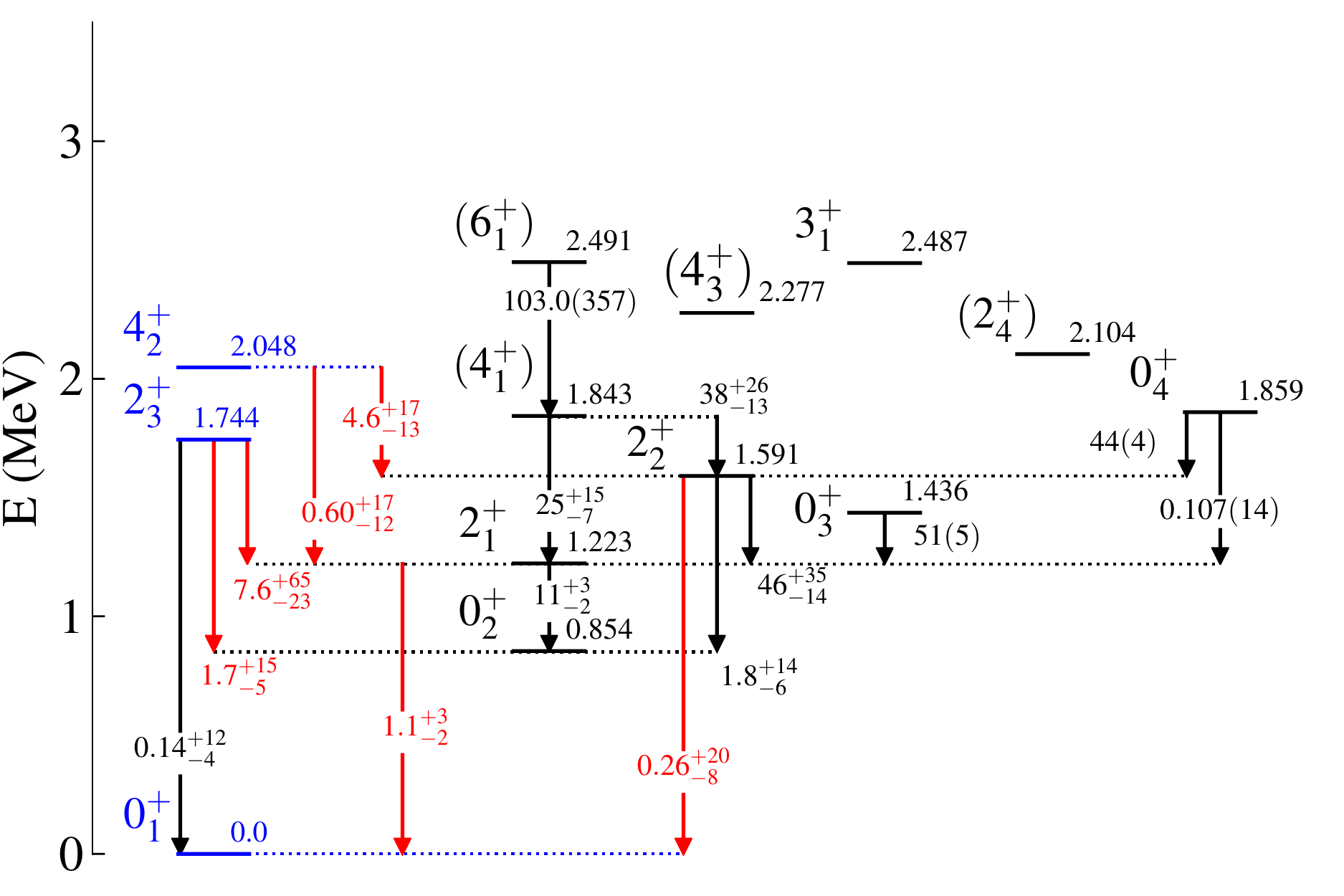}
\put (70,60) {\large (a) {\bf $^{98}$Zr exp}}
\end{overpic}
\begin{overpic}[width=0.49\linewidth]{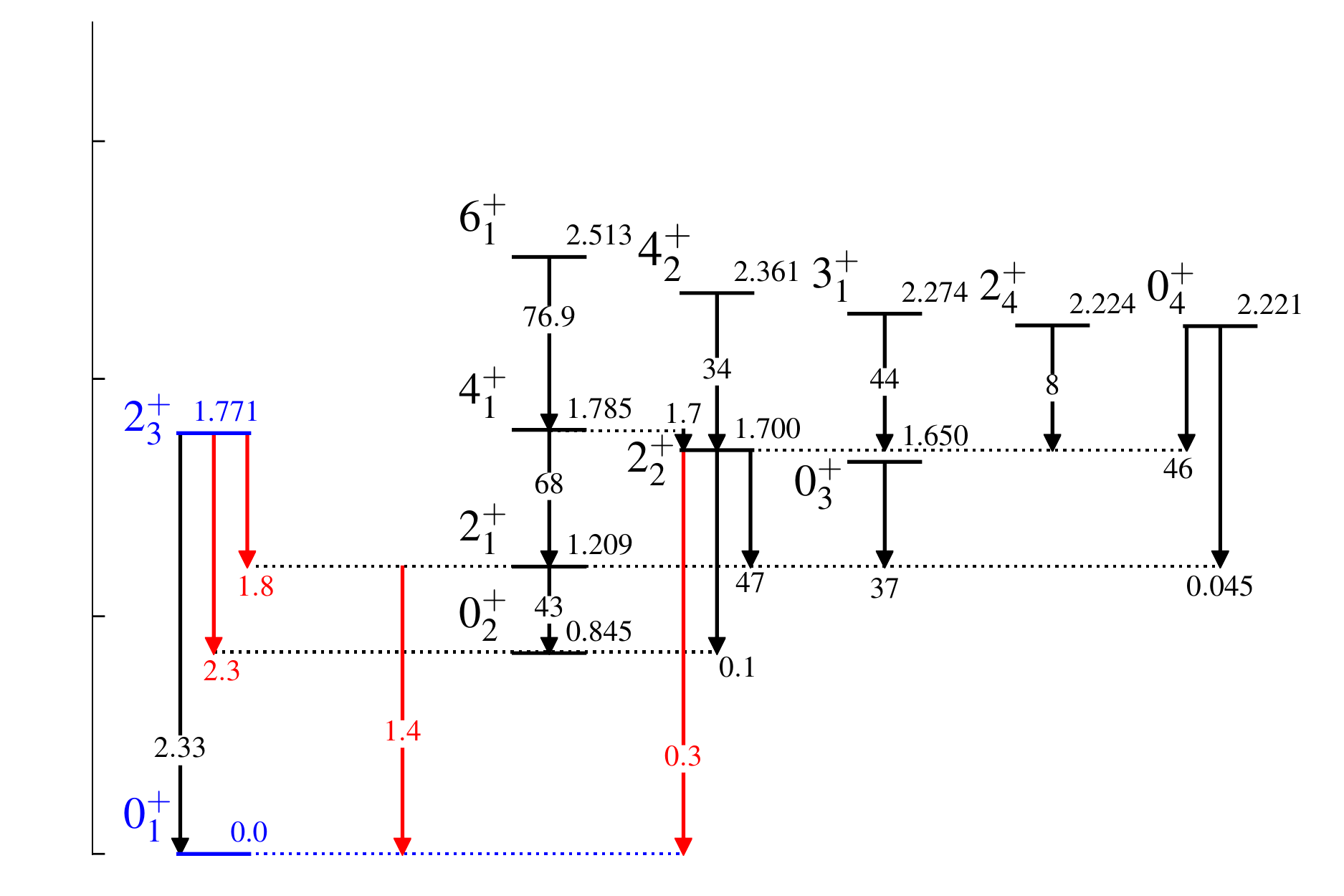}
\put (70,60) {\large (b) {\bf $^{98}$Zr calc}}
\end{overpic}\\
\begin{overpic}[width=0.49\linewidth]{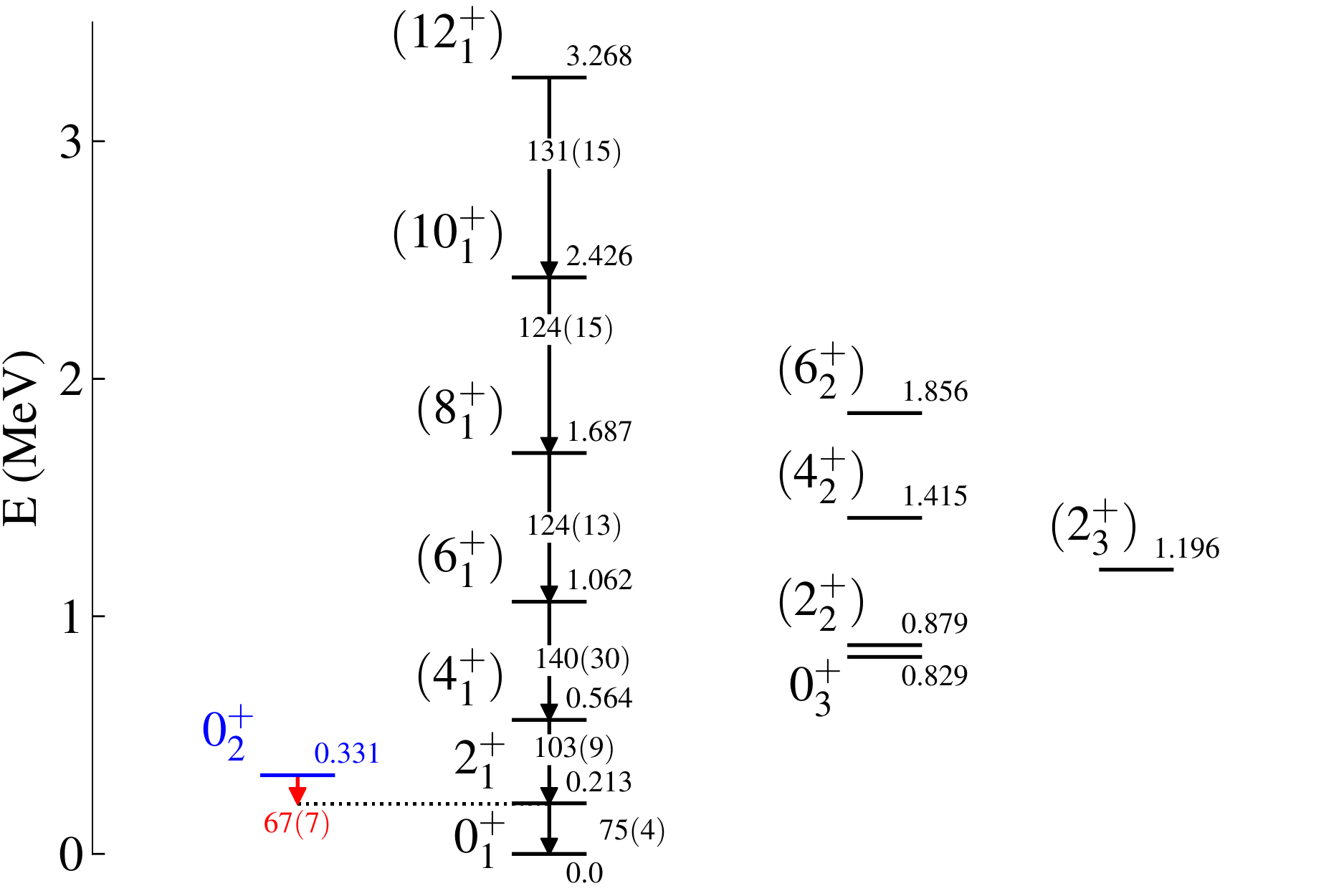}
\put (70,60) {\large (c) {\bf $^{100}$Zr exp}}
\end{overpic}
\begin{overpic}[width=0.49\linewidth]{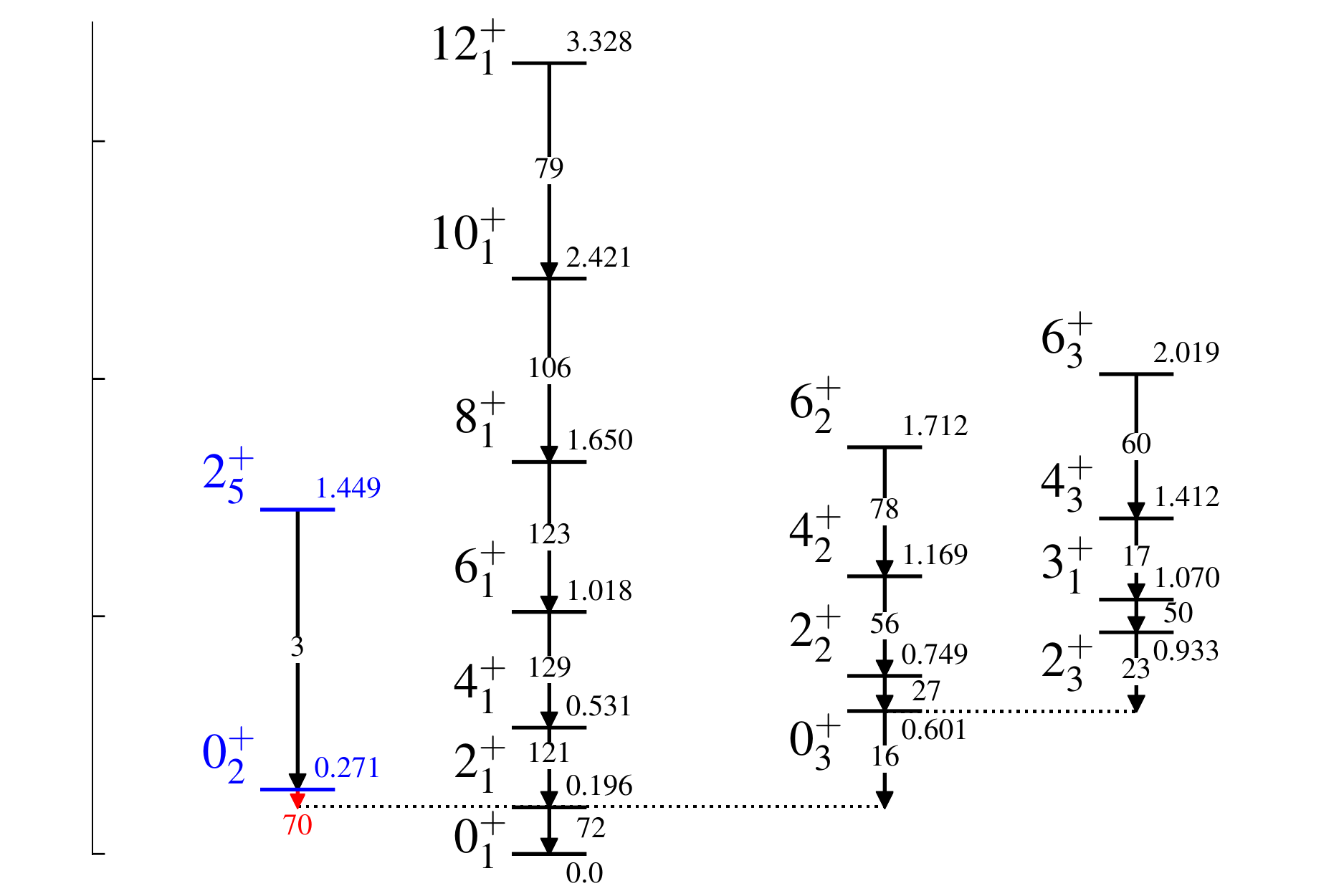}
\put (70,60) {\large (d) {\bf $^{100}$Zr calc}}
\end{overpic}\\
\begin{overpic}[width=0.49\linewidth]{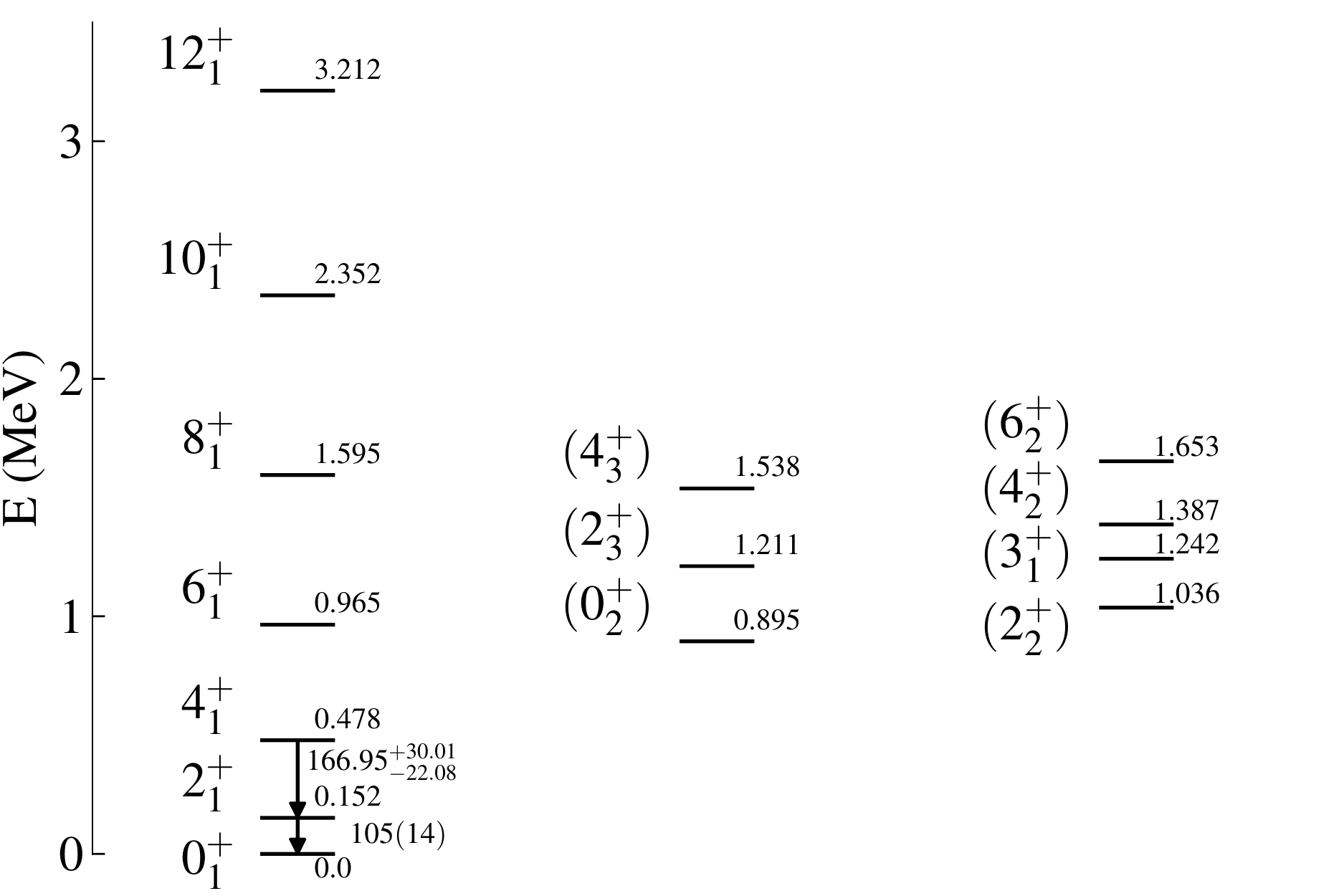}
\put (70,60) {\large (e) {\bf $^{102}$Zr exp}}
\end{overpic}
\begin{overpic}[width=0.49\linewidth]{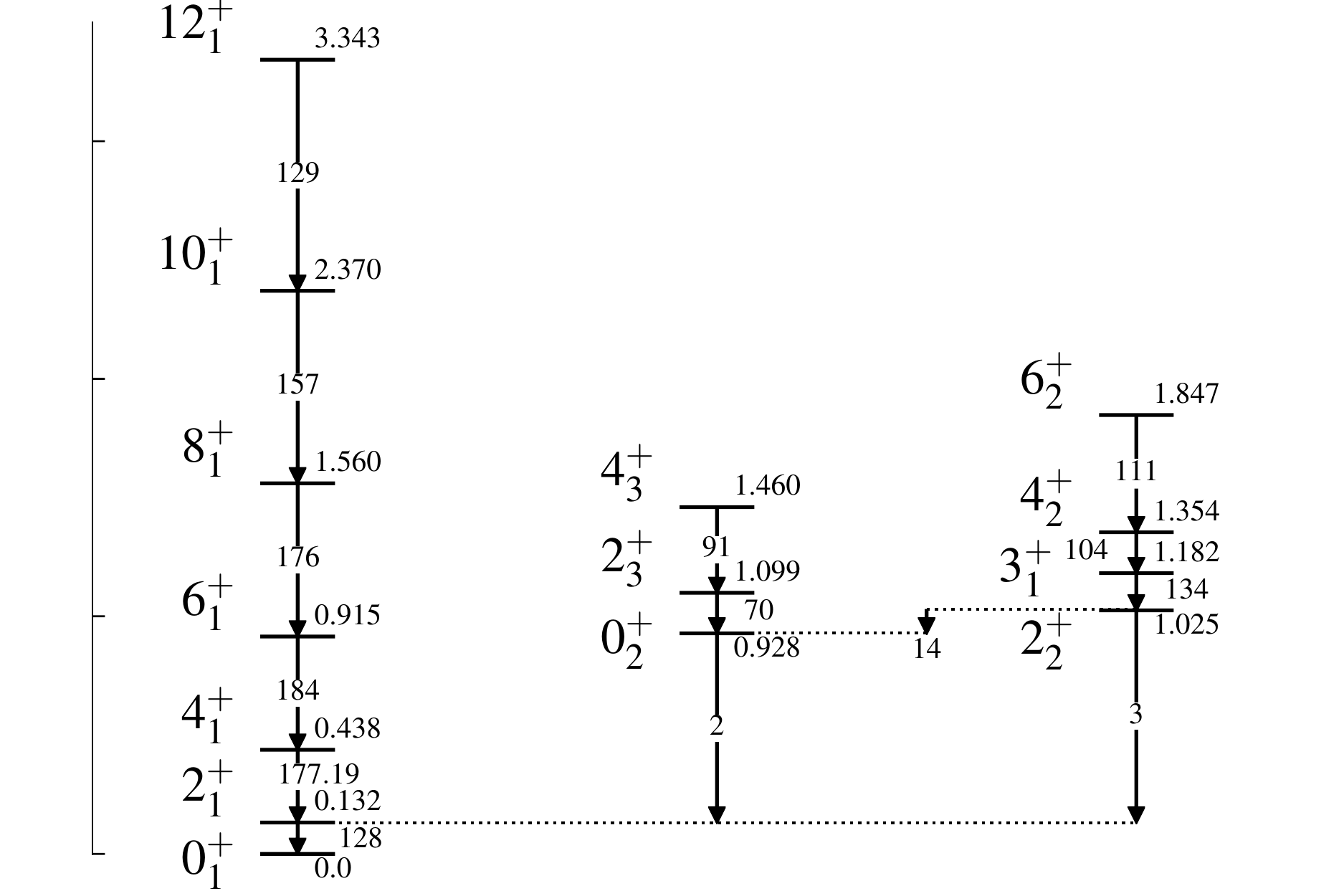}
\put (70,60) {\large (f) {\bf $^{102}$Zr calc}}
\end{overpic}
\caption{Experimental and calculated energy levels in MeV and $E2$ transition rates in W.u. Notation as in \cref{fig:92-96Zr-scheme}. Data are taken from \cite{NDS.164.1.2020,Karayonchev2020} for $^{98}$Zr, from \cite{NDS.109.297.2008,Ansari2017} for $^{100}$Zr and from \cite{NDS.110.1745.2009} for $^{102}$Zr.
\label{fig:98-102Zr-scheme}}
\end{figure*}
The spectrum of $^{98}$Zr, shown in Figs.~\eqref{fig:98-102Zr-scheme}(a) and \eqref{fig:98-102Zr-scheme}(b), exhibits coexistence of two configurations with weak mixing between them. Here the spectrum is divided into sectors of configuration~$A$ normal states (in blue, left), which are considered to be spherical, and configuration~$B$ intruder states (in black, right), which are considered to be weakly deformed (or quasispherical).
The experimental strong $E2$ rates $\be{0}{3}{2}{1}\!=\!51(5)$,  $\be{2}{2}{2}{1}\!=\!46^{+35}_{-14}$, $\be{4}{1}{2}{1}\!=\!25^{+15}_{-7}$,  $\be{0}{4}{2}{2}\!=\!44(4)$ and $\be{6}{1}{4}{1}\!=\!103.0(357)$~W.u. and weak $\be{0}{4}{2}{1}\!=\!0.107(14)$, $\be{2}{2}{0}{2}\!=\!1.8^{+14}_{-6}$~W.u. conform with the quasispherical interpretation for configuration~$B$. 
The experimental $E2$ rates with $\be{2}{3}{0}{2}\!=\!1.7^{+1.5}_{-0.5}$, $\be{2}{1}{0}{1}\!=\!1.1^{+3}_{-2}$, $\be{2}{2}{0}{1}\!=\!0.26^{+20}_{-8}$ and $\be{2}{3}{2}{1}\!=\!7.6^{+65}_{-23}$~W.u. conform with the interpretation of  $0^+_1$ and $2^+_3$ as normal $A$ configuration states with seniority-like single-particle character, weakly mixed with intruder $B$ configuration states. The experimental $E2$ rates $\be{2}{3}{2}{1}\!=\!7.6^{+65}_{-23}$ W.u. deviates from the calculated value of 1.8 W.u., however, a merely 1\% decrease of the parameter $\epsilon^{(A)}_d$ in the Hamiltonian~\eqref{eq:ham_a} results in a calculated value of 6.1 W.u. for this transition, without affecting significantly the remaining transitions in Figs.~\eqref{fig:98-102Zr-scheme}(a) and \eqref{fig:98-102Zr-scheme}(b). As mentioned in the Appendix, the experimental $4^+_2$ state is excluded from the IBM model space; however, the observed transition rates involving it, $\be{4}{2}{2}{1}\!=\!0.6^{+0.17}_{-0.12}$~W.u. and $\be{4}{2}{2}{2}\!=\!4.6^{+1.7}_{-1.3}$~W.u., support its assignment as a configuration~$A$ single-particle state, weakly mixed with configuration~$B$.

For the spectrum of $^{100}$Zr, shown in Figs.~\eqref{fig:98-102Zr-scheme}(c) and \eqref{fig:98-102Zr-scheme}(d), the spherical configuration~$A$ has now become excited and includes the calculated $0^+_2$ and $2^+_5$ states. The ground state band is associated with configuration~$B$. The close proximity of $0^+_2$ ($A$ configuration) and $0^+_1$ ($B$ configuration) suggests that $^{100}$Zr is near the critical-point of the Type~II QPT. Nevertheless, our description of energy levels and $B(E2)$ values is excellent.
One recognizes different ground, $\beta$ and $\gamma$ bands for configuration~$B$, with band heads $0^+_1,0^+_3,2^+_3$ respectively. The transition $\be{0}{2}{2}{1}\!=\!67(7)$ W.u. between the two configurations is described by the calculation (70 W.u.) to a very good agreement. For the calculated configuration-$A$ spherical $2^+_5$ state, one needs more experimental data such as $E2$ transitions to determine its exact location. We stress that the parameters employed were not optimized for this particular isotope, but rather were determined from a combined fit on the data of \textit{all} the isotopes in the chain and were varied smoothly (except $\Delta_p$) in a well defined manner (see \cref{app:fitting} for more details).

\begin{table}[h] 
\begin{center}
\caption{\label{tab:X5}
\small
Energies and $B(E2)$ values normalized to $E(2^+_1)\!=\!1$ and $\be{2}{1}{0}{1}\!=\!1$, respectively, for the X(5) critical-point symmetry~\cite{Iachello2001} and for the experimental values of $^{102}$Zr. The $0^+_{s=2},2^+_{s=2},4^+_{s=2}$ states correspond to the $0^+_2,2^+_2,4^+_2$ states of the X(5) model and to the experimental $0^+_2,2^+_3,4^+_3$ states of $^{102}$Zr.}
\begin{tabular}{lcc}
\hline
		       &  X(5)  & $^{102}$Zr exp \\
\hline
$E(4^+_1)$     &  2.91 &  3.15 \\
$E(6^+_1)$     &  5.45 &  6.36 \\
$E(8^+_1)$     &  8.51 & 10.51 \\
$E(10^+_1)$    & 12.07 & 15.49 \\
$E(0^+_{s=2})$ &  5.67 &  5.89 \\
$E(2^+_{s=2})$ &  7.48 &  7.98 \\
$E(4^+_{s=2})$ & 10.72 & 10.13 \\
\hline
$\be{4}{1}{2}{1}$  & 1.58 & 1.59 \\
\hline 
\end{tabular}
\end{center}
\end{table}

The spectrum of $^{102}$Zr, shown in \cref{fig:98-102Zr-scheme}(e)-(f), exhibits clear ground-, $\beta$- and $\gamma$-rotational bands for configuration~$B$, while the spherical states of configuration~$A$ seem to lie higher in energy. The measured $E2$ transition, $\be{2}{1}{0}{1}=105(14)$~W.u. is reproduced reasonably well by the calculation (128~W.u.). This isotope appears to have features of the so-called X(5)~critical-point symmetry~\cite{Iachello2001}, similar to the case encountered for neutron number 90 \cite{Krucken2002, Casten2001, Dewald2003, Tonev2004, Caprio2002} for the Nd-Sm-Gd-Dy isotopes, where the symmetry changes from U(5) to SU(3).
As seen in \cref{tab:X5}, the normalized energies and $E2$ transition rates agree well with those of X(5). For the $B(E2)$ ratio involving the $4^+_1 \rightarrow 2^+_1$ transition, the empirical value for $^{102}$Zr is in perfect agreement with the X(5) value 1.58.
\paragraph*{Wave functions.}
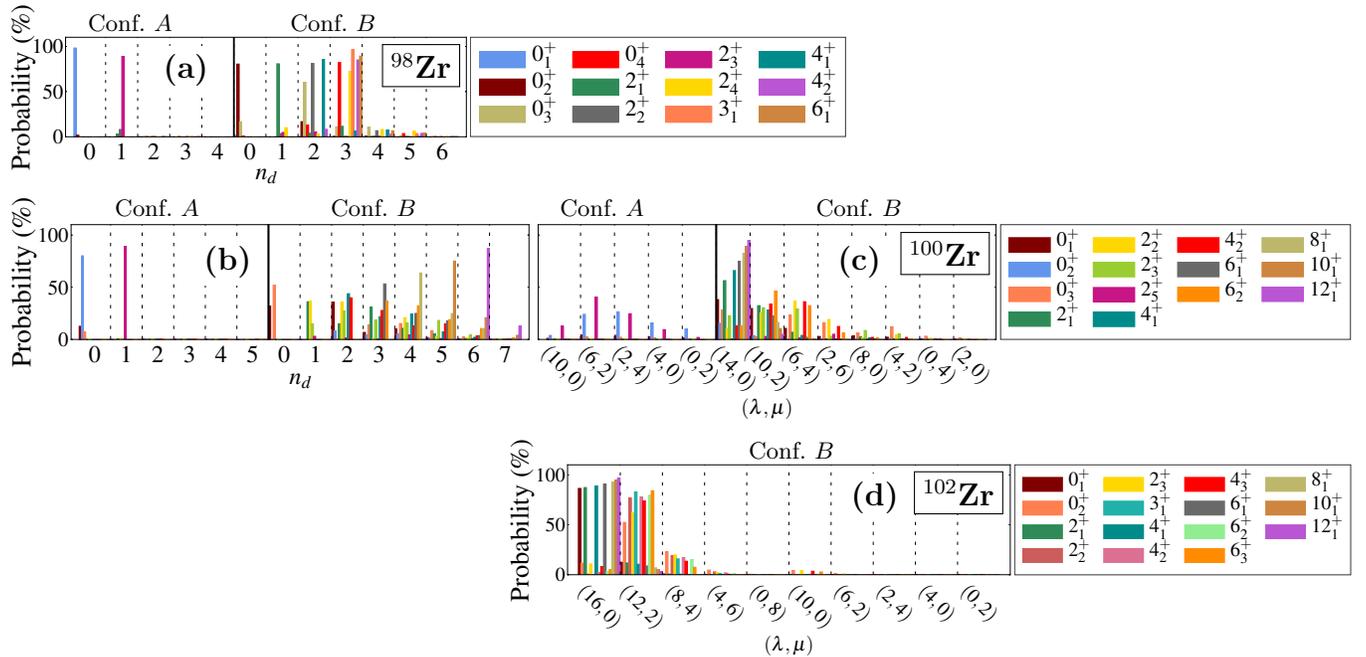
\begin{figure*}[thb]
\begin{overpic}[width=0.625\linewidth,left]{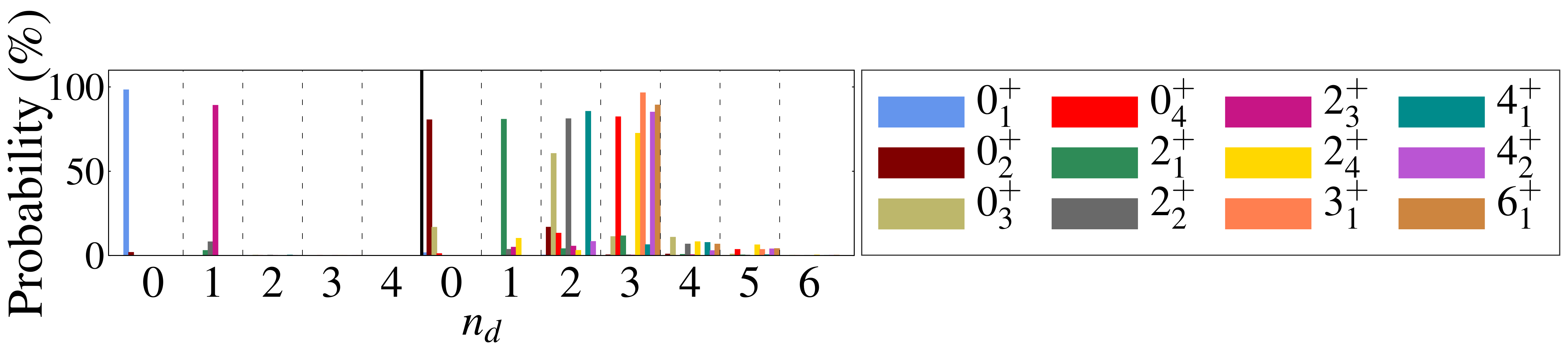}
\put (27.8,8.5) {\fcolorbox{black}{white}{\large {\bf $^{98}$Zr}}}
\put (11,8.5) {\colorbox{white}{\large {\bf (a)}}}
\put (6,12) {\small Conf.~$A$}
\put (21,12) {\small Conf.~$B$}
\end{overpic}\\
\begin{overpic}[width=1\linewidth]{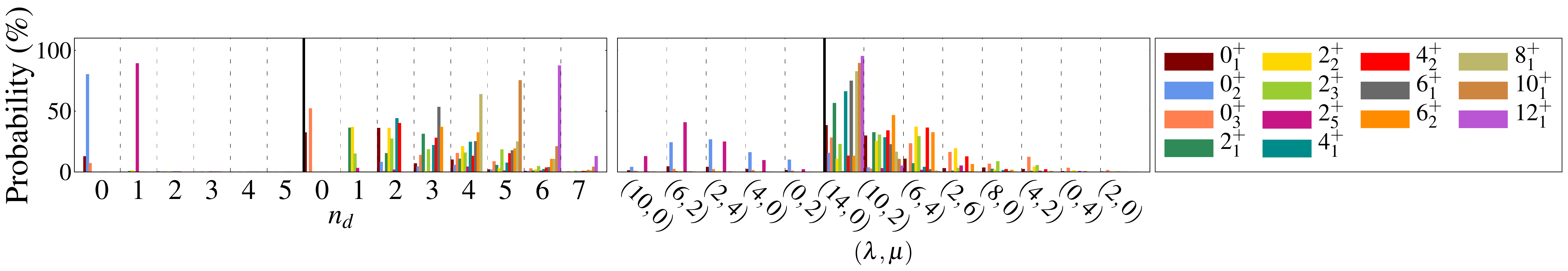}
\put (66.25,11.75) {\fcolorbox{black}{white}{\large {\bf $^{100}$Zr}}}
\put (14,11.5) {\colorbox{white}{\large {\bf (b)}}}
\put (61,11.5) {\colorbox{white}{\large {\bf (c)}}}
\put (8,15.5) {\small Conf.~$A$}
\put (24,15.5) {\small Conf.~$B$}
\put (41,15.5) {\small Conf.~$A$}
\put (60,15.5) {\small Conf.~$B$}
\end{overpic}\vspace{0.01\textheight}
\begin{overpic}[width=0.63\linewidth,right]{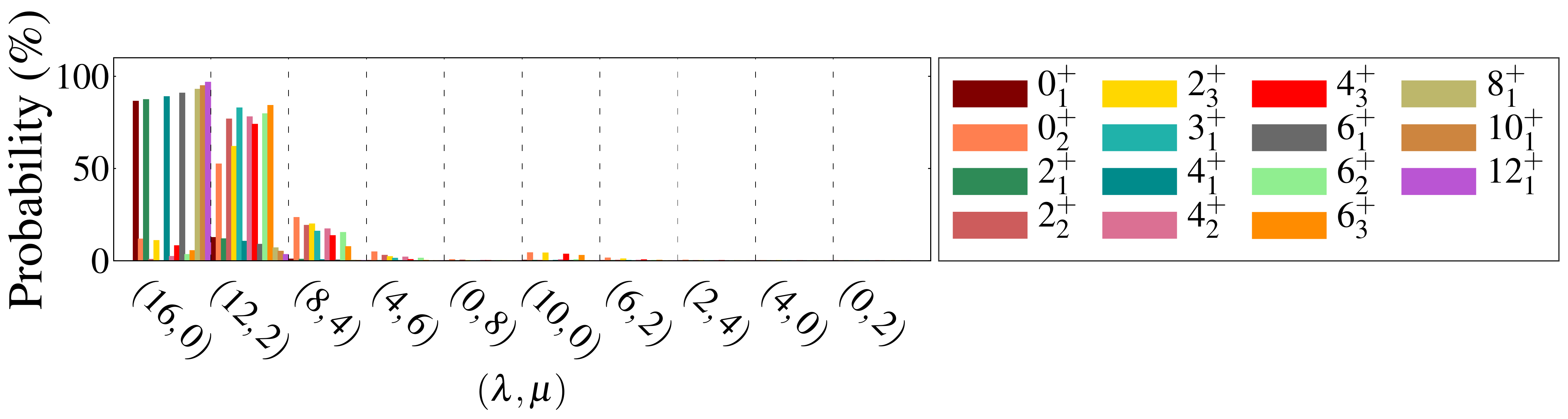}
\put (67.25,11.5) {\fcolorbox{black}{white}{\large \large {\bf $^{102}$Zr}}}
\put (62,11.5) {\colorbox{white}{\large {\bf (d)}}}
\put (55,15) {\small Conf.~$B$}
\end{overpic}
\caption{
U(5) $n_d$ decomposition for (a) $^{98}$Zr and (b) $^{100}$Zr, within configurations $A$ and $B$. SU(3) $(\lambda,\mu)$ decomposition for (c) $^{100}$Zr, within configurations $A$ and $B$ and for (d) $^{102}$Zr, only within configuration $B$. For each isotope, the order of the histograms is as in \cref{fig:92-96Zr-decomp}.
\label{fig:98-102Zr-decomp}}
\end{figure*}
For $^{98}$Zr, the $0^+_1$ state belongs to configuration $A$ and has a small configuration $B$ component, with $b^2\!=\!1.8\%$. 
For $^{100}$Zr, the $0^+_1$ state changes its configuration $B$ content and has $b^2\!=\!87.2\%$. The latter configuration-change is a clear evidence of Type II QPT, where $^{100}$Zr lies near the critical point. 
For $^{102}$Zr, the $0^+_1$ is almost pure configuration $B$ with $b^2\!=\!98.4\%$.
The $2^+_1$ state changed to configuration $B$ already in $^{98}$Zr, with $b^2\!=\!97.1\%$, as was pointed out in \cite{Witt2018}. For $^{102}$Zr, it is a pure configuration $B$ state with $b^2\!=\!99.9\%$.
The $0^+_2$ state is almost purely configuration $B$ lowest state in $^{98}$Zr, with $b^2\!=\!98.2\%$. For $^{100}$Zr, the $0^+_2$ becomes the lowest $A$ configuration state with $b^2\!=\!19.8\%$ and for $^{102}$Zr it becomes the first excited state within configuration $B$ with $b^2\!\approx\!100\%$.

The U(5) $n_d$ decomposition for $^{98}$Zr is given in panel (a) of \cref{fig:98-102Zr-decomp}. We see that the $0^+_1,2^+_3$ states have a single dominant $n_d$-component ($n_d\!\approx\!0,1$, respectively) in the configuration~$A$ side, which identifies them as spherical.
For states that belong to configuration~$B$, one can still see dominant single-$n_d$ components, with large $n_d\approx0,1,2,3$ components for the states $(0^+_2),(2^+_1),(4^+_1,2^+_2,0^+_3)$ and $(6^+_1,4^+_2,3^+_1,2^+_4,0^+_4)$, respectively. These components, however, are less dominant compared to the $^{92-96}$Zr case, \cref{fig:92-96Zr-decomp}. The calculation therefore suggests these states are weakly deformed or quasispherical.
For $^{100}$Zr, one observes dominant U(5) $n_d\!=\!0,1$ components for the configuration-$A$ $0^+_2$ and $2^+_5$ states in \cref{fig:98-102Zr-decomp}(b), while the configuration-$B$ states are spread amongst several $n_d$ values on the right side of the panel.
An SU(3) decomposition in \cref{fig:98-102Zr-decomp}(c) exhibits $(\lambda,\mu)$ components for configuration-$B$ states that become more dominant as $L$ increases. The reason for the latter is the decrease in the number of possible states to mix with in the IBM-CM model space. 
For $^{102}$Zr, \cref{fig:98-102Zr-decomp}(d) shows only the decomposition of configuration $B$, since all the indicated states belong to it. One sees that most states have a single dominant SU(3) component. Specifically, the ground-band states, $0^+_1,~2^+_1,~4^+_1,~6^+_1,~8^+_1,~10^+_1,~12^+_1$ have about $90\%$ dominant $(\lambda,\mu)=(2N+4,0)=(16,0)$ component. 
Altogether, the calculated decompositions, shown in \cref{fig:98-102Zr-decomp}, suggest the occurrence of IQPTs in this region. This involves a U(5) to SU(3) Type I QPT within configuration~$B$, along with a Type II QPT driving a change in structure of yrast states from configuration~$A$ to configuration~$B$.
\subsection{The $^{104-110}$Zr region: SU(3)-SO(6) crossover}\label{sec:104-110zr-region}
\begin{figure*}[htb]
\begin{overpic}[trim=0 0 110 0, clip, height=0.30\linewidth]{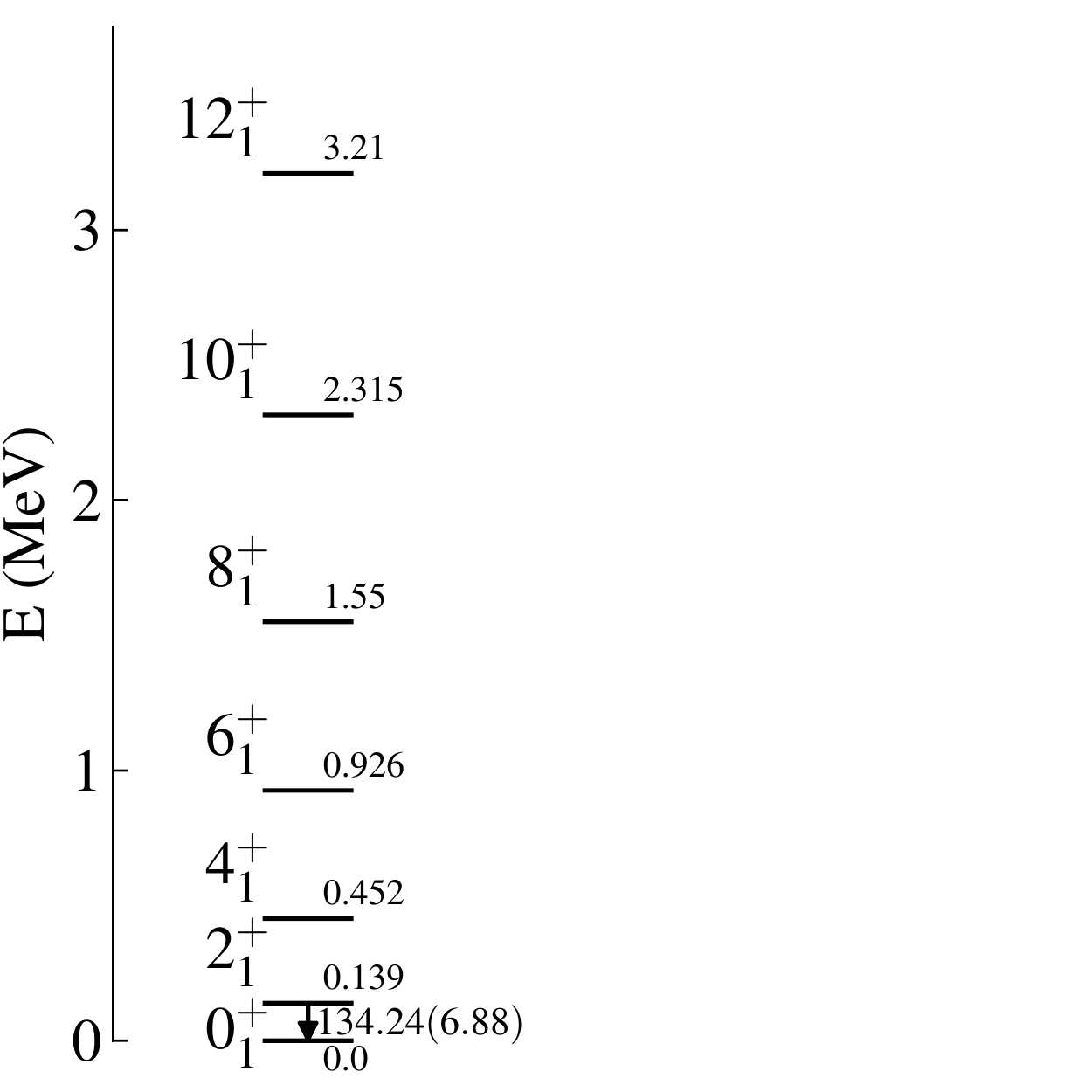}
\put (35,90) {\large {\bf $^{104}$Zr exp}}
\put (60,80) {\large {\bf (a)}}
\end{overpic}
\begin{overpic}[height=0.3\linewidth]{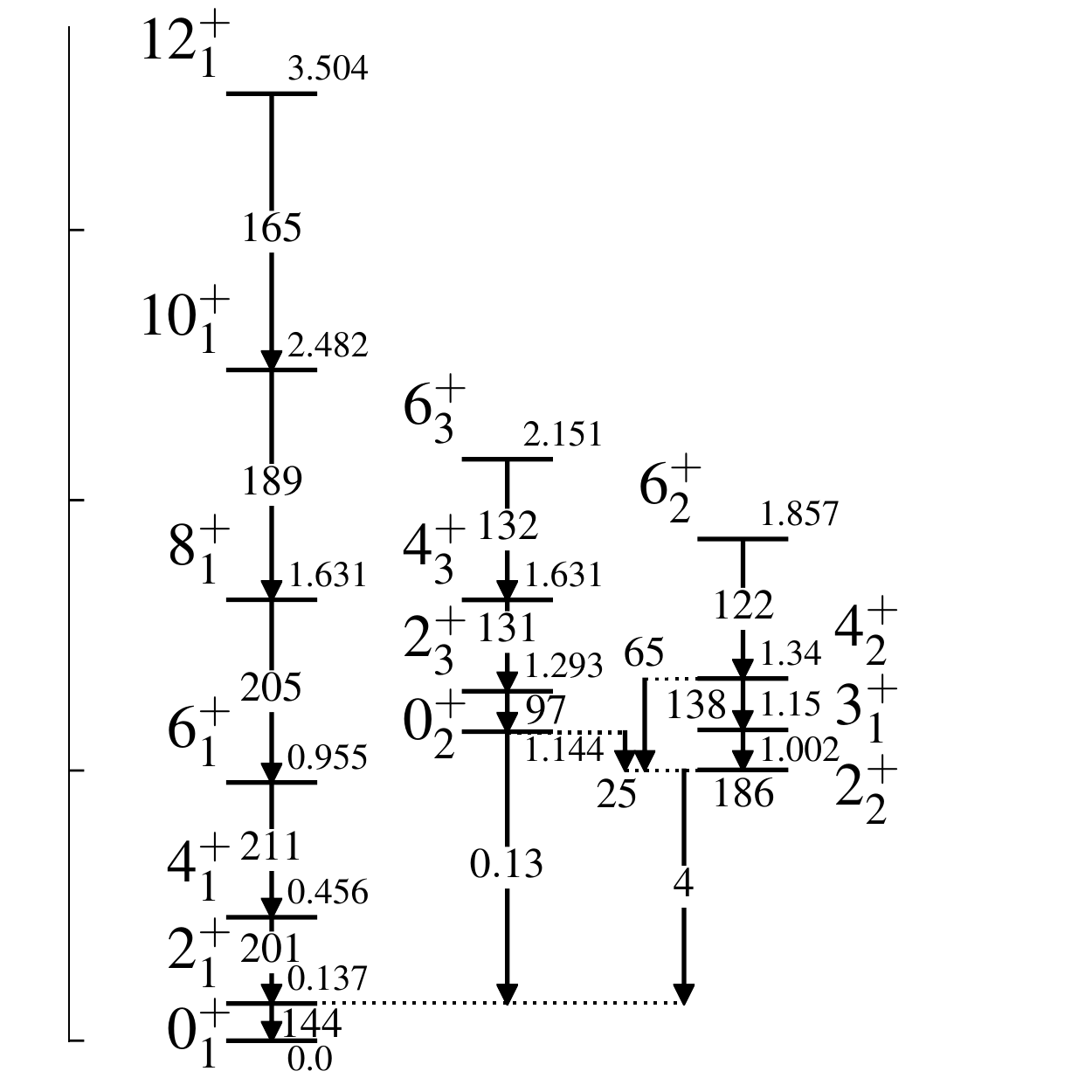}
\put (40,90) {\large {\bf $^{104}$Zr calc}}
\put (65,80) {\large {\bf (b)}}
\end{overpic}
\begin{overpic}[trim=34 0 110 0, clip, height=0.3\linewidth]{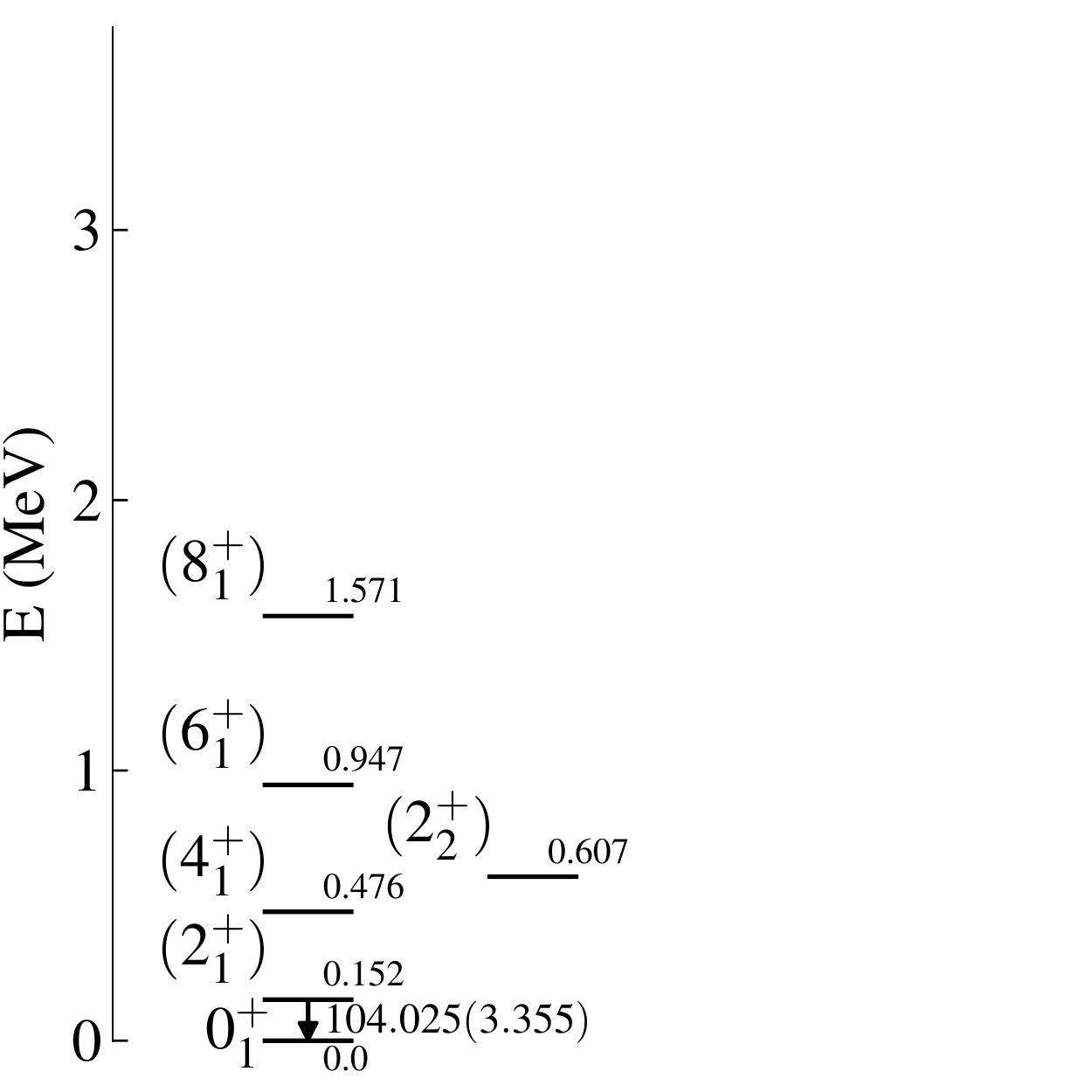}
\put (15,90) {\large {\bf $^{106}$Zr exp}}
\put (40,80) {\large {\bf (c)}}
\end{overpic}
\begin{overpic}[height=0.3\linewidth]{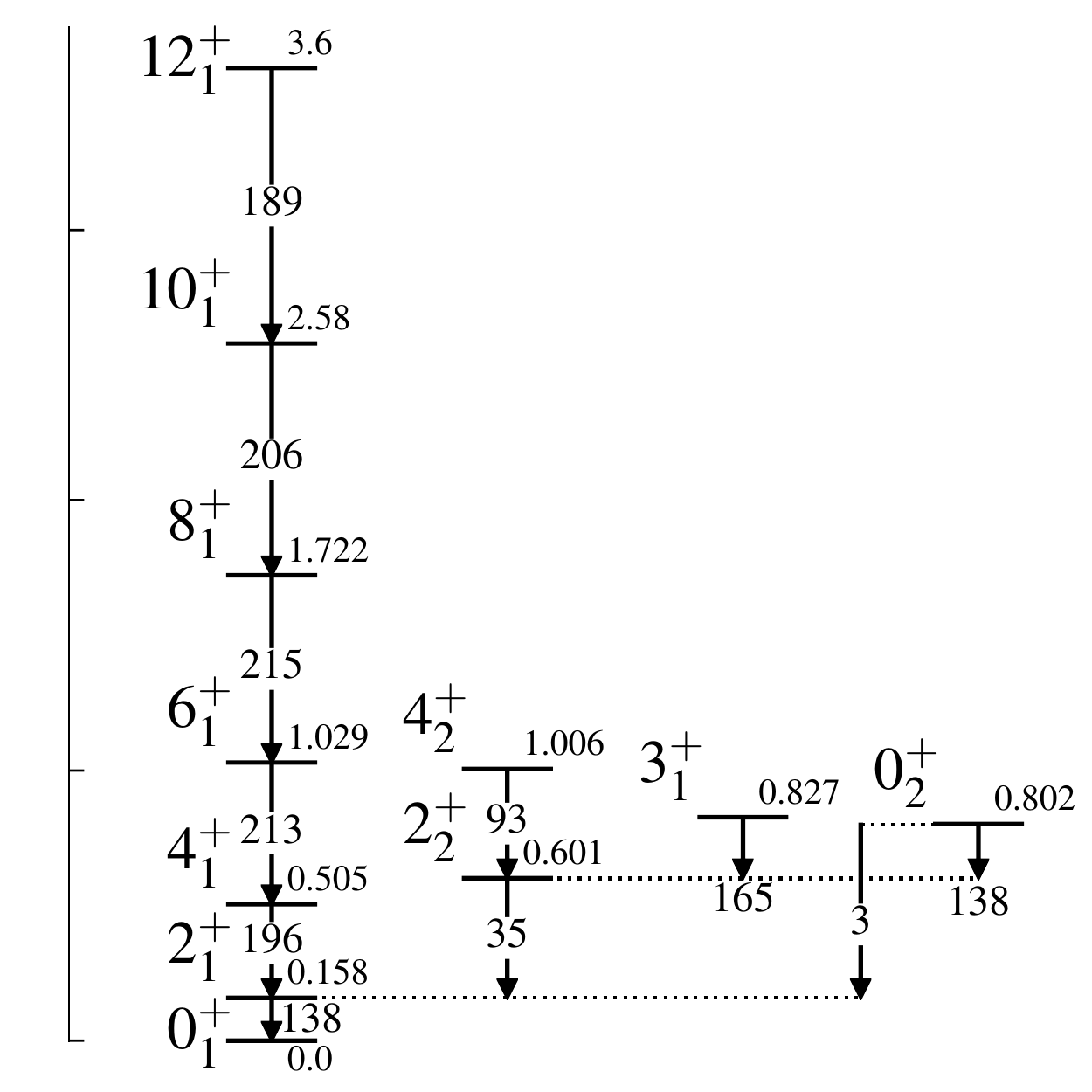}
\put (40,90) {\large {\bf $^{106}$Zr calc}}
\put (65,80) {\large {\bf (d)}}
\end{overpic}\\
\begin{overpic}[trim=0 0 110 0, clip, height=0.3\linewidth]{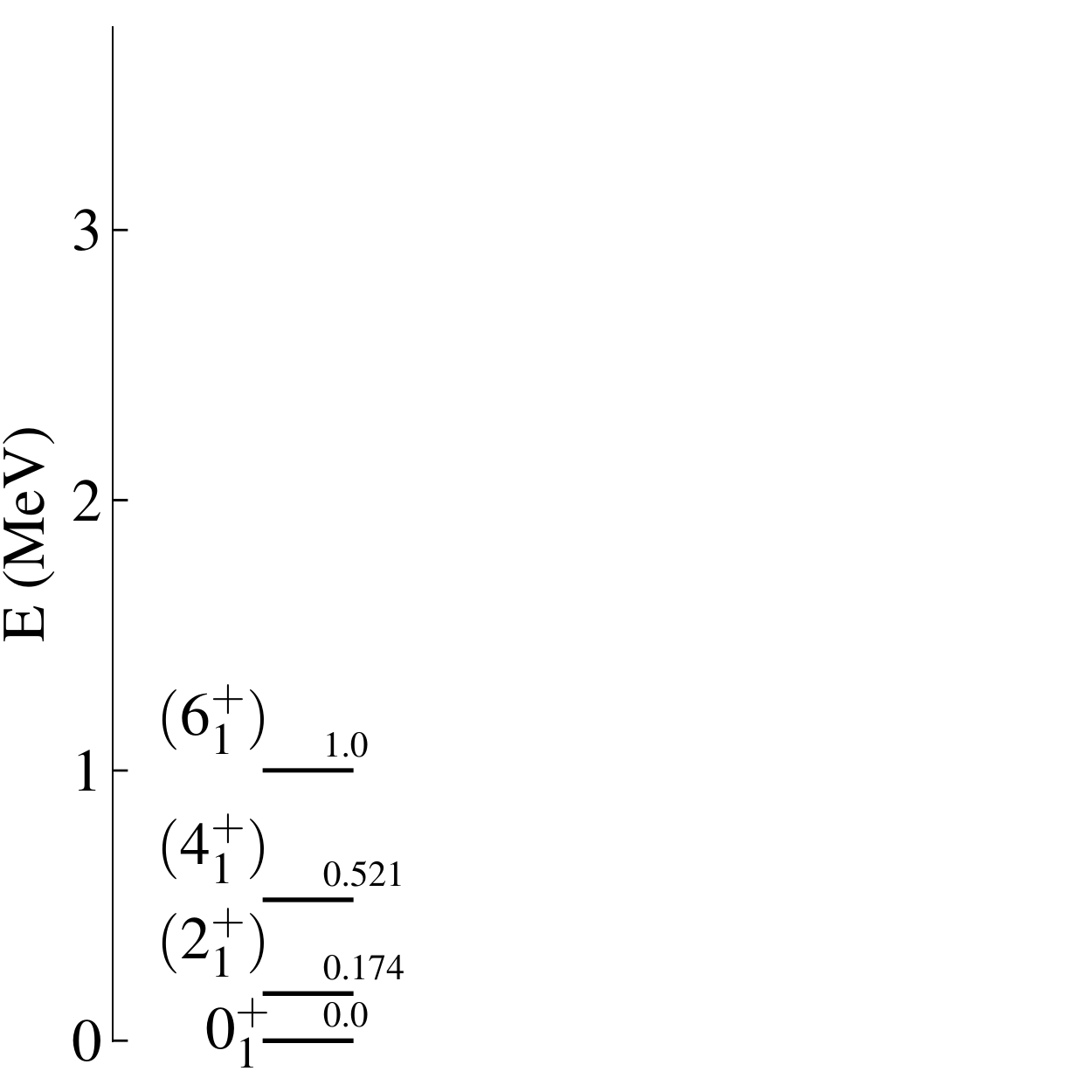}
\put (35,90) {\large {\bf $^{108}$Zr exp}}
\put (60,80) {\large {\bf (e)}}
\end{overpic}
\begin{overpic}[height=0.3\linewidth]{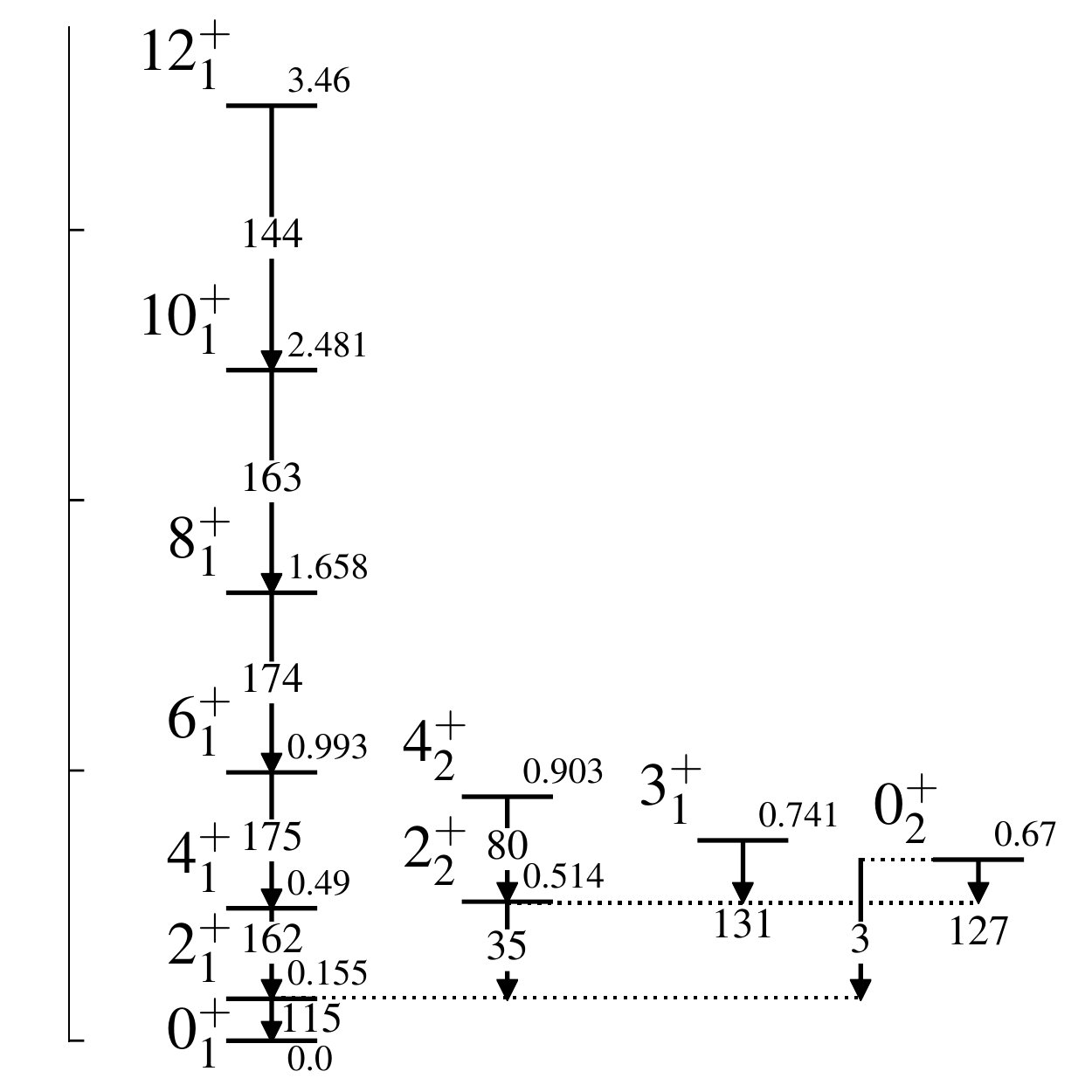}
\put (40,90) {\large {\bf $^{108}$Zr calc}}
\put (65,80) {\large {\bf (f)}}
\end{overpic}
\begin{overpic}[trim=34 0 110 0, clip, height=0.3\linewidth]{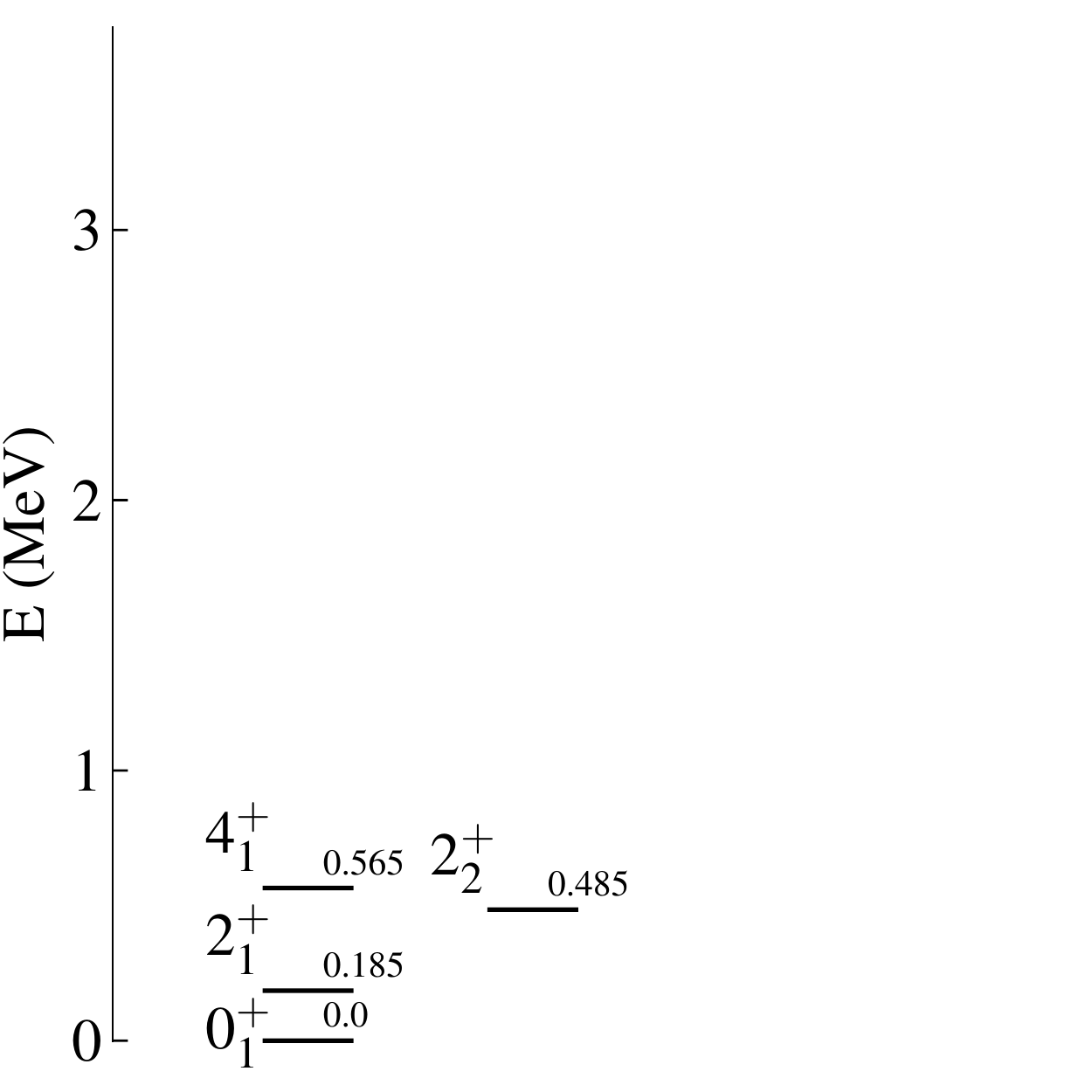}
\put (15,90) {\large {\bf $^{110}$Zr exp}}
\put (40,80) {\large {\bf (g)}}
\end{overpic}
\begin{overpic}[height=0.3\linewidth]{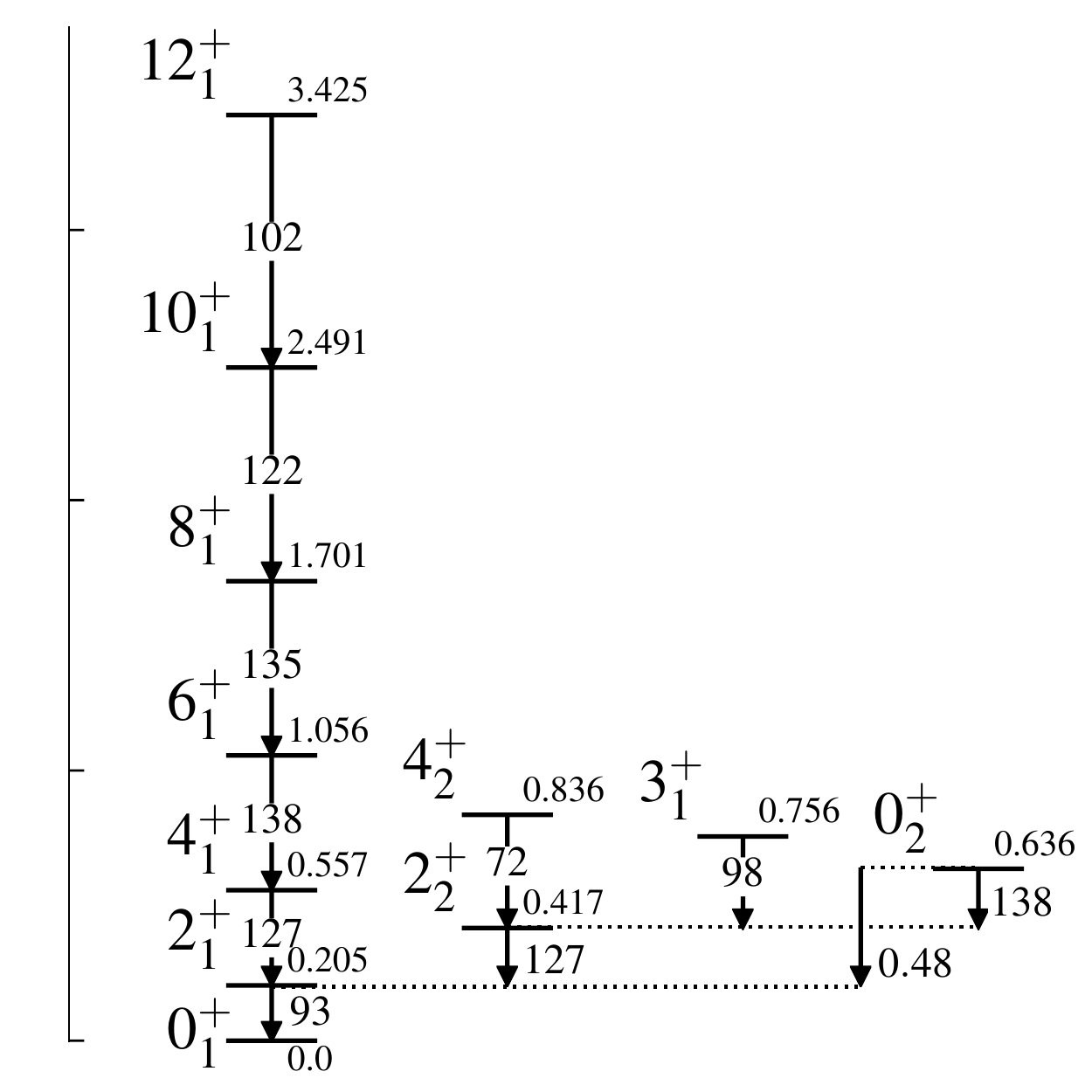}
\put (40,90) {\large {\bf $^{110}$Zr calc}}
\put (65,80) {\large {\bf (h)}}
\end{overpic}
\caption{Experimental and calculated energy levels in MeV and $E2$ transition rates in W.u. All levels belong to configuration~$B$. Data are taken from \cite{NDS.108.2035.2007,Browne2015b} ($^{104}$Zr), \cite{NDS.109.943.2008, Browne2015b} ($^{106}$Zr), \cite{NDS.90.135.2000} ($^{108}$Zr), and \cite{NDS.113.1315.2012,Paul2017} ($^{110}$Zr).
\label{fig:104-110Zr-scheme}}
\end{figure*}
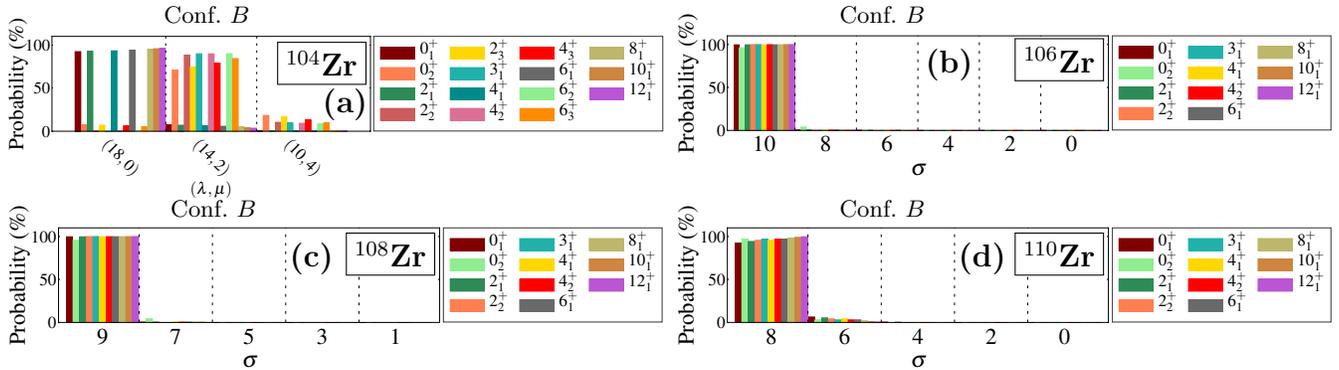
\begin{figure*}[bht]
\begin{overpic}[width=0.49\linewidth]{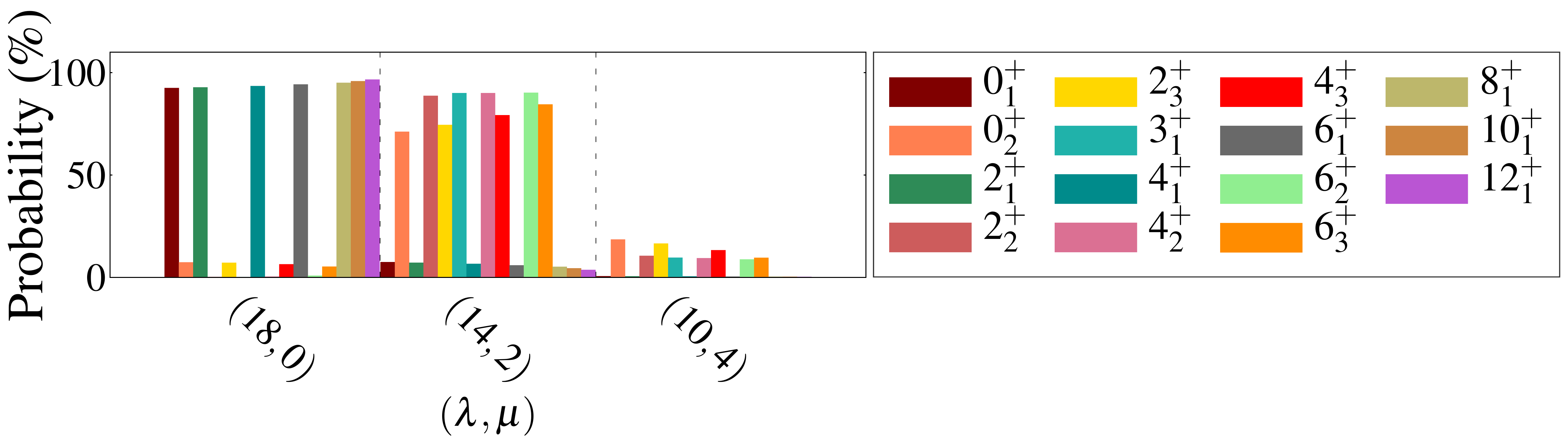}
\put (41.1,19) {\fcolorbox{black}{white}{\large {\bf $^{104}$Zr}}}
\put (48,13.5) {\large {\bf (a)}}
\put (24,27) {\small Conf.~$B$}
\end{overpic}
\begin{overpic}[width=0.49\linewidth,raise=0.01\textheight]{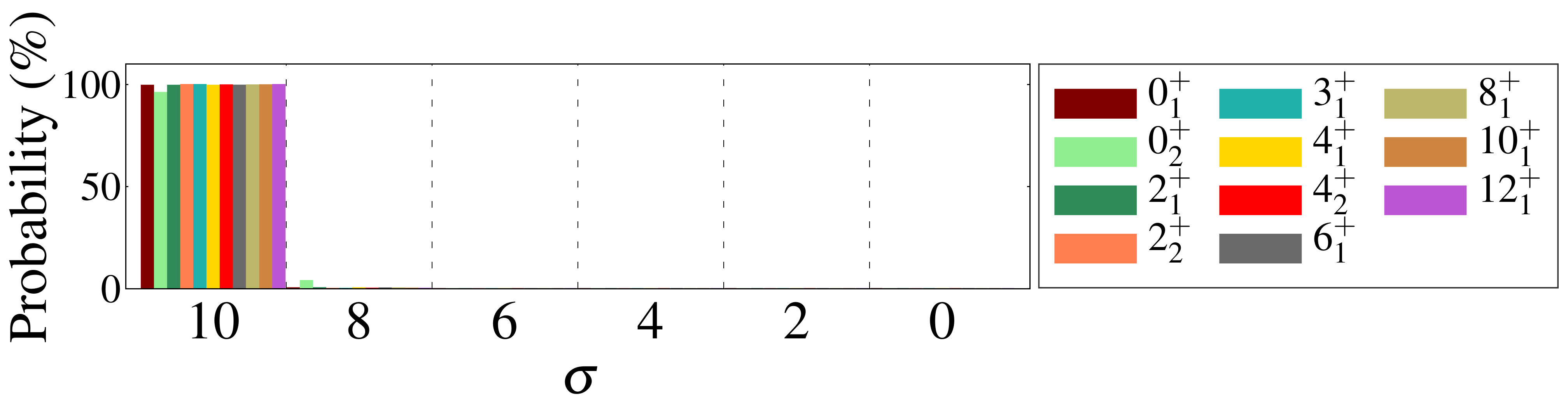}
\put (51.5,19.55) {\fcolorbox{black}{white}{\large {\bf $^{106}$Zr}}}
\put (38,19.55) {\large {\bf (b)}}
\put (25,27) {\small Conf.~$B$}
\end{overpic}\\
\begin{overpic}[width=0.49\linewidth]{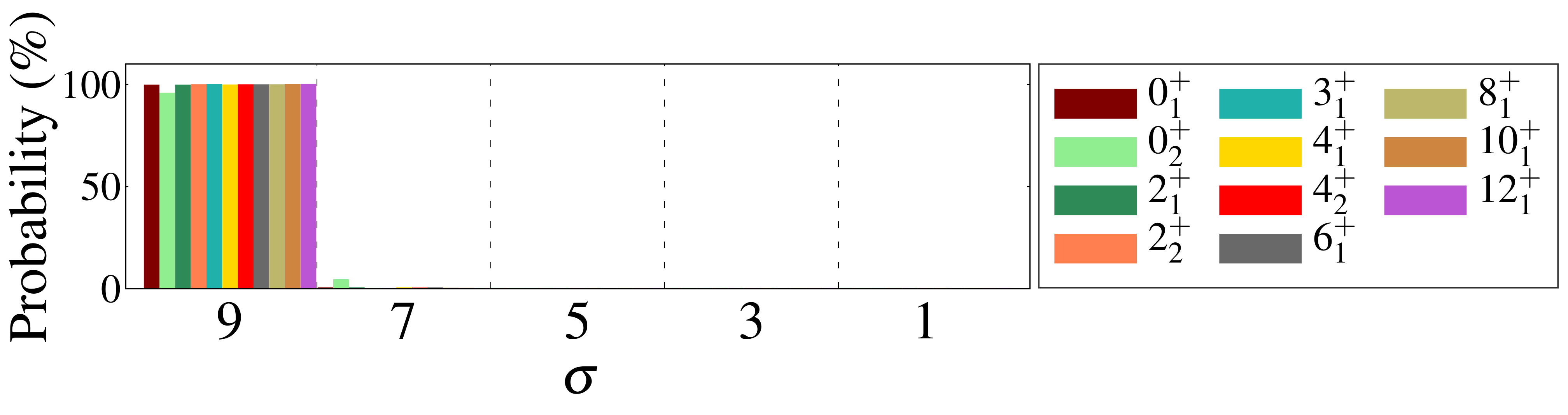}
\put (51.5,16.5) {\fcolorbox{black}{white}{\large {\bf $^{108}$Zr}}}
\put (43,16.5) {\large {\bf (c)}}
\put (25,24) {\small Conf.~$B$}
\end{overpic}
\begin{overpic}[width=0.49\linewidth]{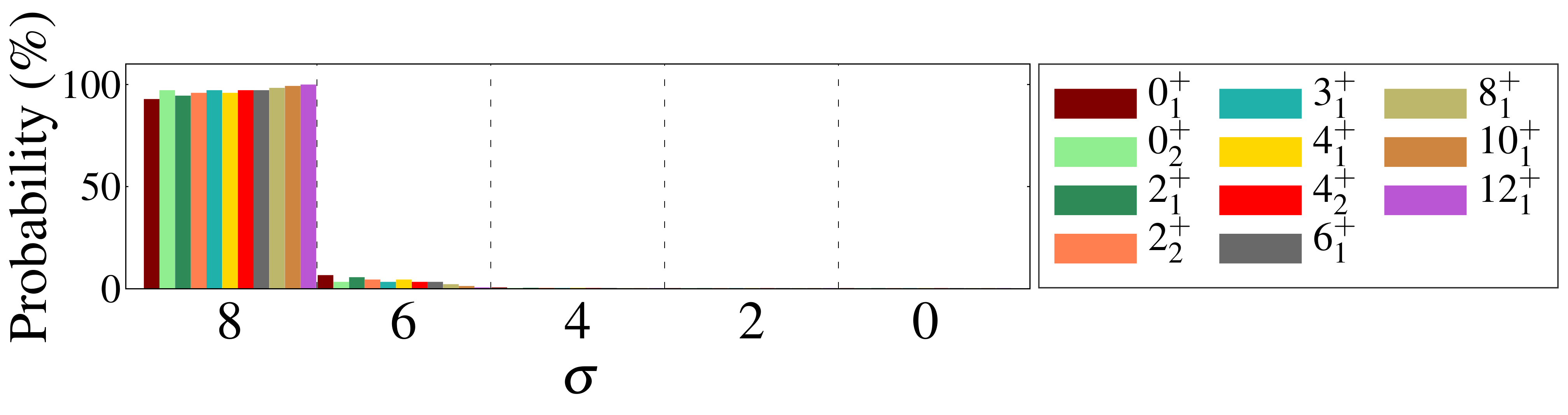}
\put (51.5,16.5) {\fcolorbox{black}{white}{\large {\bf $^{110}$Zr}}}
\put (43,16.5) {\large {\bf (d)}}
\put (25,24) {\small Conf.~$B$}
\end{overpic}
\caption{
SU(3) $(\lambda,\mu)$ and SO(6) $\sigma$ decomposition of eigenstates of the Hamiltonian~\eqref{eq:ham-cm} for $^{104}$Zr and for $^{106-110}$Zr, respectively. Each panel represents a single isotope and is divided into two parts: the decomposition within configuration~$A$ (left) and within configuration~$B$ (right). For $^{104}$Zr, only probabilities larger than 5\% are shown. For each isotope, the order of the histograms is as in \cref{fig:92-96Zr-decomp}.
\label{fig:104-110Zr-decomp}}
\end{figure*}

For the isotope $^{104}$Zr the deformation is expected to increase, since it is near mid-shell. This is reflected in the experimental value 134.24(6.88) W.u. for the $2^+_1\to0^+_1$ transition rates, as seen in \cref{fig:104-110Zr-scheme}(a). The calculated spectrum, shown in \cref{fig:104-110Zr-scheme}(b) suggests the existence of additional $\beta$ and $\gamma$bands with strong intra-band and weak inter-band transitions.
For $^{106}$Zr, considered in Figs.~\eqref{fig:104-110Zr-scheme}(c) and \eqref{fig:104-110Zr-scheme}~(d), a weaker experimental $B(E2; 2^+_1 \rightarrow 0^+_1) = 104.025(3.355)$ W.u. indicates a decrease in deformation. Furthermore, a low $2^+_2$ state at energy 607 keV, close to the $4^+_1$ at 476 keV, suggests a crossover from SU(3) to SO(6) symmetry (axial to nonaxial). A $2^+$ state close in energy to the $4^+_1$ is not observed in the spectrum of $^{108}$Zr, shown in \cref{fig:104-110Zr-scheme}(e). However, such $2^+_2$ state is seen in $^{110}$Zr at energy 485 keV, next to the $4^+_1$ at 565 keV [see Figs.~\eqref{fig:104-110Zr-scheme}(g) and \eqref{fig:104-110Zr-scheme}(h)]. Consequently, the calculation for these isotopes suggests these low-lying states are part of an SO(6) multiplet.
\paragraph*{Wave functions.}
For $^{104-110}$Zr, all states shown in \cref{fig:104-110Zr-scheme} are almost pure configuration $B$ states, with ${b^2\!\gtrsim\!99\%}$. Therefore, we concentrate on decompositions of the $B$ configuration part of the wave function, \cref{eq:wf}.
For $^{104}$Zr, we show in Fig.~\eqref{fig:104-110Zr-decomp}(a) the SU(3) $(\lambda,\mu)$-decomposition, \cref{eq:decomp-su3}. The $0^+_1,~2^+_1,~4^+_1,~6^+_1,~8^+_1,~10^+_1,~12^+_1$ states have about $93\%$ dominant $(\lambda,\mu)=(2N+4,0)=(18,0)$ component. For $^{106-110}$Zr, Figs.~\eqref{fig:104-110Zr-decomp}(b) and \eqref{fig:104-110Zr-decomp}(d) depict the SO(6) $\sigma$-decomposition, \cref{eq:decomp-so6}, for which a single dominant component ($\sigma\!=\!N+2$) is apparent for all isotopes. These states also have good SO(5) symmetry (see \cref{sec:evo-sym} below for more details).
The change in configuration $B$ from dominant $(\lambda,\mu)$ components in $^{104}$Zr to dominant $\sigma$-components in $^{106-110}$Zr, suggests that a crossover from SU(3) to SO(6) occurs in this region.

\section{Results: Evolution of wave functions and order parameters}\label{sec:results-evo}
\subsection{Evolution of configuration content}\label{sec:evo-conf}
\begin{figure}[t]
\includegraphics[width=1\linewidth]{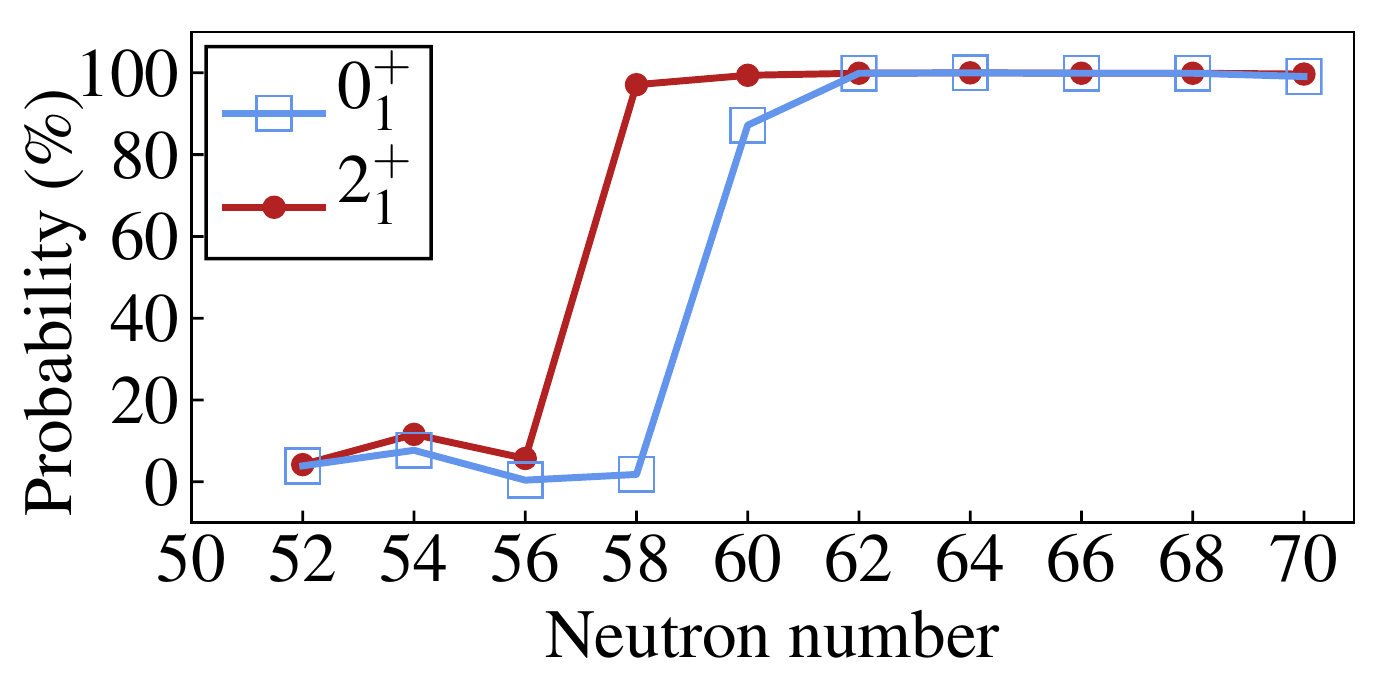}
\caption{Percentage of the wave functions within the intruder B-configuration [the $b^2$ probability in Eq.~\eqref{eq:wf}], for the ground ($0^+_1$) and excited ($2^+_1$) states in $^{92-110}$Zr.\label{fig:mixing}}
\end{figure}
Information on configuration changes for each isotope, can be inferred from the evolution of the probabilities $a^2$ or $b^2$, \cref{eq:decomp-cm}, of the states considered. Figure \eqref{fig:mixing} shows the percentage of the wave function within the $B$ configuration for the ground state ($0^+_1$) and first-excited state ($2^+_1$) as a function of neutron number across the Zr chain. 
The rapid change in structure of the $0^+_1$ state from the normal $A$ configuration in $^{92-98}$Zr (small $b^2$ probability) to the intruder $B$ configuration in $^{100-110}$Zr (large $b^2$ probability) is clearly evident, signaling a Type II QPT, mentioned in \cref{sec:98-102zr-region}.
The configuration change appears sooner in the $2^+_1$ state, which changes to
configuration $B$ already in $^{98}$Zr, in line with \cite{Witt2018}.
Outside a narrow region near neutron number 60, where the crossing occurs, the two configurations are weakly mixed and the states retain a high level of purity, especially for neutron number larger than 60.

\subsection{Evolution of symmetry content}\label{sec:evo-sym}
\definecolor{GreenNoam}{rgb}{0,0.5,0}
\definecolor{GrayNoam}{rgb}{0.8,0.8,0.9003921568627451}
\begin{figure*}[t]
\includegraphics[width=1\linewidth]{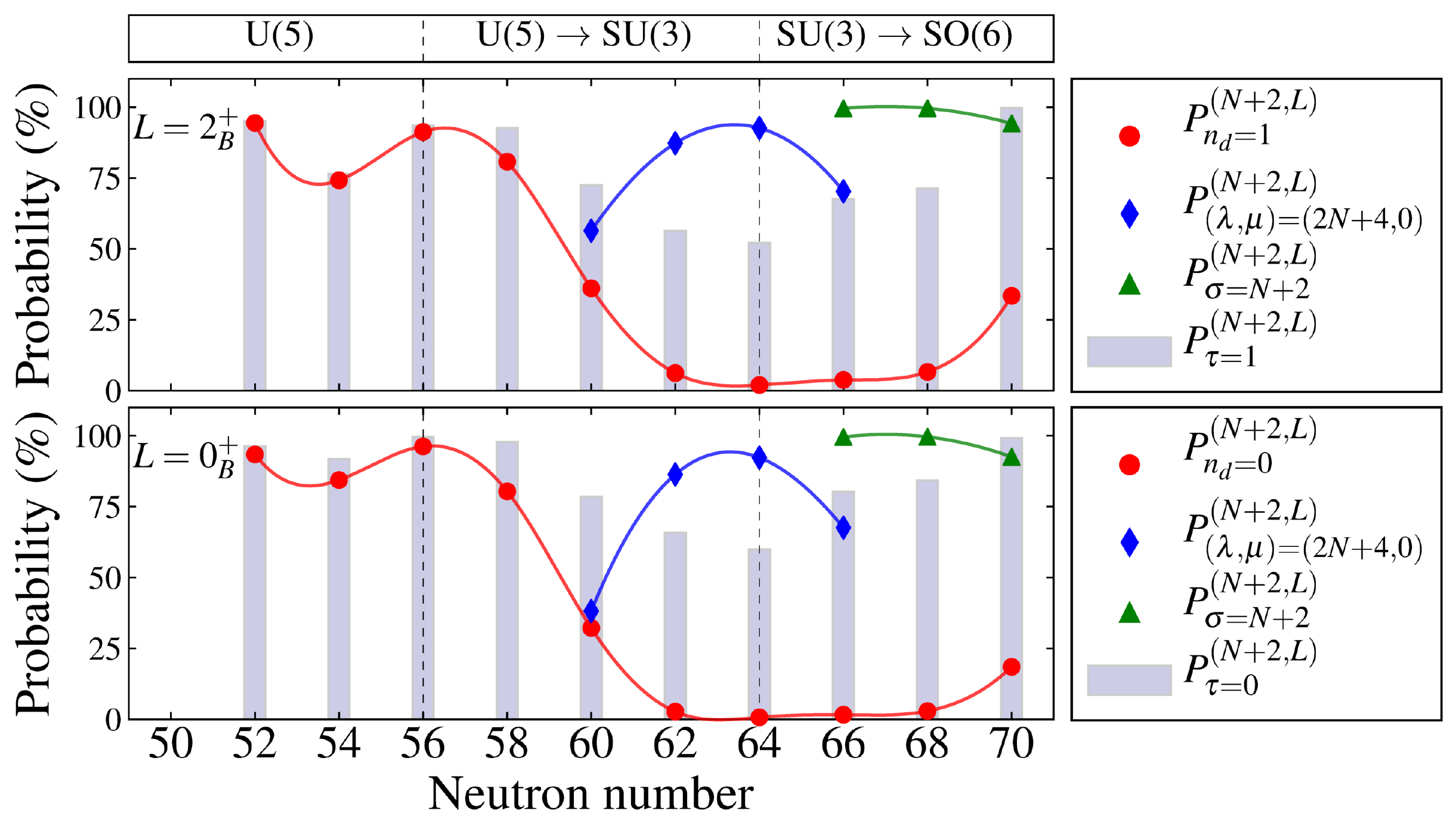}
\caption{Evolution of symmetries for the lowest $0^+$ and $2^+$ states of configuration $B$ along the Zr chain. Shown are the probabilities of selected components of U(5)~(${\color{red}\CIRCLE}$), SU(3)~(${\color{blue}\blacklozenge}$), SO(6)~(${\color{GreenNoam}\blacktriangle}$) and SO(5)~(${\color{GrayNoam}\rule[0pt]{13pt}{5pt}}$), obtained from \cref{eq:decomp-ds}. For neutron numbers 52--58 (60--70), $0^+_B$ corresponds to the experimental $0^+_2$ ($0^+_1$) state. For neutron numbers 52--56 (58--70), $2^+_B$ is the experimental $2^+_2$ ($2^+_1$) state.
\label{fig:decomp-evol}}
\end{figure*}
It is also interesting to see the changes in symmetry of the lowest $0^+$ and $2^+$ states within configuration~$B$, which undergoes a Type I QPT. \cref{fig:decomp-evol} depicts such evolution along the Zr chain. 
For the $0^+$ state (bottom panel), the red dots represent the percentage of the U(5) $n_d\!=\!0$ component in the wave function, \cref{eq:decomp-u5}. For neutron number 52--60, this component is large ($\approx90\%$) and at 60 it drops drastically ($\approx30\%$). This drop implies that additional $n_d$ components are present in the wave function, hence this state becomes deformed. For neutron number larger than 60, the $n_d\!=\!0$ component drops to zero almost and slightly rises again at 70, indicating the state is strongly deformed.
For neutron numbers 60--66, we also depict in blue diamonds the percentage of the SU(3) $(\lambda,\mu)=(2N+4,0)$ component, \cref{eq:decomp-su3}. For neutron number 60, it is moderately small ($\approx35\%$).
At neutron number 62, this $(\lambda,\mu)$ component jumps ($\approx85\%$), and it increases at 64 ($\approx92\%$), where deformation is maximal. This serves a clear evidence for a U(5)-SU(3) Type I QPT. At neutron number 66, the indicated $(\lambda,\mu)$ component is lowered and one sees in \cref{fig:decomp-evol}, by the green triangles, the percentage of the SO(6) $\sigma=N+2$ component, \cref{eq:decomp-so6}. This component becomes dominant for 66--70 ($\approx99\%$), suggesting a crossover from SU(3) to SO(6).

In order to further understand the phase transition from spherical [U(5)] to axially deformed [SU(3)] and the subsequent crossover to $\gamma$-unstable deformed [SO(6)], it is instructive to examine also the evolution of SO(5) symmetry in comparison with U(5), along the Zr chain. To recall, the SO(5) quantum number $\tau$ is valid in both the U(5) and SO(6) DS limits, but is broken in SU(3) DS. On the other hand, the U(5) quantum number $n_d$ is valid in the U(5) DS, but is broken in both the SU(3) and SO(6) DS limits. Accordingly, for a given $L$-state at the U(5) limit, both the $n_d$ and $\tau$ probabilities, $P^{(N+2,L)}_{n_d}$ and $P^{(N+2,L)}_{\tau}$, are maximal (100\%). In the U(5) to SU(3) transition, both probabilities decrease, while in the SU(3) to SO(6) crossover, $P^{(N+2,L)}_{n_d}$  remains small, but  $P^{(N+2,L)}_{\tau}$ increases towards its maximal value at the SO(6) limit.
This is precisely the pattern exhibited by the $n_d=0$ probability (red dots) and $\tau=0$ probability (gray histograms) in \cref{fig:decomp-evol}, for the lowest $0^+$ and $2^+$ states within configuration $B$.

Considering the $0^+_B$ state. For neutron numbers 52--56, $P^{(N+2,L=0^+_B)}_{\tau=0} \approx P^{(N+2,L=0^+_B)}_{n_d=0}$, meaning that the state is composed mainly of a single ($n_d\!=\!0,\tau\!=\!0$) component, appropriate for a spherical state. For neutron number 58, $P^{(N+2,L=0^+_B)}_{n_d=0} < P^{(N+2,L=0^+_B)}_{\tau=0}$,  implying the presence of additional components with ($n_d\neq 0, \tau=0$). For neutron numbers 60--64, both $n_d=0$ and $\tau=0$ probabilities decrease, satisfying $P^{(N+2,L=0^+_B)}_{n_d=0} \ll P^{(N+2,L=0^+_B)}_{\tau=0} < 100\%$, implying admixtures of components with ($n_d\neq 0, \tau\neq 0$), appropriate for an axially-deformed state.
For neutron number above 64, $P^{(N+2,L=0^+_B)}_{n_d=0}$ remains small but $P^{(N+2,L=0^+_B)}_{\tau=0}$ increases towards its maximum value at 70, appropriate for a crossover to $\gamma$-unstable structure with good SO(5) symmetry.

A very similar trend is observed for the $2^+_B$ state. For neutron numbers 52--58, it is dominated by a single ($n_d\!=\!1,\tau\!=\!1$) component. For neutron number 60, $P^{(N+2,L=2^+_B)}_{n_d=1} < P^{(N+2,L=2^+_B)}_{\tau=1}$, for 62--64, $P^{(N+2,L=2^+_B)}_{n_d=0} \ll P^{(N+2,L=2^+_B)}_{\tau=0} < 100\%$, implying admixtures of components with ($n_d\!\neq\!1, \tau\!\neq\!1$), and for neutron numbers 66--70, $P^{(N+2,L=2^+_B)}_{n_d=0}$ remains small
but  $P^{(N+2,L=2^+_B)}_{\tau=1}$ increases towards its maximum value at 70.

The similarity between the trends of the $0^+_B$ and $2^+_B$ states is particularly interesting since, as shown in \cref{fig:mixing}, the $2^+_1$ changes its configuration content from $A$ to $B$ already at neutron number 58, rather than 60 for the $0^+_1$ state. This is a good example of how the two types of QPTs, I and II, progress simultaneously without interrupting one another, and support the occurrence of intertwined QPTs.

\subsection{Evolution of order parameters}\label{sec:evo-order}
\begin{figure}[t]
\centering
\includegraphics[width=0.87\linewidth]{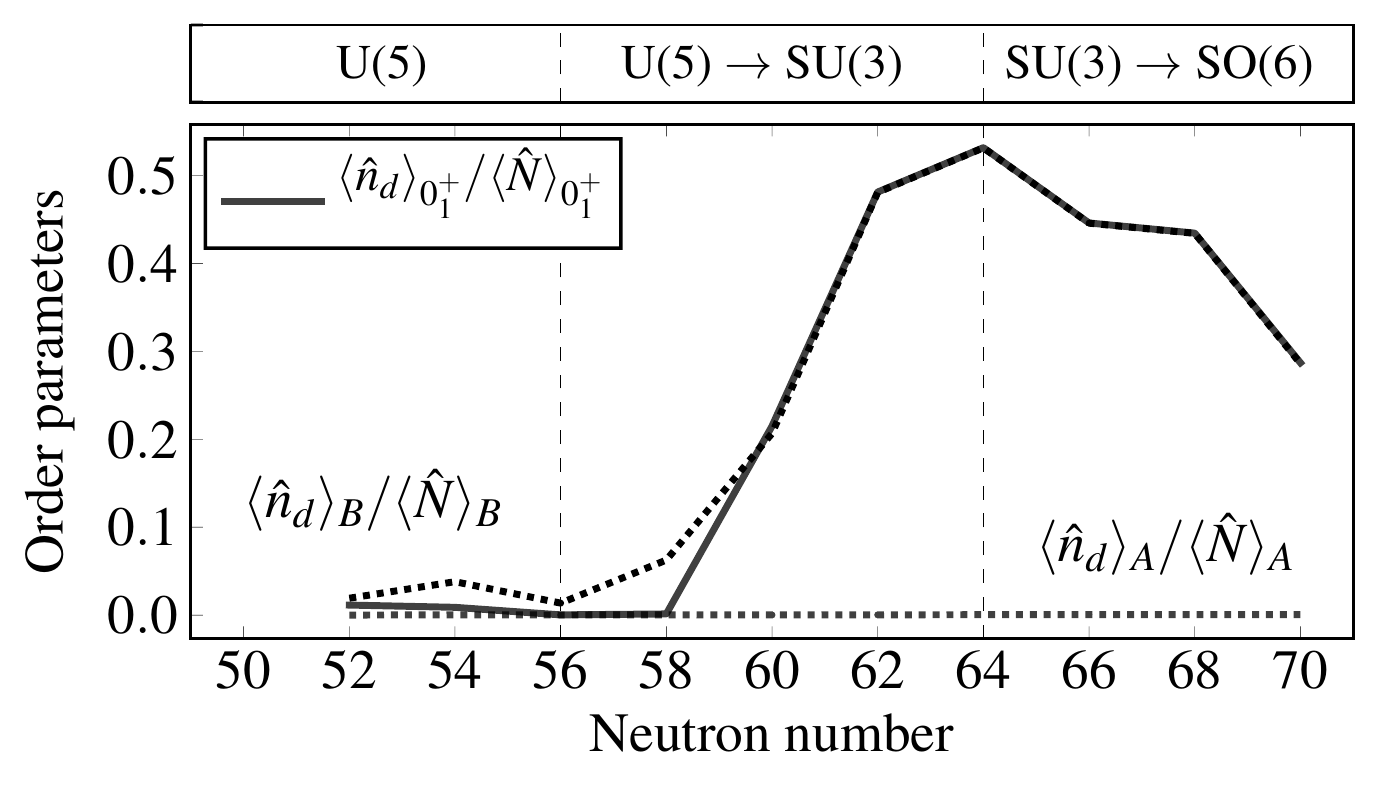}
\caption{Evolution of order parameters along the Zr chain. The latter are the calculated expectation values of $\hat n_d$ in the total ground state wave function $\ket{\Psi; L=0^+_1}$, \cref{eq:wf} (solid line), and in its $A$ and $B$ components (dotted lines), normalized by the respective boson numbers $\braket{\hat N}_{0^+_1} = a^2N + b^2(N+2)$, $\braket{\hat N}_A = N, \braket{\hat N}_B = N+2$.
\label{fig:nd}}
\end{figure}

Figures~\eqref{fig:mixing} and \eqref{fig:decomp-evol} above, exemplify in a clear manner the simultaneous occurrence of Type I and II QPTs, respectively. However, in order to encapsulate both types, it is instructive to examine the behavior of the order parameters, \cref{eq:order-param-cm}.
\cref{fig:nd} shows the evolution along the Zr chain of the individual order parameters, $\braket{\hat n_d}_A$ and $\braket{\hat n_d}_B$ (in dashed lines) and  $\braket{\hat n_d}_{0^+_1}$ (in solid line), normalized by the respective boson numbers, 
$\braket{\hat{N}}_A\!=\!N$, 
$\braket{\hat{N}}_B\!=\!N\!+\!2$,
$\braket{\hat{N}}_{0^{+}_1}\!=\!a^2N + b^2(N\!+\!2)$.
$\braket{\hat n_d}_{0^+_1}$ is close to $\braket{\hat n_d}_A$ for neutron numbers 52--58 and coincides with $\braket{\hat n_d}_B$ at 60 and above, consistent with a high degree of purity with respect to configuration-mixing.
Configuration~$A$ appears to be spherical for all neutron numbers considered. 
In contrast, configuration~$B$ is weakly deformed for neutron numbers 52--58 and becomes more deformed above 58. One can see a clear jump in $\braket{\hat n_d}_{0^+_1}$ between neutron numbers 58 and 60, changing from configuration~$A$ to configuration~$B$, indicating a first-order configuration-changing phase transition (Type II QPT). A further increase in $\braket{\hat n_d}_{0^+_1}$ at neutron numbers 60--64 indicates a U(5)-SU(3) shape-phase transition within configuration $B$ (Type I QPT), and, finally, there is a decrease at neutron number 66, due in part to the crossover from SU(3) to SO(6) and in part to the shift in configuration~$B$ from boson particles to boson holes after the middle of the major shell 50--82. 
These findings further support the occurrence of two configurations that are weakly mixed and interchange their roles in the ground state while their individual shapes evolve gradually with neutron number, i.e., intertwined Type I and II QPTs.

\section{Results: Classical analysis}\label{sec:pes}
\begin{figure*}[]
\centering
\begin{overpic}[width=0.19\linewidth]{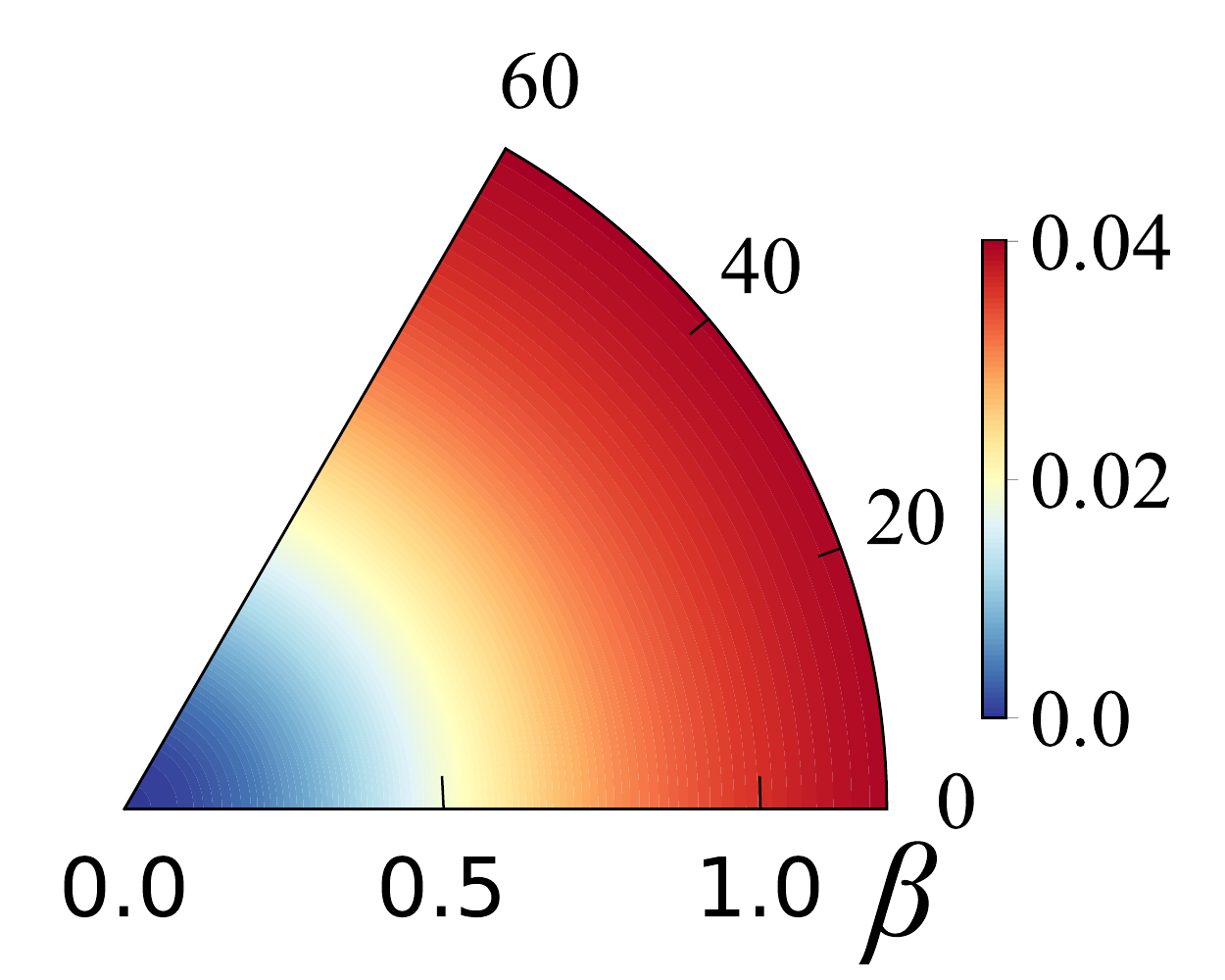}
\put (0,67.5) {\large(a)}
\put (0,50) {\large $^{92}$Zr}
\put (60,70) {$\gamma$(deg)}
\end{overpic}
\begin{overpic}[width=0.19\linewidth]{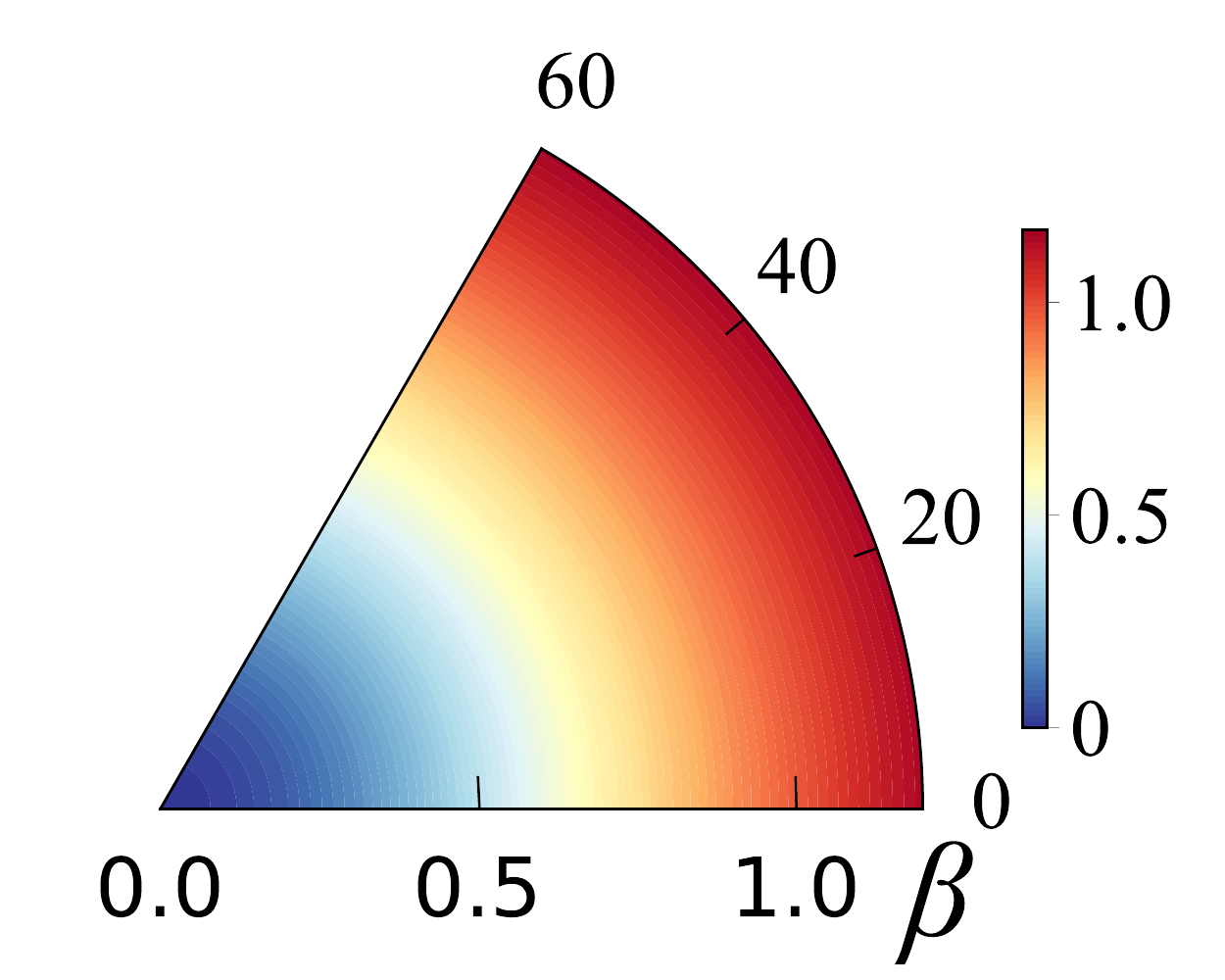}
\put (0,67.5) {\large(b)}
\put (0,50) {\large $^{94}$Zr}
\put (60,70) {$\gamma$(deg)}
\end{overpic}
\begin{overpic}[width=0.19\linewidth]{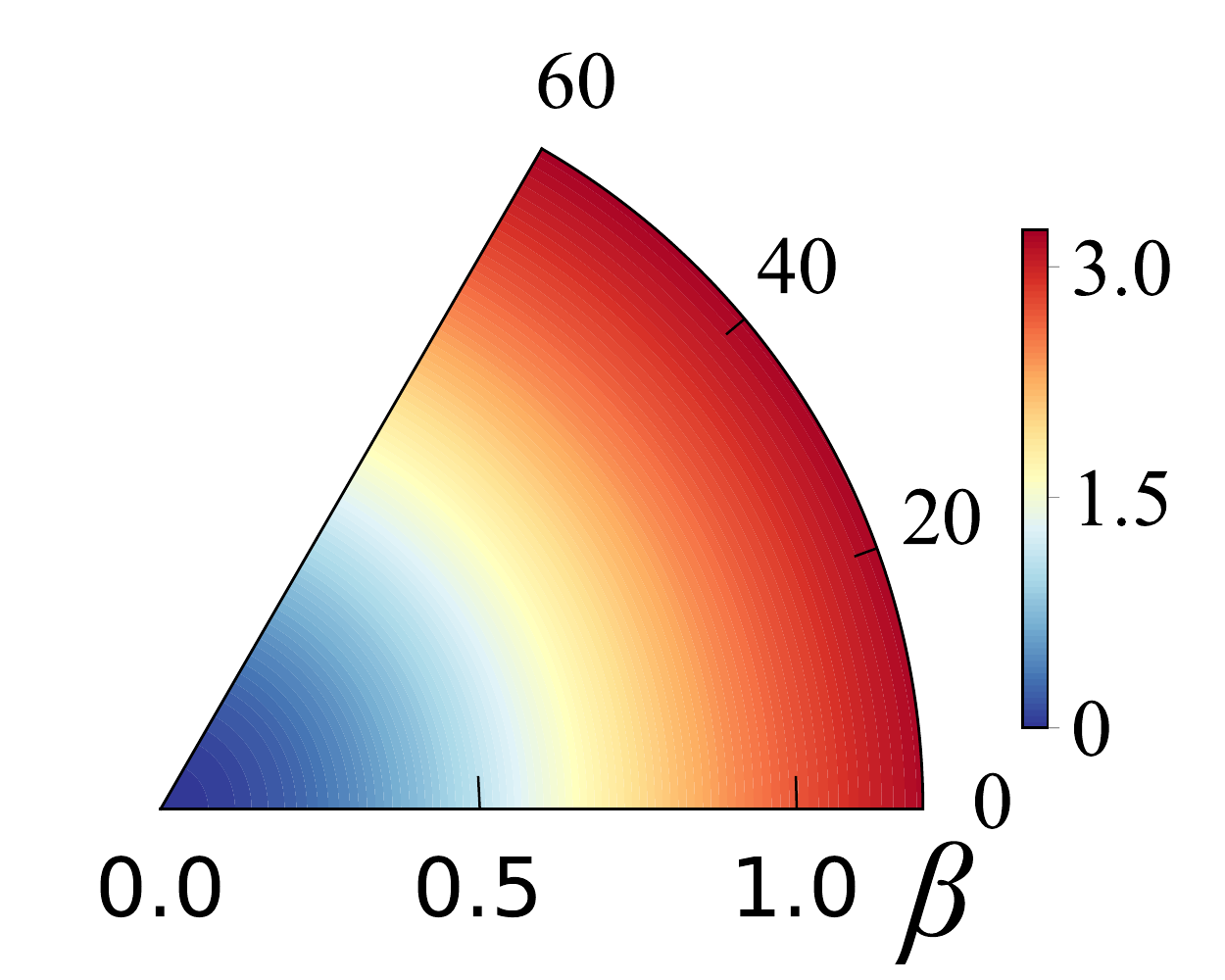}
\put (0,67.5) {\large(c)}
\put (0,50) {\large $^{96}$Zr}
\put (60,70) {$\gamma$(deg)}
\end{overpic}
\begin{overpic}[width=0.19\linewidth]{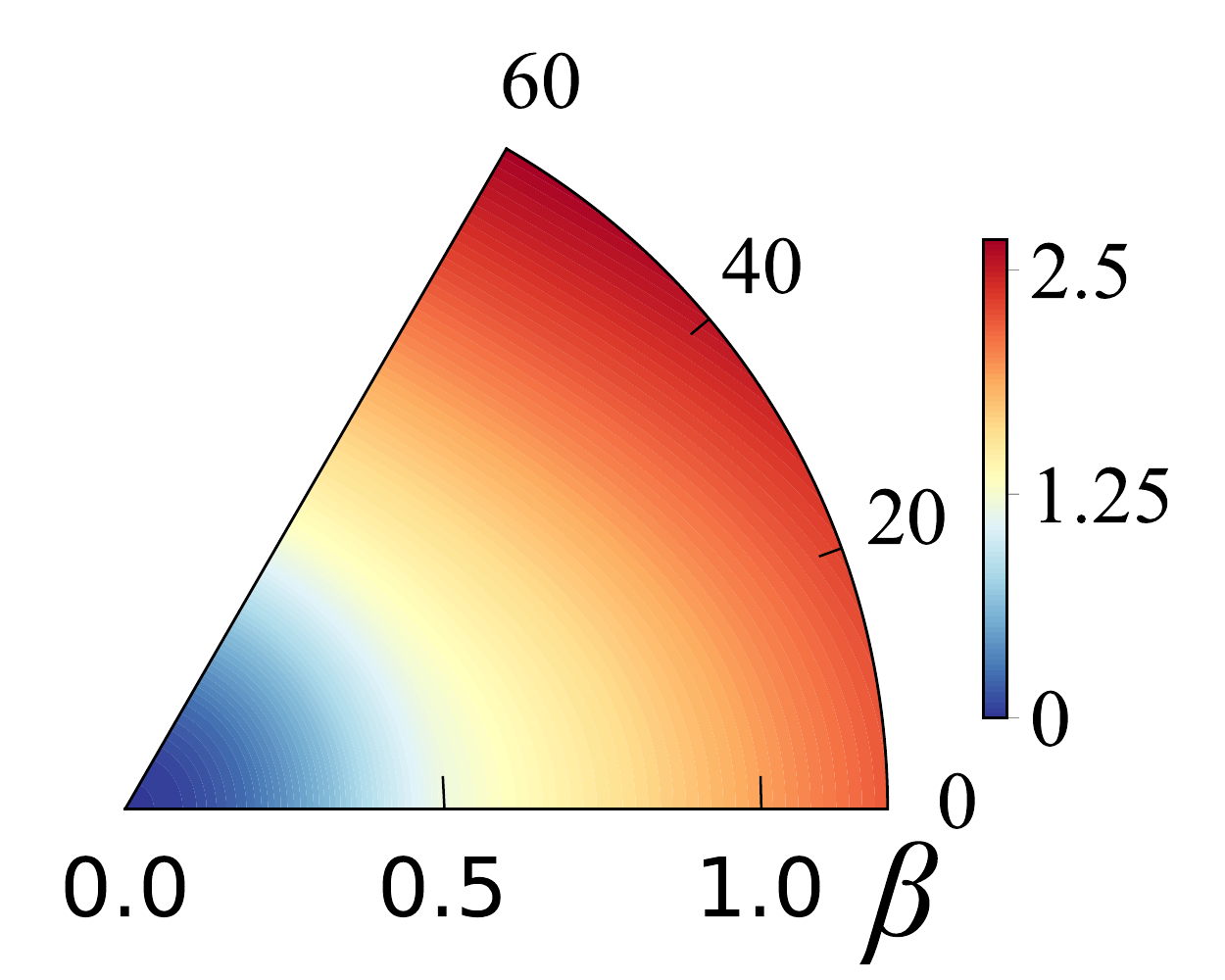}
\put (0,67.5) {\large(d)}
\put (0,50) {\large $^{98}$Zr}
\put (60,70) {$\gamma$(deg)}
\end{overpic}
\begin{overpic}[width=0.19\linewidth]{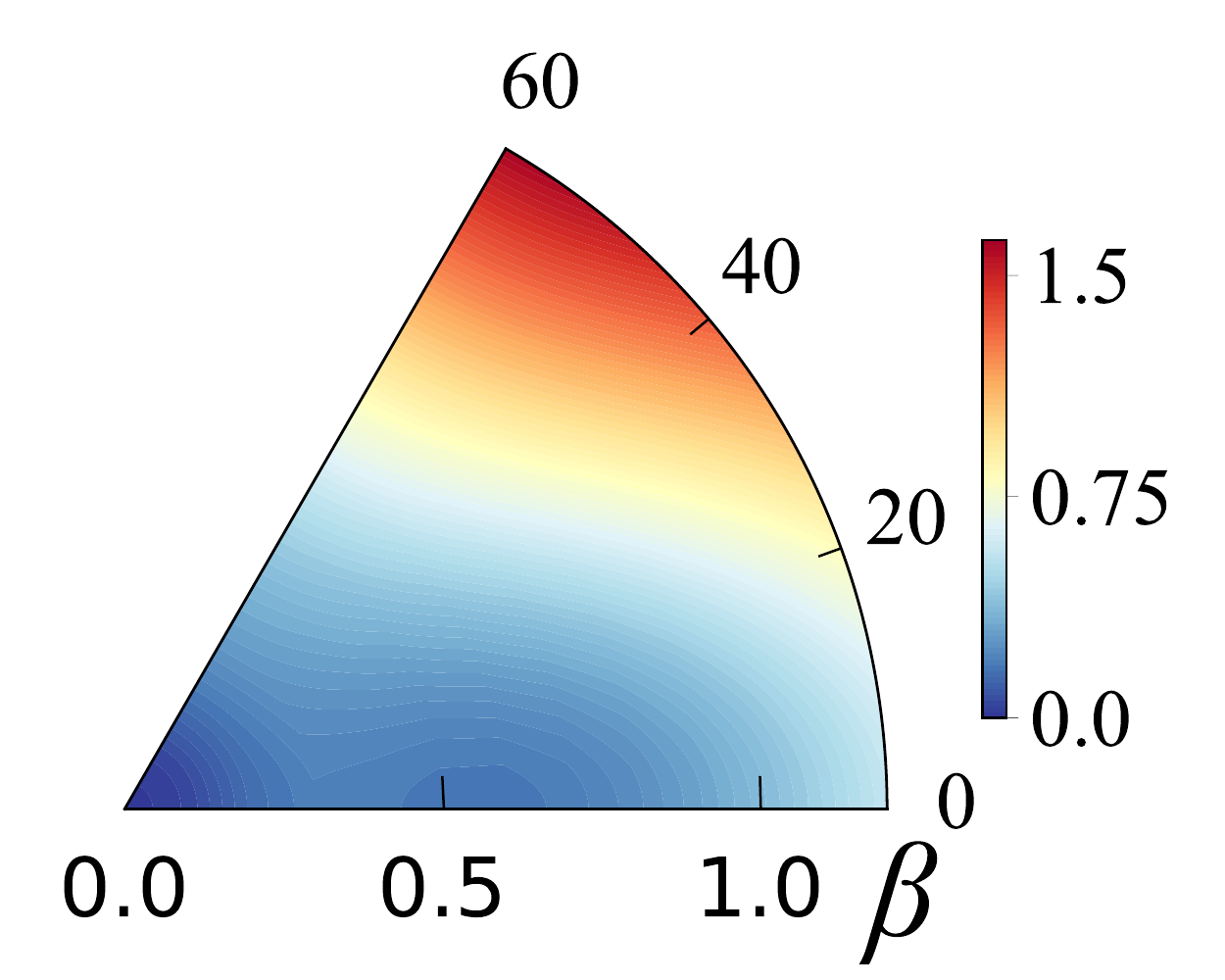}
\put (0,67.5) {\large(e)}
\put (0,50) {\large $^{100}$Zr}
\put (60,70) {$\gamma$(deg)}
\end{overpic} \\ 
\begin{overpic}[width=0.19\linewidth]{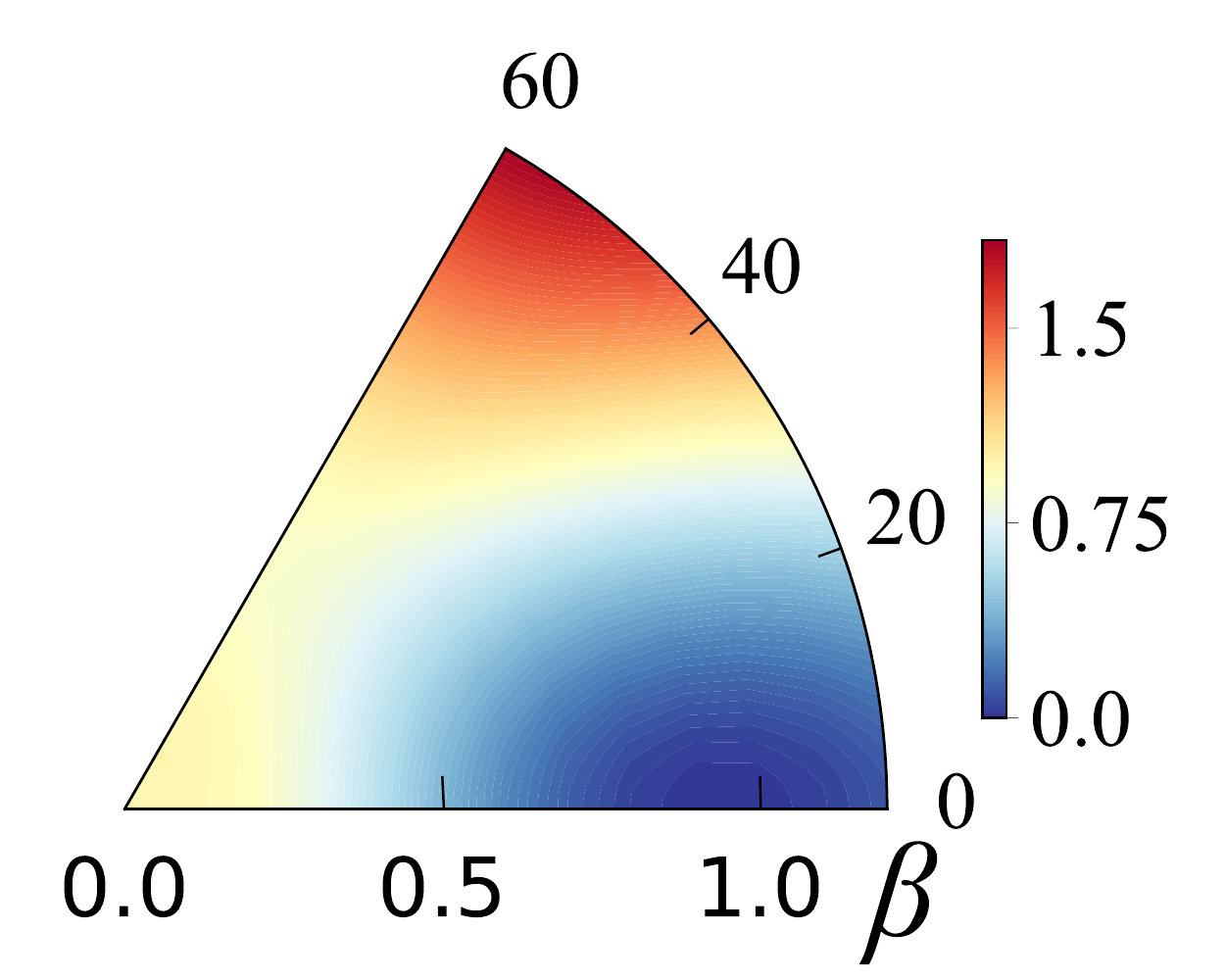}
\put (0,67.5) {\large(f)}
\put (0,50) {\large $^{102}$Zr}
\put (60,70) {$\gamma$(deg)}
\end{overpic}
\begin{overpic}[width=0.19\linewidth]{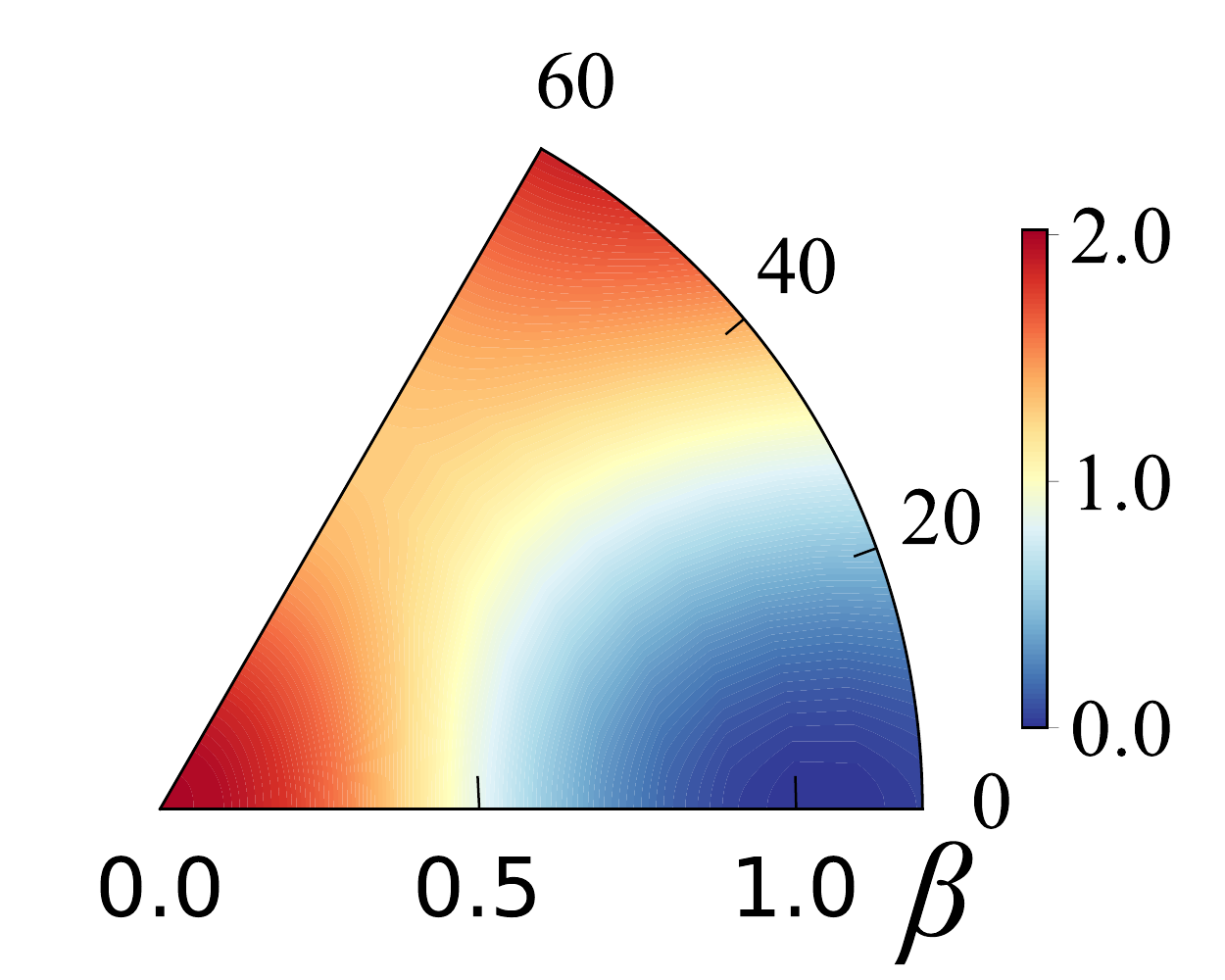}
\put (0,67.5) {\large(g)}
\put (0,50) {\large $^{104}$Zr}
\put (60,70) {$\gamma$(deg)}
\end{overpic}
\begin{overpic}[width=0.19\linewidth]{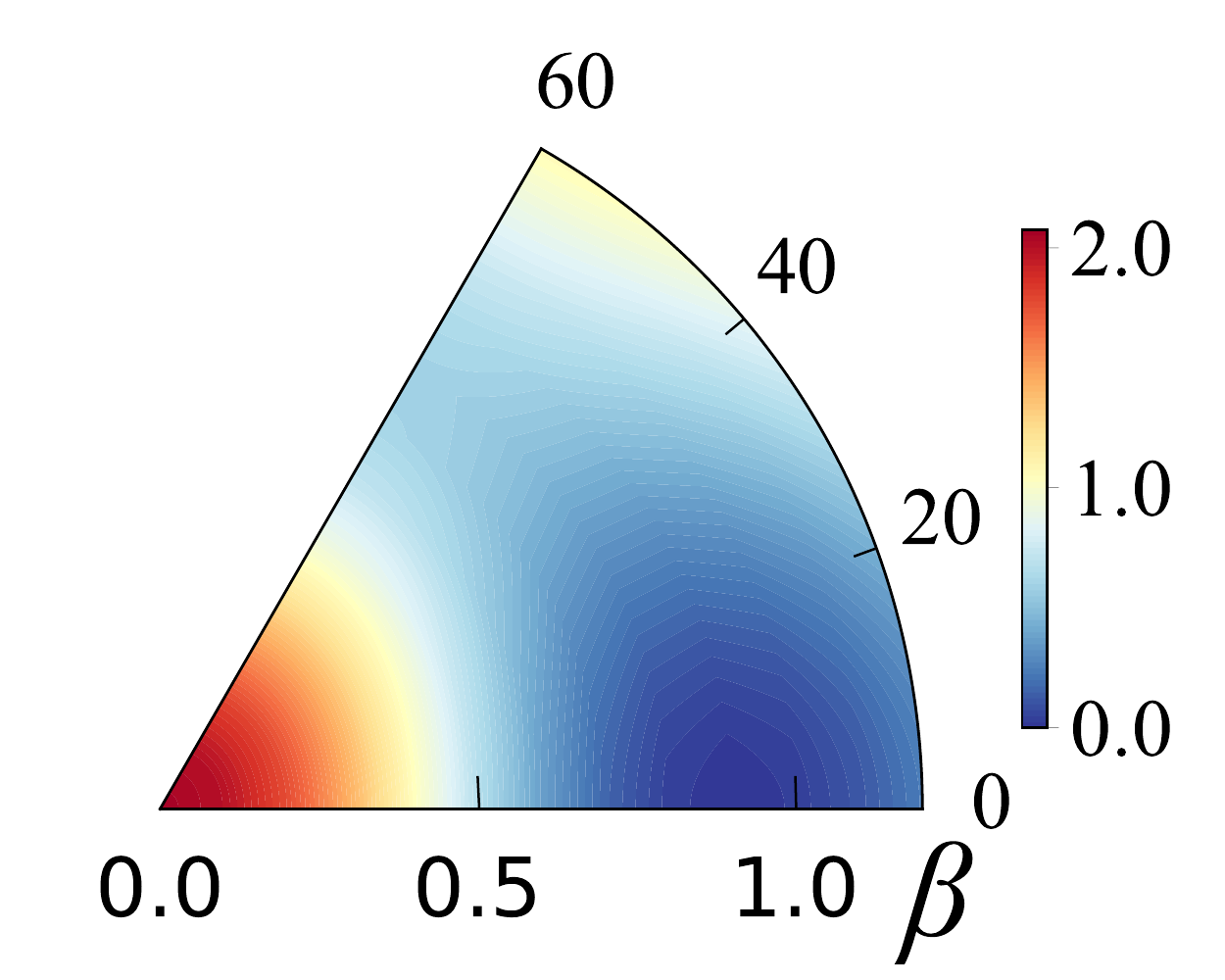}
\put (0,67.5) {\large(h)}
\put (0,50) {\large $^{106}$Zr}
\put (60,70) {$\gamma$(deg)}
\end{overpic}
\begin{overpic}[width=0.19\linewidth]{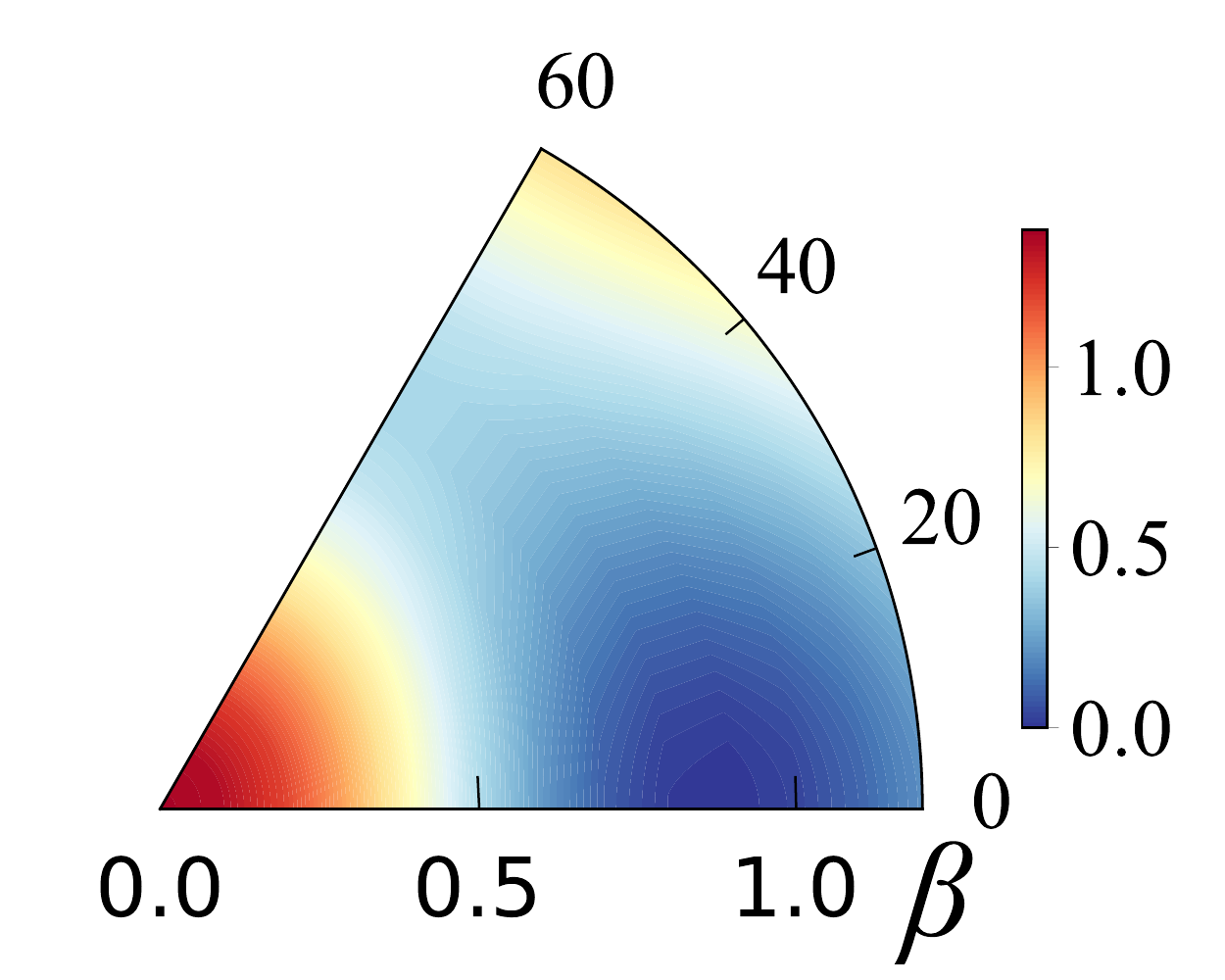}
\put (0,67.5) {\large(i)}
\put (0,50) {\large $^{108}$Zr}
\put (60,70) {$\gamma$(deg)}
\end{overpic}
\begin{overpic}[width=0.19\linewidth]{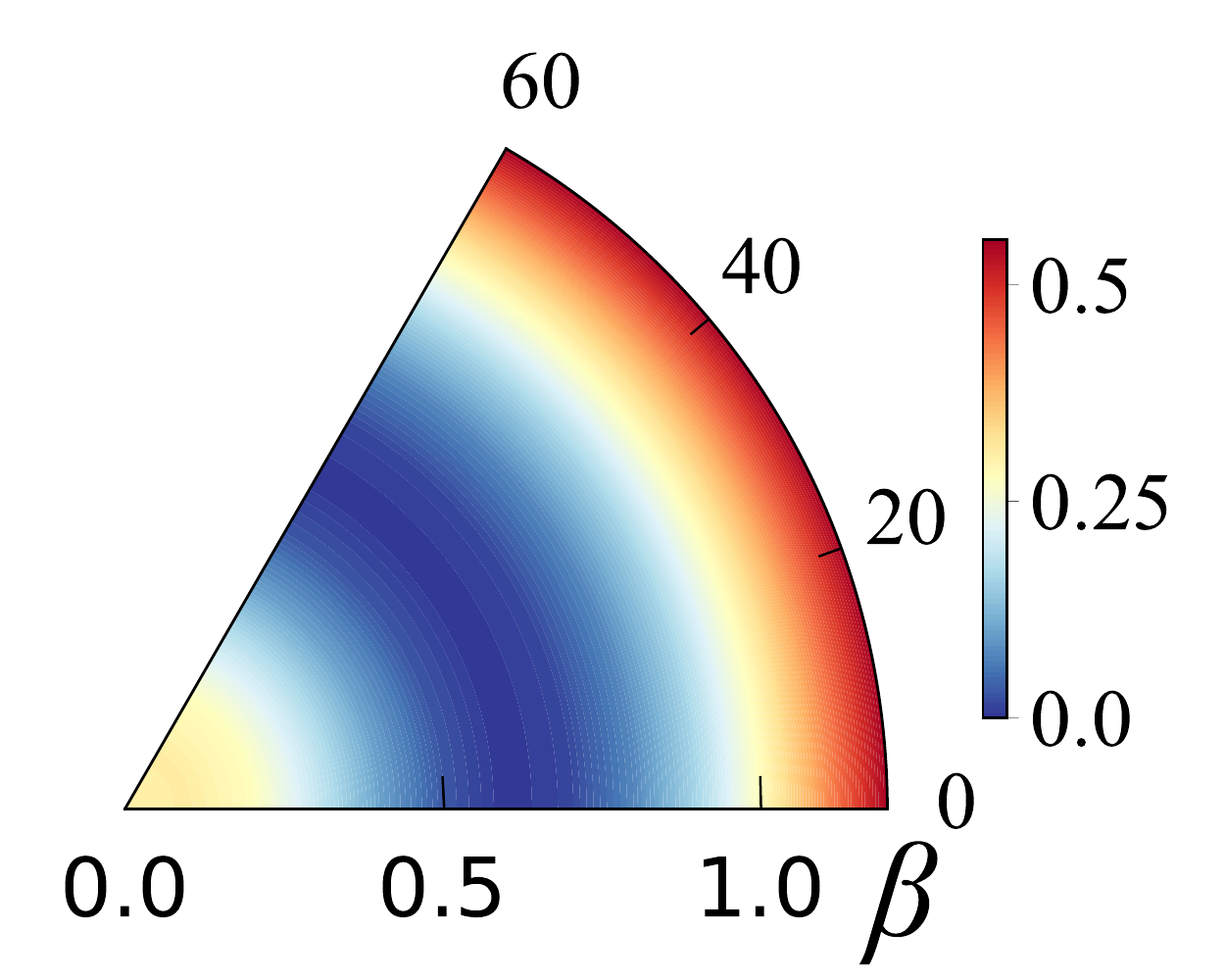}
\put (0,67.5) {\large(j)}
\put (0,50) {\large $^{110}$Zr}
\put (60,70) {$\gamma$(deg)}
\end{overpic}
\caption{\label{Eminus}
\small
Contour plots in the $(\beta ,\gamma )$ plane of the lowest eigen-potential 
surface, $E_{-}(\beta ,\gamma )$, for the $^{92-110}$Zr isotopes.}
\end{figure*}
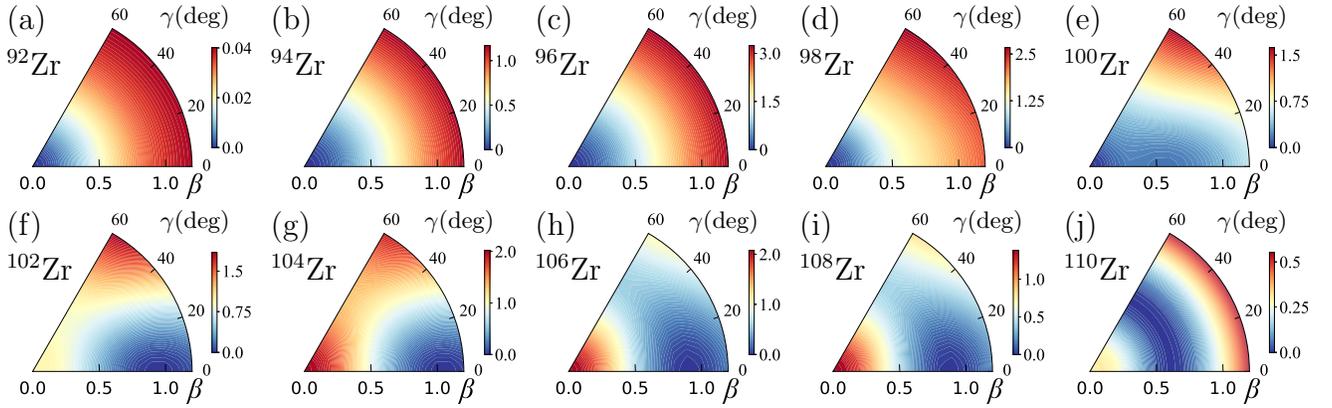
In addition to the quantum analysis, the algebraic method can perform also a classical analysis. In Fig.~\ref{Eminus}, we show the calculated lowest eigenpotential $E_{-}(\beta,\gamma)$, which is the lowest eigenvalue of the two-by-two matrix~\eqref{eq:surface-mat}, with elements given in \cref{eq:surface-elements} for the entire chain of isotopes. These classical potentials confirm the quantum results, as they show a transition from spherical ($^{92-98}$Zr), to a double-minima potential at $^{100}$Zr, to prolate axially deformed ($^{102-104}$Zr), and finally to $\gamma$-unstable ($^{106-110}$Zr). At $^{100}$Zr, $E_{-}(\beta,\gamma)$ exhibits two minima, one at $(\beta,\gamma)=(0,0)$ and one at $(\beta,\gamma)=(0.5617,0)$, separated by a saddle point at $(\beta,\gamma)=(0.3127,0)$ that serves as a barrier. 
In the limit of ($\beta\!\to\!\infty$, $\gamma\!=\!0$) the lowest eigenpotential has the value of 2.9~MeV, while the height of the barrier is 0.3~MeV, i.e., the potential is flat-bottomed. We further note that in the classical calculation the global minimum is the spherical one, rather than the deformed one as in the quantum analysis [see Figs.~\eqref{fig:98-102Zr-scheme}(c) and \eqref{fig:98-102Zr-scheme}(d)]. This demonstrates the difficulties in describing the dynamics near the critical-point by mean field methods.

The classical analysis above and the quantum analysis of \cref{sec:results-indi,sec:results-evo} suggest coexisting Type~I and Type~II QPTs, which is the defining property of IQPTs. 

\section{Results: Evolution of observables along the Zr chain}\label{sec:evo-obs}
In order to understand the change in structure of the Zr isotopes, it is insightful to examine the evolution of observables along the chain. The observables include energy levels, two-neutron separation energies, $E2$ and $E0$ transition rates, isotope shifts, and magnetic moments.
\subsection{Energy levels}\label{sec:evo-energy}
\begin{figure}[b]
\begin{overpic}[width=1\linewidth]{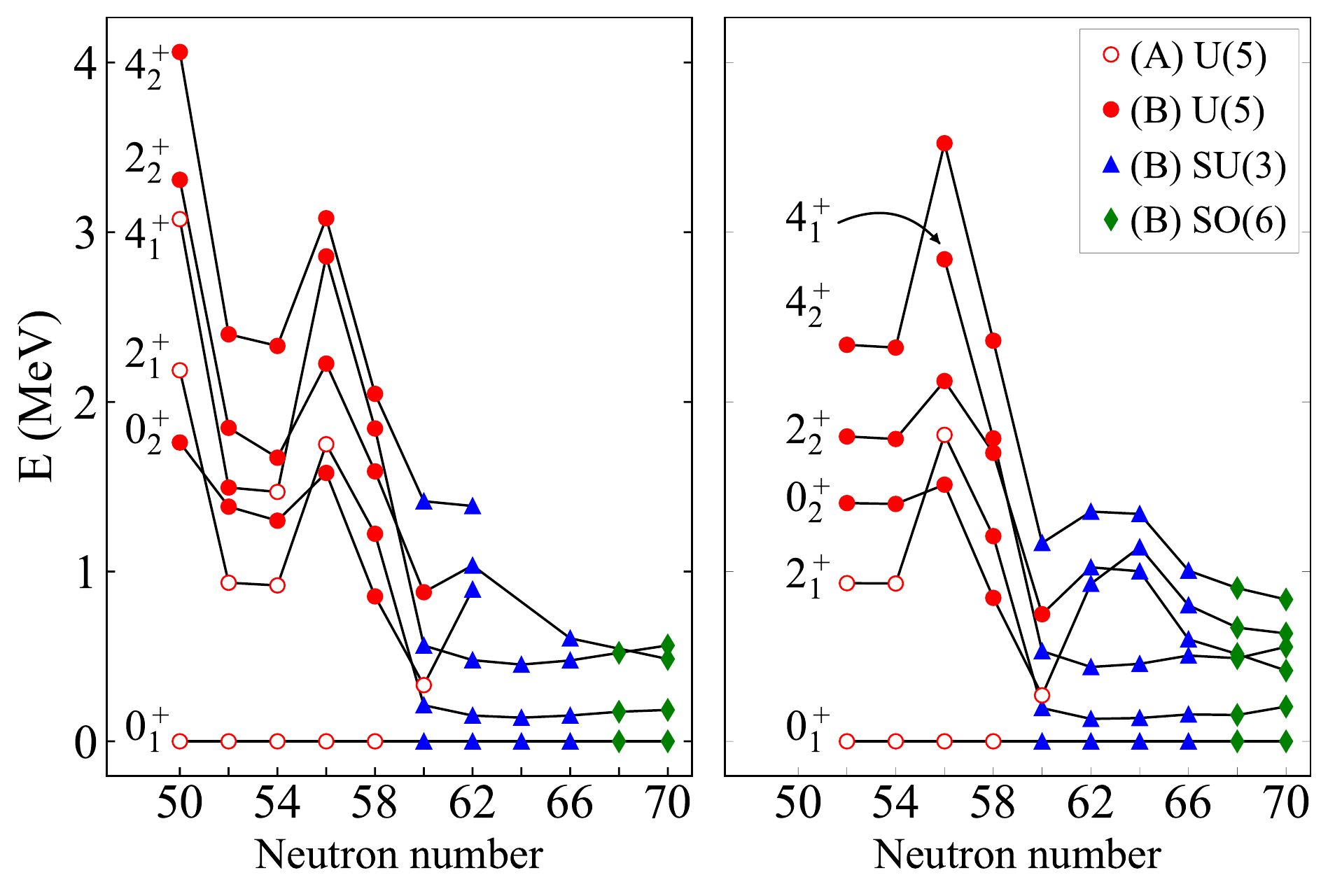}
\put (33.5,61) {\large (a) {\bf Exp}}
\put (56,61) {\large (b) {\bf Calc}}
\end{overpic}
\caption{Comparison between (a)~experimental and
(b)~calculated energy levels
$0_{1}^{+},2_{1}^{+},4_{1}^{+},0_{2}^{+},2_{2}^{+},4_{2}^{+}$.
Empty (filled) symbols indicate a state dominated by the normal $A$ configuration (intruder $B$ configuration), with assignments based on \cref{eq:decomp-cm}. The shape of the symbol [$\circ,\triangle,\diamondsuit$], indicates the closest dynamical
symmetry [U(5), SU(3), SO(6)] to the level considered, based on \cref{eq:decomp-ds}.
Note that the calculated values start at neutron number 52, while the experimental values include the closed shell at 50. Data are taken from \cite{NDS.113.2187.2012} ($^{92}$Zr), \cite{NDS.107.2423.2006} ($^{94}$Zr), \cite{NDS.109.2501.2008} ($^{96}$Zr), \cite{NDS.164.1.2020} ($^{98}$Zr), \cite{NDS.109.297.2008} ($^{100}$Zr), \cite{NDS.110.1745.2009} ($^{102}$Zr), \cite{NDS.108.2035.2007} ($^{104}$Zr), \cite{NDS.109.943.2008} ($^{106}$Zr), \cite{NDS.90.135.2000} ($^{108}$Zr), \cite{NDS.113.1315.2012,Paul2017} ($^{110}$Zr).
\label{fig:levels}}
\end{figure}

In \cref{fig:levels}, we show a comparison between experimental and calculated 
levels, along with assignments to configurations based on \cref{eq:decomp-cm} and to the closest dynamical symmetry based on the decompositions of \cref{eq:decomp-ds}, for each state. One can see here a rather complex structure. 
In the region between neutron numbers 50 and 58, there appear to be two configurations, one spherical (seniority-like), $A$, and one weakly deformed, $B$, as evidenced by the ratio $R_{4/2}$ in each configuration, $R^{(A)}_{4/2}\!=\!1.6,~1.6,~1.76,~1.2$ and $R^{(B)}_{4/2}\!=\!2.2,~2.8,~2,~2.7$, for neutron numbers 52, 54, 56, and 58, respectively. The value $R^{(B)}_{4/2}\!=\!2.8$ for $^{94}$Zr is somewhat larger, possibly as a consequence of fluctuations due to the subshell closure at neutron number 56.
At neutron number 58, there is a pronounced drop in energy for the states of configuration~$B$, suggesting a slight increase in deformation, where the $2^+_1$ becomes already a configuration~$B$ state. At neutron number 60, the two configurations exchange their roles, indicating a Type II QPT. This is evident from \cref{fig:mixing}, showing the exchange in the decomposition of the ground state $0^+_1$ from the $A$ configuration ($a^2\!=\!98.2\%$) in $^{98}$Zr to the $B$ configuration ($b^2\!=\!87.2\%$) in $^{100}$Zr.
At this stage, configuration~$B$ appears also to be close to the critical-point of a U(5)-SU(3) QPT, as evidenced by the low value of the excitation energy of the $0^+_3$ state in $^{100}$Zr [see \cref{fig:98-102Zr-scheme}(c)], which is the first excited $0^+$ state of the $B$ configuration ($b^2\!=\!92.9\%$). As pointed out in  \cref{sec:98-102zr-region}, the spectrum of states of the next isotope, $^{102}$Zr, resembles that of the X(5) critical-point symmetry \cite{Iachello2001}.

\begin{figure*}[t]
\includegraphics[width=0.300442758\linewidth]{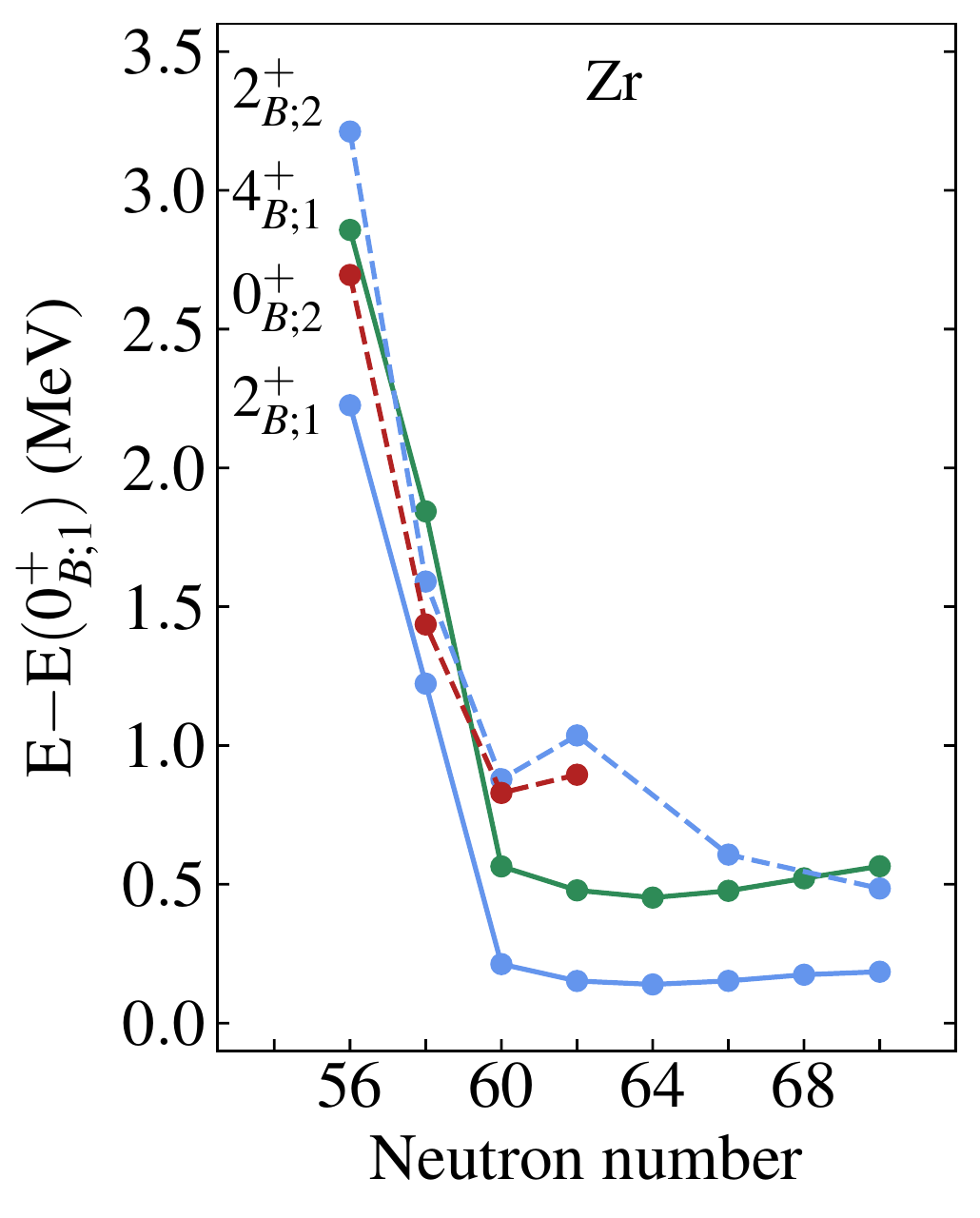}
\includegraphics[width=0.300442758\linewidth]{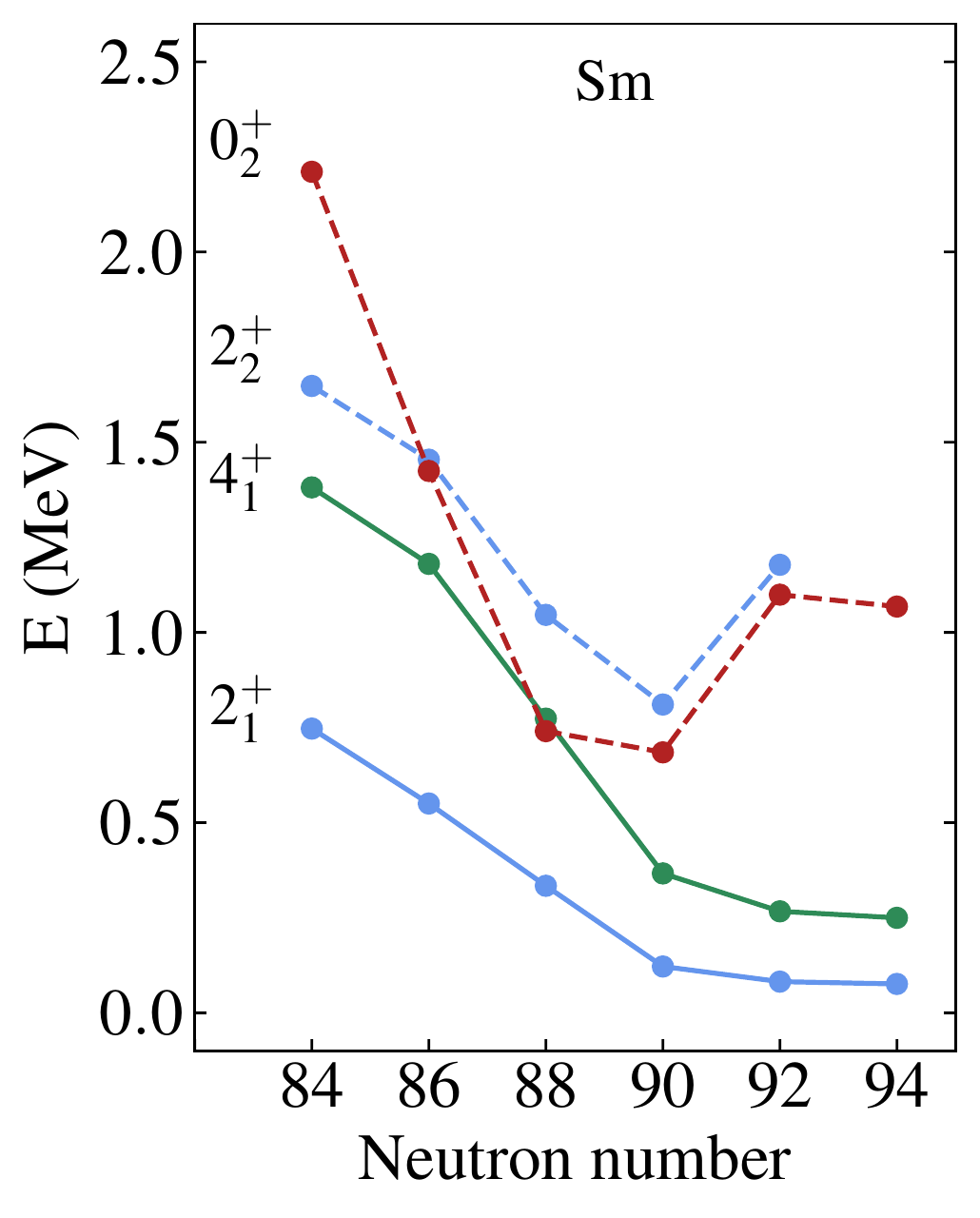}
\includegraphics[width=0.25\linewidth]{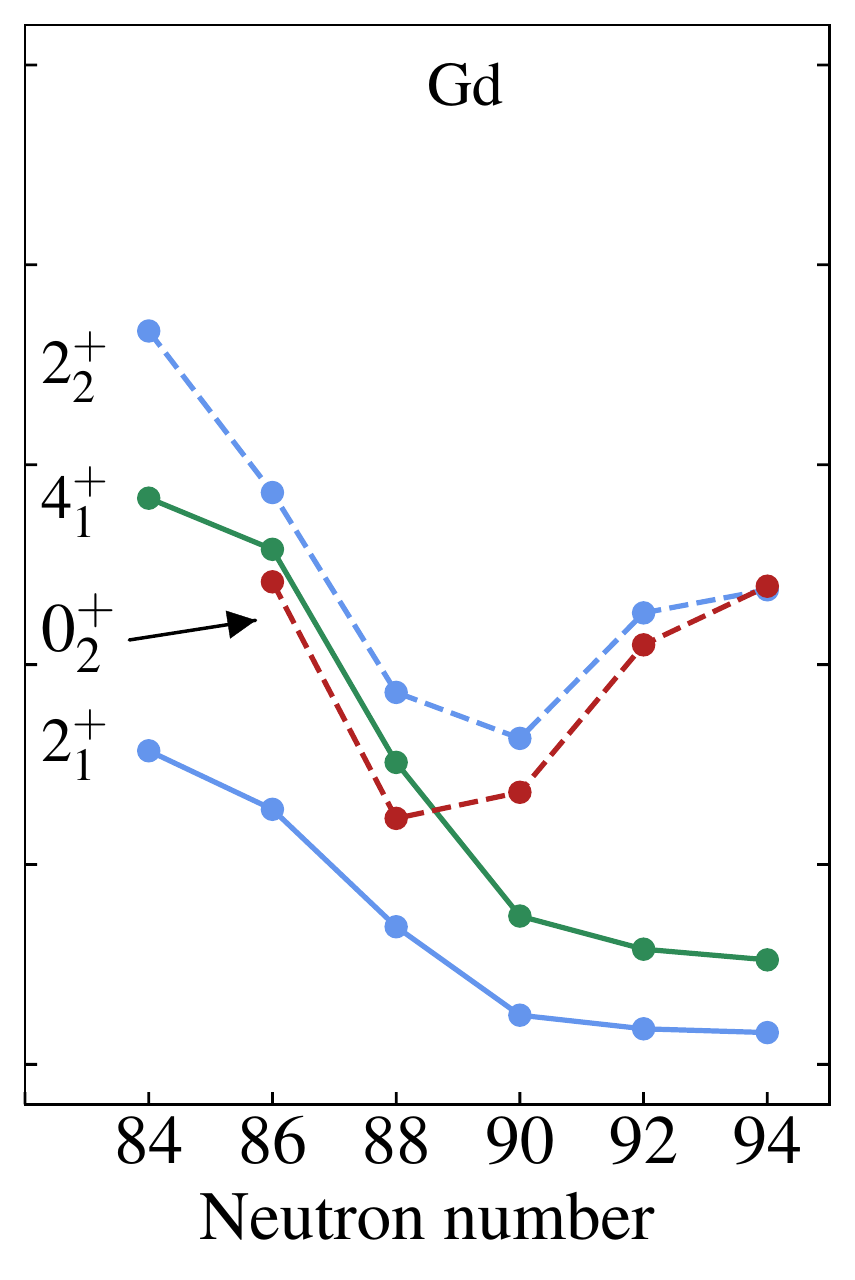}
\caption{Experimental energy levels in MeV, comparing the evolution with neutron number of states $L^+_{B;i}$ of the $B$ configuration in the Zr isotopes to the situation encountered in the Sm and Gd isotopes. Note that the energy of states in the $B$ configuration of Zr are with respect to the lowest state ($L=0^+_{B;1}$) in that configuration. Data taken from \cite{ensdf}.
\label{fig:X5}}
\end{figure*}

Beyond neutron number 60, the intruder configuration~$B$ becomes progressively strongly deformed. This is evidenced for neutron number 62, by the small value of the excitation energy of the state $2_1^+$, $E(2^+_1)\!=\!151.78$~keV and by the ratio $R^{(B)}_{4/2}\!=\!3.15$, where the first excited $0^+$ state within configuration~$B$ is now the $0^+_2$ state and serves as the bandhead of a $\beta$ band [see \cref{fig:98-102Zr-scheme}(e)]. For neutron number 64, the energy of the $2^+_1$ state is even smaller, $E(2^+_1)\!=\!139.3$~keV, and the ratio $R^{(B)}_{4/2}\!=\!3.24$ larger, suggesting further increase in deformation. At still larger neutron numbers 66--70, the ground state band becomes $\gamma$-unstable (or triaxial) as evidenced by the close energy of the $2^+_2$ and $4_1^+$ states in $^{106,110}$Zr, discussed in \cref{sec:104-110zr-region}, a signature of the SO(6) symmetry.  In this region, the ground state configuration undergoes a~crossover from SU(3)~to~SO(6).

The trend in energies of configuration $B$ for neutron numbers 56--70 is in part similar to the case of $_{62}$Sm and $_{64}$Gd isotopes~\cite{IachelloArimaBook,Scholten1978}, as depicted in \cref{fig:X5}.
One can see a lowering of the $4^+_1,2^+_2,0^+_2$ states while the $0^+_2$ state rises up again at neutron number 90 to become a $\beta$ band head member, a situation very similar to the trend of the states within configuration~$B$ of Zr isotopes. One minor difference is in the second excited $2^+$ state within configuration~$B$, $2^+_{B;2}$, which becomes degenerate with the $4^+_1$ state at neutron numbers 66--70, due to the discussed SU(3)-SO(6) crossover. However, a major difference occurs in the onset of deformation. While for Type I QPT (single configuration in Sm-Gd) the onset is gradual and the behavior smooth, for Type II QPT (two configurations in Zr) the onset is abrupt.
\subsection{Two neutron separation energy}\label{sec:evo-s2n}
In the IBM, two-neutron separation energies $S_{2n}$ can be written as~\cite{IachelloArimaBook},
\begin{equation}
S_{2n} = -\tilde A - \tilde B N_v \pm S^{\rm def}_{2n} - \Delta_n ~,
\end{equation}
where $N_v$ is half the number of valence particles and
$S^{\rm def}_{2n}$ is the contribution of the deformation,
obtained by the expectation value of the Hamiltonian in the ground state~$0^+_1$.
\begin{figure}[b]
\includegraphics[width=1\linewidth]{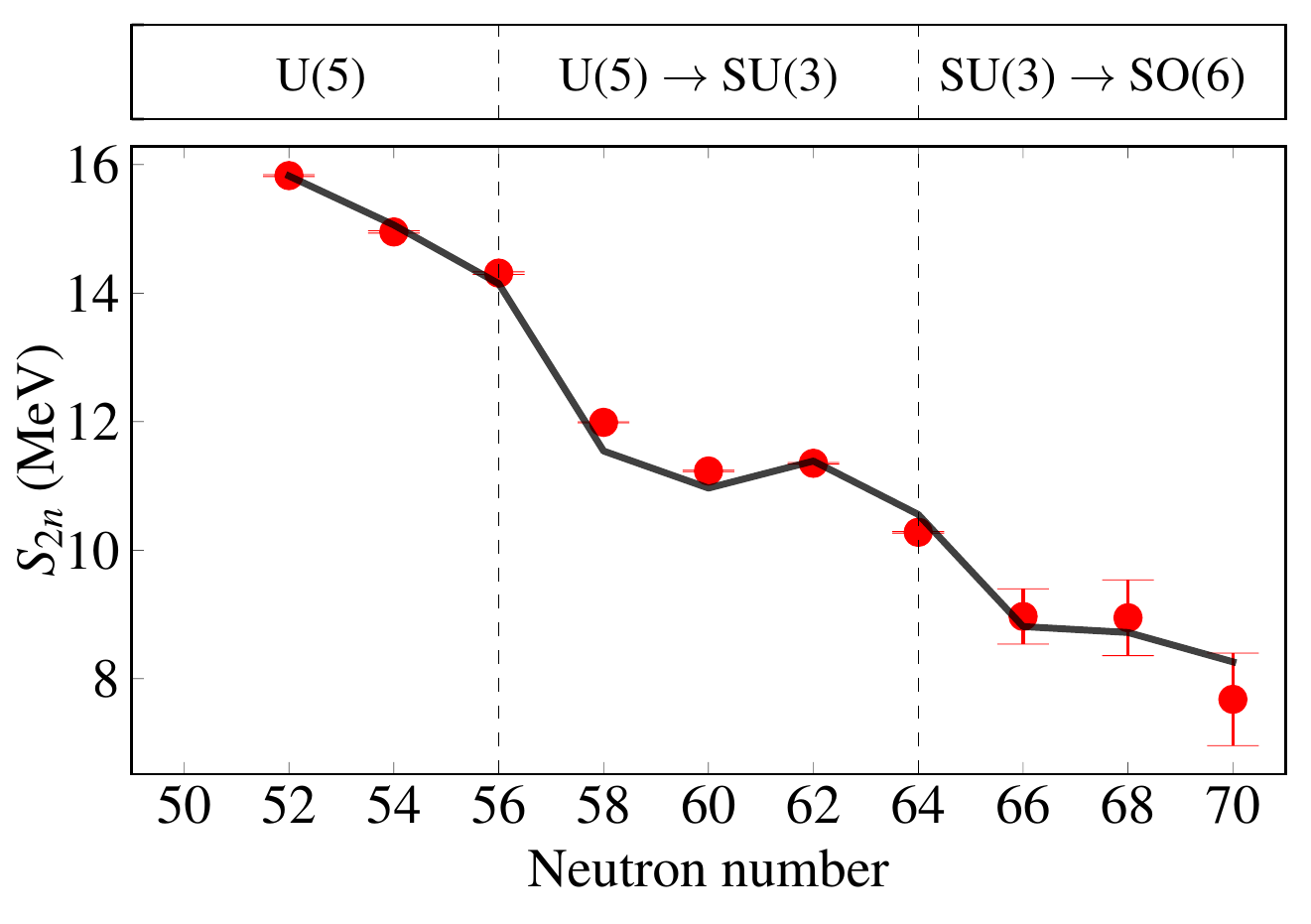}
\caption{Evolution of two-neutron separation energies, $S_{2n}$, in MeV along the Zr chain. Data are taken from AME2016~\cite{Wang2017}.\label{fig:S2n}}
\end{figure}
The $+$ sign applies to particles and the $-$ sign to holes. $\Delta_n $ takes into account the neutron subshell closure at 56, $\Delta_n = 0 $ for 50--56 and $ \Delta_n = 2 $ MeV for 58--70. The value of $ \Delta_n $ is adapted from Table XII of~\cite{Barea2009} and $\tilde{A}\!=\!-16.5,\,\tilde{B}\!=\!0.758$ MeV are determined by a fit to the binding energies of $^{92,94,96}$Zr. The calculated $ S_{2n}$, shown in Fig.~\ref{fig:S2n}, is in agreement with the empirical results and displays a complex behavior. Between neutron numbers 52 and 56 it is a straight line, as the ground state is spherical (seniority-like)
configuration~$A$. After 56, it first goes down due to the subshell closure at~56, then it flattens as expected from a first-order QPT (see, for example the same situation in the $_{62}$Sm isotopes~\cite{Scholten1978}). After 62, it goes down again due to the increase of deformation and finally it flattens as expected from a \mbox{crossover from SU(3) to SO(6)}.
\subsection{$E2$ transition rates}\label{sec:evo-e2}
\definecolor{GreenNoam}{rgb}{0,0.5,0}
\begin{figure}[t]
\centering
\includegraphics[width=1\linewidth]{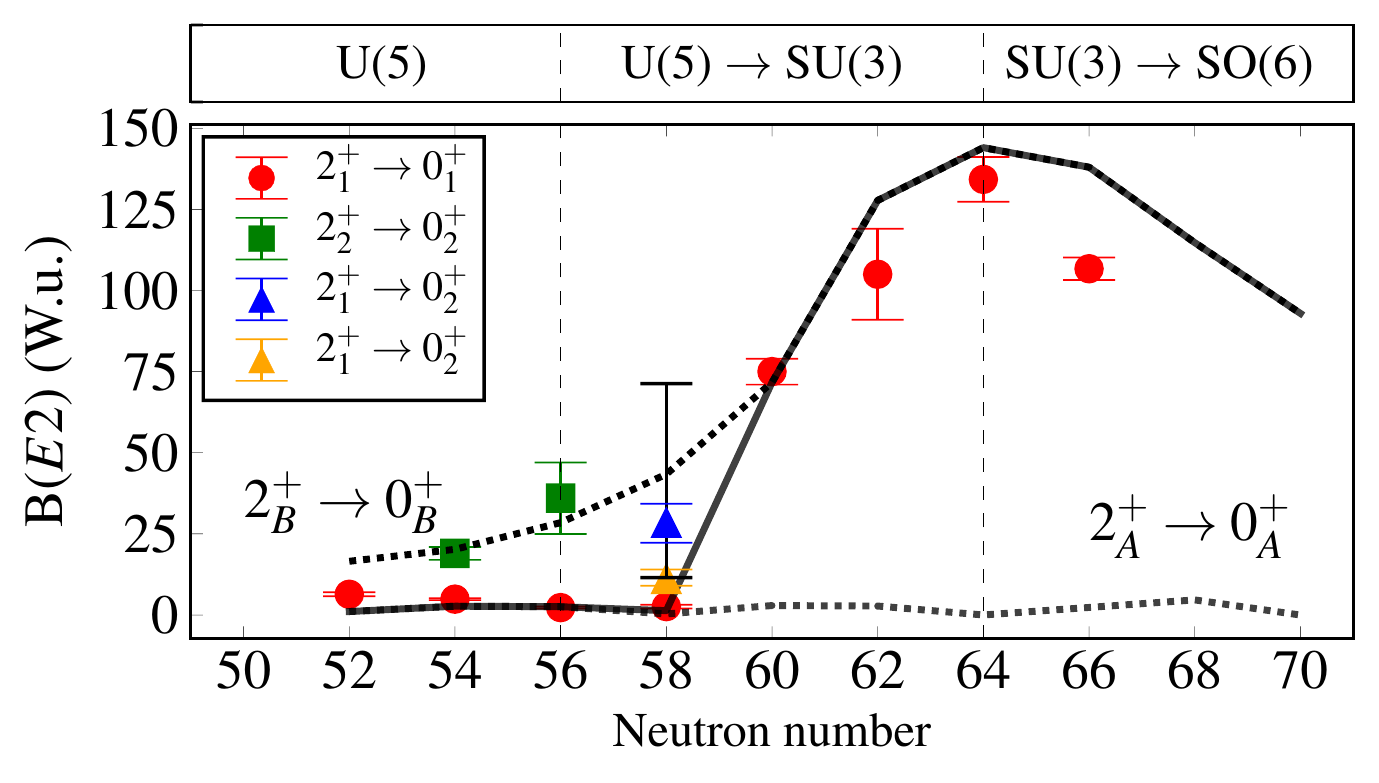}
\caption{$B(E2)$ values in W.u. for $2^+\rightarrow0^+$ transitions in the Zr chain. The solid line (symbols ${\color{red}\CIRCLE},~{\color{GreenNoam}\blacksquare},~{\color{blue}\blacktriangle},~{\color{orange}\blacklozenge}$) denote calculated results (experimental results). Dotted lines denote calculated $E2$ transitions within a configuration. 
The data for $^{94}$Zr, $^{96}$Zr, $^{100}$Zr, $^{102}$Zr and ($^{104}$Zr, $^{106}$Zr) are taken from \cite{Chakraborty2013}, \cite{Kremer2016}, \cite{Ansari2017}, \cite{NDS.110.1745.2009}, \cite{Browne2015b}, respectively.
For $^{98}$Zr (neutron number 58), the experimental values are from \cite{Karayonchev2020} ({\color{orange}$\blacklozenge$}), from \cite{Singh2018} ({\color{blue}$\blacktriangle$}), and the upper and lower limits (black bars) are from \cite{Ansari2017,Witt2018}.
\label{fig:be2}}
\end{figure}
The above conclusions are stressed by an analysis of other observables, 
in particular, $B(E2)$ values. As shown in \cref{fig:be2}, the calculated $B(E2)$'s agree with the empirical values and follow the same trends as the respective order parameters (see \cref{fig:nd}). The calculated $2^+_A\to 0^+_A$ transition rates coincide with the empirical $2^+_1\to 0^+_1$ rates for neutron numbers 52--56. The calculated $2^+_B\to 0^+_B$ transition rates coincide with the empirical $2^+_2\to 0^+_2$ rates for neutron numbers 52--56, with the empirical $2^+_1\to 0^+_2$ rates at neutron number 58 and with the empirical $2^+_1\to 0^+_1$ rates at neutron numbers 60--64. The large jump in $\be{2}{1}{0}{1}$ between neutron number 58 and 60 reflects the passing through a critical-point, common to a Type II QPT involving a crossing of two configurations and a spherical to deformed U(5)-SU(3) Type I QPT within configuration~$B$. The further increase in $\be{2}{1}{0}{1}$ for neutron numbers 60--64 is as expected for a U(5)-SU(3) QPT (see Fig.~2.20 in~\cite{IachelloArimaBook}) and reflects an increase in the deformation in a spherical to deformed shape-phase transition within configuration~$B$. The subsequent decrease from the peak at neutron number 64 towards 70 is in accord with the aforementioned SU(3) to SO(6) crossover (see Fig.~2.22 in~\cite{IachelloArimaBook}).

\subsection{Isotope shift and $E0$ transitions}\label{sec:evo-iso-e0}
\begin{figure}[t]
\centering
\includegraphics[width=0.86\linewidth]{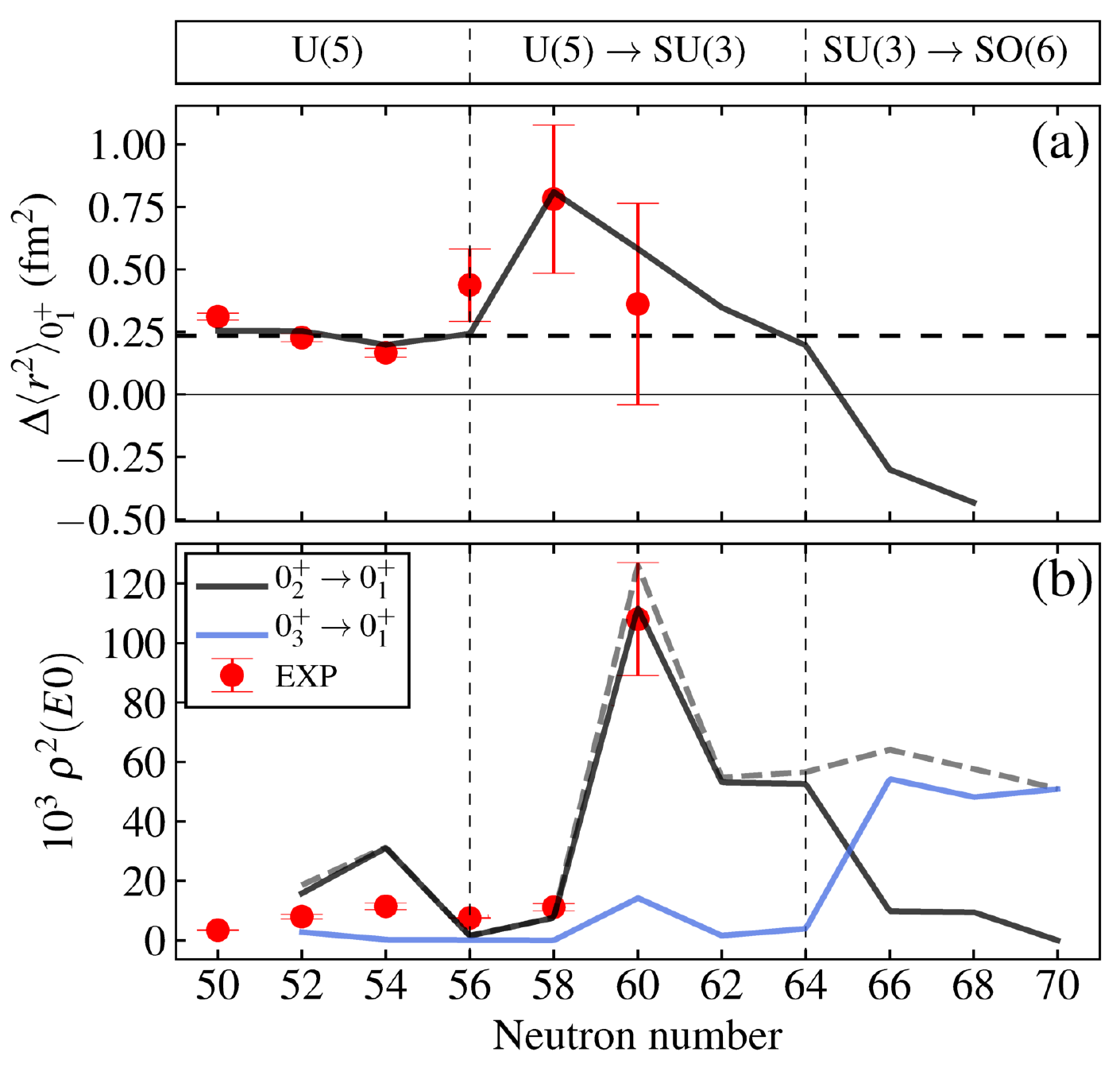}
\caption{Evolution of isotope shifts and $E0$ transitions along the Zr chain.
Symbols (solid lines) denote experimental data (calculated results).
(a) Isotope shift, $\Delta\braket{\hat{r}^{2}}_{0^{+}_1}$ in fm$^{2}$. Data
taken from~\cite{Angeli2013}. The horizontal dashed line at $0.235$
fm$^{2}$ represents the smooth behavior in
$\Delta \braket{\hat{r}^{2}}_{0^{+}_1}$ due to the $A^{1/3}$ increase of
the nuclear radius. (b) $E0$ strength squared. The gray dashed line denotes the calculated sum of both transitions and is approximately constant for neutron numbers 62--70.\label{fig:iso-e0}}
\end{figure}
Further evidence for the indicated structural changes occurring in the Zr chain can be obtained from analyzing the isotope shift  
$\Delta\braket{\hat r^2}_{0^+_1}
=\braket{\hat{r}^{2}}_{0^{+}_1;A+2}-\braket{\hat{r}^2}_{0^{+}_1;A}$, where 
$\braket{\hat r^2}_{0^+_1} $ is the expectation value of $\hat r^2$
in the ground state, $0^+_1$. In the IBM-CM, the charge radius operator can be written as
\begin{equation}\label{eq:charge-rad}
\hat T(r^2) = r^2_c + \alpha \hat N + \eta \hat n_d~,
\end{equation}
where $r^2_c$ is the square radius of the closed shell, $\hat N$ ($\hat n_d$) is total boson ($d$-boson) number operator~\cite{IachelloArimaBook, Zerguine2008, Zerguine2012}. 
The isotope shift depends on two parameters, $\alpha$ and $\eta$, given in units of fm$^2$. $\alpha$ represents the smooth behavior in $\Delta \braket{\hat{r}^{2}}_{0^{+}_1}$ due to the $A^{1/3}$ increase of the nuclear radius, while $\eta$ takes into account the effect of deformation. Their values are fitted to the data and yield $\alpha\!=\!0.235$ fm$^2$ and $\eta\!=\!0.12$ fm$^2$.

As seen in \cref{fig:iso-e0}(a), the calculated $\Delta\braket{\hat r^2}_{0^+_1}$ increases at the transition point and decreases afterwards, which is in accord with the expected behavior of a first-order QPT and the experimental values, although the error bars are large and no data are available beyond neutron number 60. (In the large $N$ limit, this quantity, proportional to the derivative of the order parameter  $\braket{\hat{n}_d}_{0^{+}_1}$, diverges at the critical point). 

The current calculated result is different from our previous one \cite{Gavrielov2019}. The reason is that in \cref{eq:charge-rad} we use the boson number operator, where in \cite{Gavrielov2019} we used $N_v$. The difference is at the transition point, where the expectation value of the boson operator in the ground state becomes approximately 7 for $^{100}$Zr (and 8, 9, 10 for $^{102-106}$Zr, respectively) while for $N_\nu$ it is 5 (and 6, 7, 8 for $^{102-106}$Zr, respectively). Therefore, the use of the boson number operator results in a peak at neutron number 58 rather at 60 (when using $N_\nu$).

The monopole strength for $E0$ transitions between initial $\ket{i}$ and final $\ket{f}$ states,
\begin{equation}\label{eq:rho}
\rho(E0)=\frac{\braket{f|\hat T(E0)|i}}{eR^2},
\end{equation}
can be evaluated using the $E0$ transition operator:
\begin{equation}\label{eq:e0}
\hat T(E0)=(e_nN+e_pZ)\hat T(r^2).
\end{equation}
The latter is constructed from the charge radius operator, \cref{eq:charge-rad}, in the manner suggested in \cite{Zerguine2008, Zerguine2012}.
We note that in such a case the values of $\alpha$ and $\eta$ that are used for the isotope shift operator are the same for the $E0$ transitions operator.

Similarly to $E2$ transition rates, the quantity in \cref{eq:rho} can also highlight the underlying structure of the wave functions. Figure~\eqref{fig:iso-e0}(b) depicts two calculations of the square of the monopole strength, for $0^+_2 \to 0^+_1$ (black line), compared to experimental values (red dots), and for $0^+_3 \to 0^+_1$ (blue line). One can see an intricate behavior of the data. 
At neutron numbers 52--58, the transitions are weak since the wave function of the $0^+_1$ ($0^+_2$ and $0^+_3$) state has a dominant component of configuration~$A$ $B$, in accord with the discussion in \cref{fig:92-96Zr-decomp,fig:98-102Zr-decomp,fig:mixing}.
At neutron number 60, there is an increase in strength of both transitions, reflecting the occurrence of both types of QPTs. The increase in $0^+_2 \to 0^+_1$ occurs as a consequence of the increase in mixing between the configurations: the $0^+_2$ state (which was the spherical $0^+_1$ state in $^{92-98}$Zr) is more mixed with the $0^+_1$ state. This is similar to the process presented in Ref.~\cite{Wood1999a}, in which large mixing induces large $E0$ transitions.
The increase in $0^+_3 \to 0^+_1$ (with values 0.029 14.308  1.609 for $^{98-102}$Zr, respectively) occurs since for $^{100}$Zr the $0^+_3$ state is now the first excited $0^+$ within configuration~$B$ alongside the ground state $0^+_1$, and both are deformed. 
As shown in \cite{VonBrentano2004b}, in a single configuration, an increase in deformation can give rise to an increase in the monopole strength. 
At neutron numbers 62--64 there is a decrease in $0^+_2 \to 0^+_1$ since these states are pure configuration~$B$, with no mixing. Nevertheless, the transition value is still large, again since they are deformed, consistent with the view of \cite{VonBrentano2004b}.
At neutron numbers 66--70 there is an SU(3)-SO(6) crossover and thus the $0^+_2 \to 0^+_1$ and $0^+_3 \to 0^+_1$ strengths interchange, large $0^+_3 \to 0^+_1$ transitions emerge, while $0^+_2 \to 0^+_1$ transitions are weak. Such a behavior arises from the fact that the $E0$ operator of \cref{eq:e0} is an SO(5) scalar. In the SO(6)-DS limit, $0^+_1,0^+_3$ are $\tau\!=\!0$ states while $0^+_2$ has $\tau\!=\!3$.
Finally, as noted in \cite{VonBrentano2004b}, the sum of $0^+_2 \to 0^+_1$ and $0^+_3 \to 0^+_1$ strengths remains nearly constant for neutron numbers 60--70 (as shown by a dashed line in \cref{fig:iso-e0}).

\subsection{Magnetic moments}\label{sec:evo-mag-m1}
For a single configuration, the magnetic dipole operator can be written in its simplest, one-body, form as
\begin{equation}\label{eq:m1-op-single}
\hat T(M1) = \sqrt{\frac{3}{4\pi}}\textsl{g} \hat L~,
\end{equation}
where $\hat L$ is the angular momentum operator and $g$ is the effective boson \textsl{g}-factor \cite{IachelloArimaBook}. For two mixed configurations, the magnetic dipole operator reads
\begin{equation}\label{eq:m1-op-mix}
\hat T^{(M1)} = \sqrt{\frac{3}{4\pi}}\Big(\textsl{g}^{(A)} \hat L^{(N)} + \textsl{g}^{(B)} \hat L^{(N+2)} \Big)~,
\end{equation}
where $\hat L^{(N)}\!=\! \hat P^\dagger_{N} \hat L \hat P_N$ is the angular momentum operator projected onto the $N$ boson space and $g^{(A)}$ and $g^{(B)}$ are the coefficients. The magnetic moment $\mu_L$ of a state as in \cref{eq:wf} is then given by
\begin{equation}\label{eq:magnetic}
\mu_L = \Big(a^2 g^{(A)} + b^2 g^{(B)} \Big) L,
\end{equation}
with $a^2 + b^2 \!=\!1$. Similarly to the case of $\hat T(E2)$ in \cref{eq:te2}, also here we do not include two-body terms in \cref{eq:m1-op-single}.
\begin{figure}[t]
\includegraphics[width=1\linewidth]{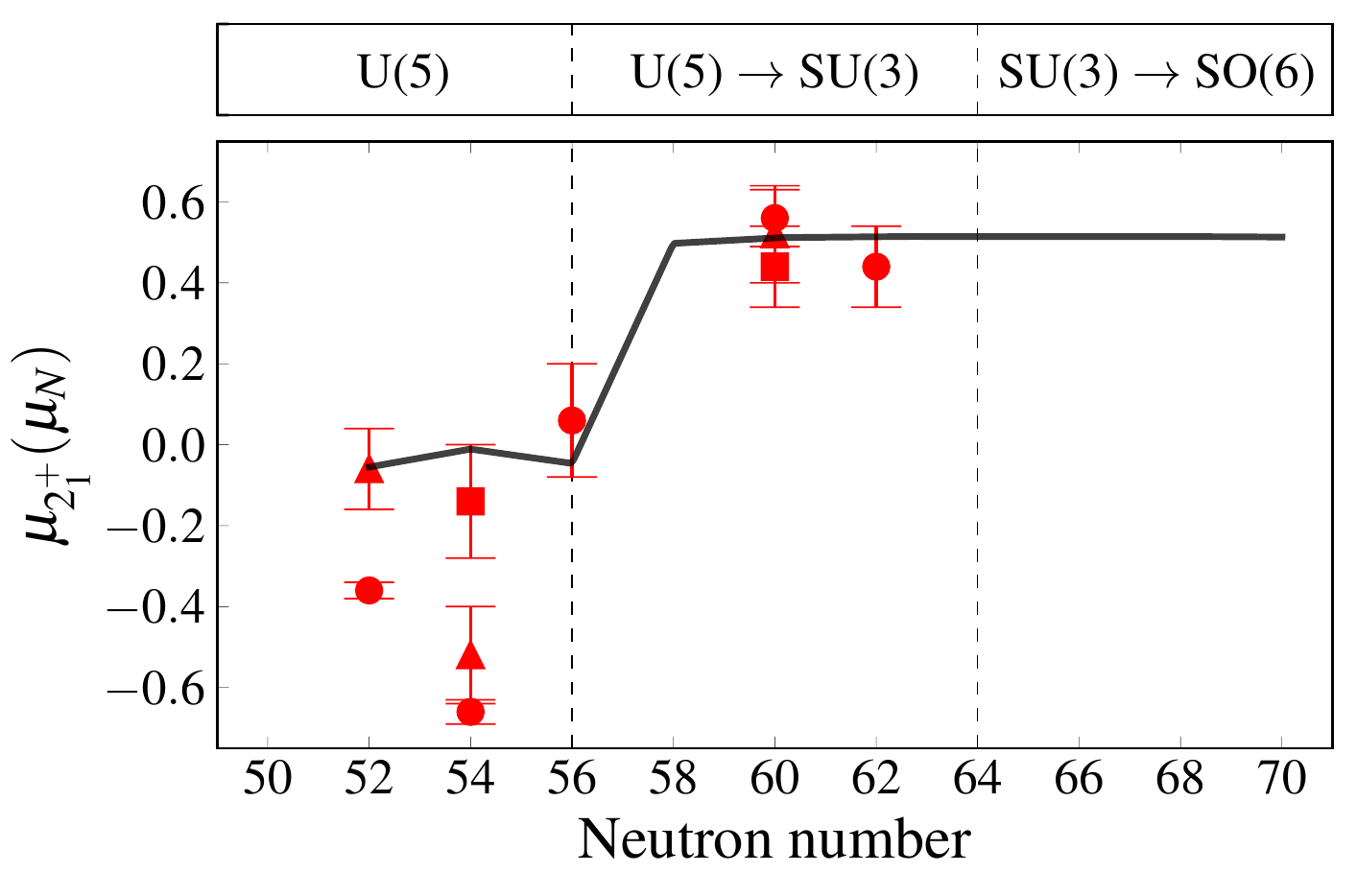}
\caption{Evolution of magnetic moments $\mu_{2^+_1}$ in units of $\mu_N$, along the Zr chain. Symbols denote experimental data, taken from \cite{Stone2005}. Solid line denotes calculated results, based on \cref{eq:magnetic}.
\label{fig:magnetic}}
\end{figure}

Experimental and calculated magnetic moments for the $2^+_1$ state in Zr isotopes are shown in \cref{fig:magnetic}. The calculated values are based on \cref{eq:magnetic}, with $\textsl{g}^{(A)}$ and $\textsl{g}^{(B)}$ taken as constants for simplicity. ${\textsl{g}^{(A)}\!=\!-0.04~\mu_N}$ is determined from the average of the experimental lower value of $^{96}$Zr and upper value of $^{94}$Zr. ${\textsl{g}^{(B)}\!=\!+0.2575~\mu_N}$ is determined from the average of the experimental lower value of $^{100}$Zr and the upper value of $^{102}$Zr. One can see an interesting trend. The empirical and calculated values are close to zero (or negative values) for neutron numbers 52--56 and are close to $+0.5~\mu_N$ for neutron numbers 58--70. The latter is close to the collective $\textsl{g}$-factor for a rigid rotor \cite{BohrMott-II}, $g_{2^+_1}\!=\!Z/A$. In general, values of $\mu_{2^+_1}$ close to zero (or negative) reflect single-particle structures, while large positive values reflect collective structures. The approximately constant trend for neutron numbers 52--56 and 58--70 suggests that the amount of mixing in the $2^+_1$ state is also approximately constant for each set of neutron numbers. This is inline with our calculations reported in \cref{sec:results-evo}, that suggest the same amount of weak mixing, approximately (see \cref{fig:mixing}). The mixing in the wave function of the $2^+_1$ state, \cref{eq:wf}, along the chain of isotopes is $a^2\!=\! 96\%,~88\%,~93\%,~3\%,~1\%$ for neutron numbers 52--60 and $a^2\!\approx\!0\%$ for neutron numbers 62-70. Consequently, for neutron numbers 52--56 (58--70) mainly the $\textsl{g}^{(A)}$ ($\textsl{g}^{(B)}$) part dominates in \cref{eq:magnetic}. The sharp increase when going from neutron number 56 to 58 reflects the fact that the calculated $2^+_1$ wave function changes its structure from being a dominant $A$ to $B$ configuration, respectively. Thus, magnetic moments can be used as a signature for identifying the amount of collectivity, the amount of mixing between different configurations and for Type II QPTs. Some of these ideas were previously suggested in \cite{De-Shalit1953} and are inline with the more recent work of Ref.~\cite{Kumbartzki2012}.
Other known experimental magnetic moment values are $\mu_{2^+_2}\!=\!+1.52(10)$ for $^{92}$Zr \cite{NDS.113.2187.2012} and $\mu_{2^+_2}\!=\!+1.76(54)$ $\mu_N$ for $^{94}$Zr \cite{NDS.107.2423.2006}. The calculated values are 0.495$\mu_N$ and 0.421$\mu_N$, respectively. The large positive values reflect the fact that the $2^+_2$ state is part of the collective $B$ configuration; however, the calculated values are too low, possibly due to the fact that for neutron numbers 52 and 54 the boson numbers are small ($N\!=\!1,2$).

\section{Comparison with other works}\label{sec:compar-works}
The Zr isotopes have been investigated by several theoretical approaches mentioned in the Introduction. 
Here we compare our results with representative large scale shell-model calculations: the Monte Carlo shell model (MCSM) \cite{Togashi2016} and the complex excited VAMPIR model (EXVAM) \cite{Petrovici2012} and with other IBM-CM calculations: mean-field based (IBM-MF) \cite{Nomura2016c} and an independent calculation \cite{GarciaRamos2019,GarciaRamos2020} similar to ours, but with a different fitting protocol, denoted henceforth by IBM-CM-2. We focus the comparison on the $^{98,100}$Zr isotopes, which lie near the critical-point of both Type I and Type II QPTs.

\subsection{The $^{98}$Zr isotope}\label{sec:comp-98zr}
\definecolor{GreenNoam}{rgb}{0,0.5,0.1}
\begin{figure}[t]
\includegraphics[width=1\linewidth]{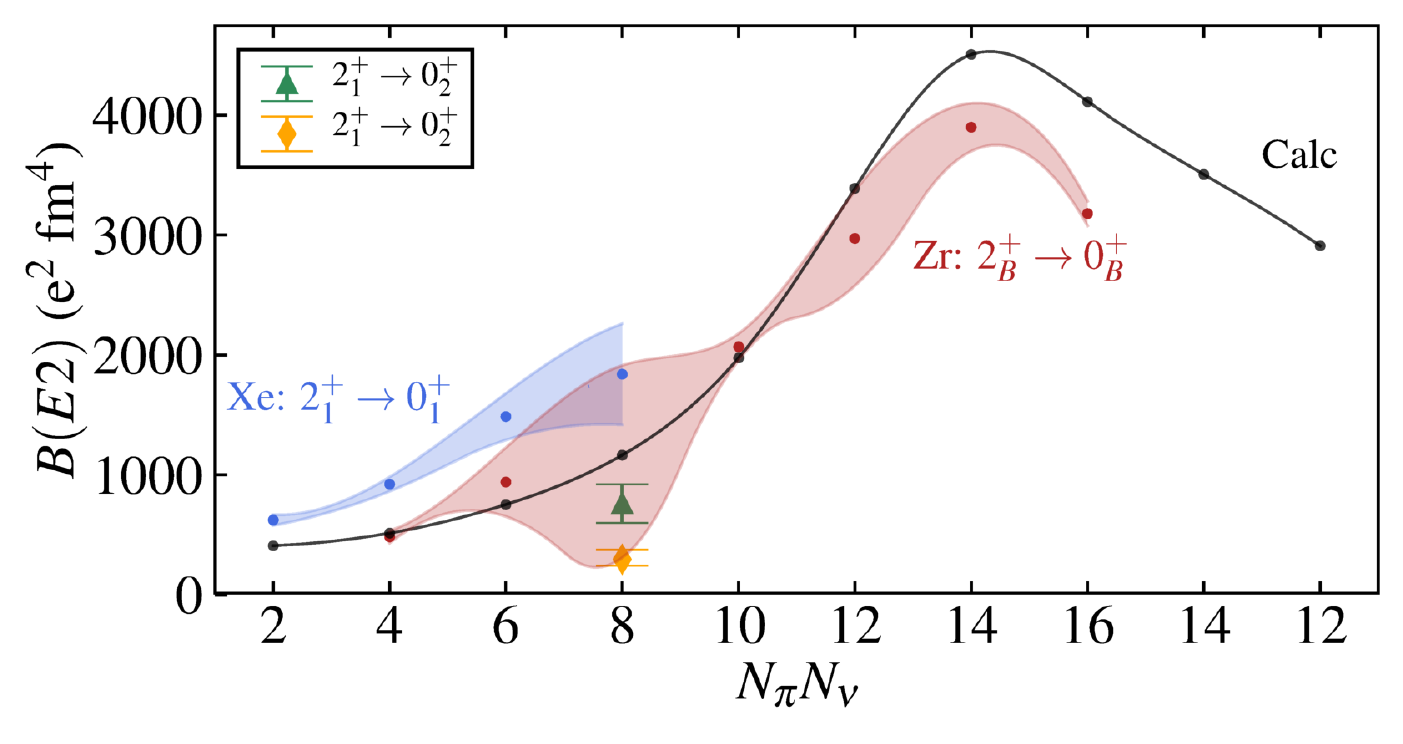}
\caption{$B(E2)$ values in $\text{e}^2\text{fm}^4$ for $2^+_B\rightarrow0^+_B$ in Zr isotopes and $2^+_1\rightarrow0^+_1$ in Xe isotopes, as a function of $N_\pi N_\nu$. Shown are calculated values (black dots connected by a line) and experimental values (blue and red dots) with errors in shaded areas. For $^{98}$Zr ($N_\pi N_\nu=8$), the experimental upper and lower limits are from \cite{Ansari2017,Witt2018} and the explicit values, {\color{GreenNoam}$\blacktriangle$} and {\color{orange}$\blacklozenge$}, are from \cite{Singh2018} and \cite{Karayonchev2020}, respectively. \label{fig:Ze-Xe}}
\end{figure}
\begin{table*}
\begin{center}
\caption{Experimental transition probabilities in W.u. for $^{98}$Zr from \cite{Karayonchev2020} (unless stated otherwise) compared to different theoretical calculations. The present calculation is denoted by \textbf{IBM-CM}.}
\label{tab:98Zr-be2}
\resizebox{\textwidth}{!}{
\begin{tabular}{lcccccc}
\hline
\multicolumn{6}{c}{$B(E2)$ [W.u.]}\\
Transition & Experiment & \textbf{IBM-CM~~} &  IBM-CM-2 \cite{GarciaRamos2019,GarciaRamos2020} & MCSM-1 \cite{Togashi2016}\footnote{Level assignments as in \cite{Ansari2017}.} &  MCSM-2 \cite{Togashi2016}\footnote{Level assignments as in \cite{Singh2018}.\label{foot:togashi-1}} & EXVAM \cite{Petrovici2012}
\\
\hline
$\be{2}{1}{0}{1}$ & $1.1^{+0.3}_{-0.2}$, 2.9(6)\footnote{From \cite{Singh2018}.\label{foot:singh}} & 1.35 & 9.6 & 0.0 & 0.0 & 42.5 \\
$\be{2}{1}{0}{2}$ & $11^{+3}_{-2}$, 28.3(6.0)\footref{foot:singh}, $<71.3$\footnote{From \cite{Witt2018}.}, $>11.5$\footnote{From \cite{Ansari2017}.}& 43.39 & 32 & 70 & 70 & 7.38 \\
$\be{2}{2}{0}{1}$ & $0.26^{+0.20}_{-0.08}$ & 0.34 & 2.5 & 0.0 & 0.0 & $-$ \\
$\be{2}{2}{0}{2}$ & $1.8^{+1.4}_{-0.6}$ & 0.06 & 47 & 2.0 & 2.0 & 1.043 \\
$\be{2}{2}{0}{3}$ & $-$ & 6.54  & 3.2 & 49 & 49 & 48 \\
$\be{2}{2}{2}{1}$ & $46^{+35}_{-14}$ & 47.22 & 0.55 & 8.7 & 8.7 & 70 \\
$\be{2}{3}{0}{1}$ & $0.14^{+0.12}_{-0.04}$ & 2.33 & 0.01 & $-$ & $-$ & $-$ \\ 
$\be{2}{3}{0}{2}$ & $1.7^{+1.5}_{-0.5}$ & 2.28 & 0.56 & $-$ & $-$ & $-$ \\ 
$\be{2}{3}{2}{1}$ & $7.6^{+6.5}_{-2.3}$ & 1.81 & 46 & $-$ & $-$ & $-$ \\ 
$\be{4}{1}{2}{1}$ & $25^{+15}_{-7}$, 43.3(8.7)\footref{foot:singh} & 68.0 & 59 & 103 & 0.6 & 77 \\
$\be{4}{1}{2}{2}$ & $38^{+26}_{-13}$, 67.5(13.5)\footref{foot:singh} & 1.68 & 67 & 0.7 & 76 & 23 \\
$\be{4}{2}{2}{1}$ & $0.6^{+0.17}_{-0.12}$ & $-$ \footnote{Outside of the IBM-CM model space. See text.\label{foot:ibm}} & 0.05 & 0.6 & 103 & 59 \\
$\be{4}{2}{2}{2}$ & $4.6^{+1.7}_{-1.3}$ & $-$ \footref{foot:ibm} & 0.11 & 76 & 0.7 & 2.1 \\			
$\be{6}{1}{4}{1}$ & 103.0(35.7)\footref{foot:singh} & 76.9 & 143 & 102 & 87 & $-$ \\
$\be{0}{3}{2}{1}$ & 58(8)\footnote{From~\cite{NDS.164.1.2020}.\label{foot:NDS.164.1.2020}} & 37 & 53 & $-$ & $-$ & 30 \\
$\be{0}{4}{2}{2}$ & 42(3)\footref{foot:NDS.164.1.2020} & 46 & 42 & $-$ & $-$ & $-$ \\
$\be{0}{4}{2}{1}$ & 0.103(8)\footref{foot:NDS.164.1.2020} & 0.045 & 0.33 & $-$ & $-$ & 0.074 \\
\hline
\end{tabular}
}
\end{center}
\end{table*}
\begin{table}[t]
\begin{center}
\caption{\label{tab:coll-factor}
\small
$N_\pi N_\nu$ values of the intruder $B$ (normal) configuration  for Zr and (Xe) isotopes.}
\resizebox{\linewidth}{!}{%
\begin{tabular}{lcccccccccc}
\hline
$N_\pi N_\nu$ & 2 & 4 & 6 & 8 & 10 & 12 & 14 & 16 & 14 & 12 \\
\hline
Zr & $^{92}$Zr & $^{94}$Zr & $^{96}$Zr & $^{98}$Zr & $^{100}$Zr & $^{102}$Zr & $^{104}$Zr & $^{106}$Zr & $^{108}$Zr & $^{110}$Zr \\
Xe & $^{134}$Xe & $^{132}$Xe & $^{130}$Xe & $^{128}$Xe & & & & & & \\
\hline 
\end{tabular}
}
\end{center}
\end{table}
Recently, absolute transition rates in $^{98}$Zr were measured in  Refs.~\cite{Singh2018,Karayonchev2020}. The results, adapted from \cite{Karayonchev2020}, are presented in \cref{tab:98Zr-be2}, with an added comparison with the EXVAM calculation. In \cref{tab:98Zr-be2}, MCSM-1 and MCSM-2 are the same MCSM calculation employing different assignment of levels (see Ref. \cite{Karayonchev2020} for more details). The IBM-CM in boldface and IBM-CM-2 are the current IBM-CM calculation and that of \cite{GarciaRamos2019,GarciaRamos2020}, respectively.

Both IBM-CM calculations consider two configurations, normal and intruder composed of (0p-0h) and (2p-2h) states, respectively. For $^{98}$Zr, the resulting $0^+_1$ state is spherical and the $0^+_2$ state is weakly deformed or quasi-spherical (see the discussion in \cref{sec:98-102zr-region}).
In contrast, the MCSM calculation considers three configurations dominated by different $n$p-$n$h proton excitations. Specifically, for $^{98}$Zr the ground state $0^+_1$ is spherical, the $0^+_2$ state is weakly deformed and the $0^+_3$ state is strongly deformed.
The EXVAM calculation finds the lowest bandhead $0^+$ states to be prolate-oblate mixed ($0^+_1$), spherical ($0^+_2$), and prolate ($0^+_3$). The assignment of the $0^+_1$ state as prolate-oblate is at variance with the MCSM and IBM-CM calculations and contrasts with the experimental data of Refs.~\cite{Witt2018, Karayonchev2020}.

The current IBM-CM calculation describes well most of the experimental transitions shown in \cref{tab:98Zr-be2} and \cref{fig:98-102Zr-scheme}(a) (for a detailed discussion, see \cite{Karayonchev2020}). However, some of the newly measured transitions, within the intruder $B$ configuration, exhibit marked differences from the calculation and one another.
Specifically, the recently measured value $\be{2}{1}{0}{2}=11^{+3}_{-2}$ W.u. \cite{Karayonchev2020} is significantly lower than the value 28.3(60) W.u. measured in Ref.~\cite{Singh2018} and conforms only with the lower (11.5 W.u.) and upper (71.3 W.u.) limits obtained in Refs.~\cite{Ansari2017} and \cite{Witt2018}, respectively. 
Our calculated value is 43 W.u., which is considerably larger than both explicitly measured values yet it lies in-between the lower and upper limits.
The calculated values of the MCSM (70 W.u.) and EXVAM (7.38 W.u.) deviate considerably from the explicit experimental values.
The IBM-CM-2 calculation \cite{GarciaRamos2019,GarciaRamos2020} can reproduce the measured value of \cite{Singh2018}, since the effective charge in the $E2$ operator was fixed by this transition.
However, the calculated $2^+_1$ state is found to have a large intruder component [$b^2=0.45$ in \cref{eq:wf}], compared to a small mixing ($b^2=0.97$) in the current calculation, which conforms with \cite{Witt2018}. 

These deviations are somewhat surprising, as we now discuss. Figure~\eqref{fig:Ze-Xe} displays the experimental $\be{2}{B}{0}{B}$ value for transitions within the $B$ configuration in the Zr isotopes, as a function of $N_\pi N_\nu$. The values for the latter product of proton and neutron boson numbers, appropriate to the Zr isotopes, are given in \cref{tab:coll-factor}. As seen, our calculated values agree with the measured values for all Zr isotopes, except for $^{98}$Zr. Furthermore, as seen in \cref{fig:Ze-Xe}, the calculated trend is similar to that of the experimental $\be{2}{1}{0}{1}$ values for the Xe isotopes, which have the same $N_\pi N_\nu$ values as the Zr isotopes (see \cref{tab:coll-factor}).
Since deformation increases with the value of $N_\pi N_\nu$ \cite{Casten1985a, Casten1985b}, we expect the $B(E2)$ values to increase when going from $N_\pi N_\nu \!=\! 6$ to 10 (neutron numbers 56 to 60 for Zr). Such a trend is present in both the MCSM and IBM-CM calculations for Zr isotopes, both giving values higher than the measured ones for $^{98}$Zr reported in \cite{Singh2018,Karayonchev2020}.
It should be noted that the indicated transitions in $_{40}$Zr isotopes involve intruder (2p-2h) states while those in $_{54}$Xe isotopes involve normal (0p-0h) states. The comparison between $^{92-98}$Zr and $^{134-128}$Xe is therefore kept only up to $^{98}$Zr. The Xe isotopes do not involve proton-neutron partner-orbitals, as in $^{100-110}$Zr.
\begin{table*}[t!]
\begin{center}
\caption{Experimental transition probabilities in W.u. for $^{100}$Zr \cite{NDS.109.297.2008} compared to different theoretical calculations. The column of the present work is denoted by \textbf{IBM-CM}.}
\label{tab:100Zr-be2}
\begin{ruledtabular}
\begin{tabular}{lccccc}
\multicolumn{6}{c}{$B(E2)$ (W.u.)}\\
Transition & Experiment & \textbf{IBM-CM} &  IBM-CM-2 \cite{GarciaRamos2019,GarciaRamos2020} & IBM-MF \cite{Nomura2016c} & MCSM \cite{Togashi2016} 
\\
\hline
$\be{2}{1}{0}{1}$   & 75(4)   & 72   & 70  	& 74  	& 91  \\
$\be{4}{1}{2}{1}$   & 103(9)  & 121  & 120 	& 102 	& 130 \\
$\be{6}{1}{4}{1}$   & 140(30) & 129  & 128 	& 112 	& $-$ \\
$\be{8}{1}{6}{1}$   & 124(13) & 123  & 122 	& 123 	& $-$ \\
$\be{10}{1}{8}{1}$  & 124(15) & 106  & 105 	& $-$ 	& $-$ \\
$\be{12}{1}{10}{1}$ & 131(15) & 79   & 79  	& $-$ 	& $-$ \\
$\be{0}{2}{2}{1}$   & 67(7)   & 70   & 64  	& 0.9 	& $-$ \\
$\be{2}{2}{0}{2}$   & $-$     & 1.52 & 1.58 & $-$ 	& 42 \\
$\be{4}{2}{2}{2}$   & $-$     & 56   & 14 	& 92 	& 59 \\
$\be{2}{3}{0}{3}$   & $-$     & 23   & 19 	& $-$ 	& 83 \\
$\be{4}{3}{2}{3}$   & $-$     & 50   & 6 	& 67 	& 118 \\
\end{tabular}
\end{ruledtabular}
\end{center}
\end{table*}

Additional discrepancies between calculated and measured values occur in $^{98}$Zr for transitions involving the $4^+_1$ state. Specifically, the experimental transition rates $\be{4}{1}{2}{1}\!=\!25^{+15}_{-7}$~W.u. \cite{Karayonchev2020} (43.3(8.7) W.u. in \cite{Singh2018}) and $\be{4}{1}{2}{2}\!=\!38^{+26}_{-13}$~W.u. \cite{Karayonchev2020} (67.5(13.5)~W.u. in \cite{Singh2018}) are strong, a situation that cannot be accommodated by the current calculation, which yields 68 and 2 W.u., respectively. The calculated values reflect the fact that both the $4^+_1$ and $2^+_2$ are members of the $n_d\approx2$ triplet of configuration~$B$ and are weakly mixed with states of configuration~$A$. In such circumstances, these states cannot be connected by strong $E2$ transitions, which follow the selection rules $\Delta n_d=\pm1$ [for small $\chi$ in the $E2$ operator \cref{eq:te2}].
As shown in Table~\ref{tab:98Zr-be2}, both the MCSM-1 and MCSM-2 encounter a similar problem and cannot accommodate simultaneously two strong transitions from the $4^+_1$ state.
In the IBM-CM-2 \cite{GarciaRamos2019,GarciaRamos2020}, the structure of the $4^+_1$ state is similar to that of the  current IBM-CM calculation; however, the $2^+_1$ and $2^+_2$ states exhibit strong normal-intruder mixing with $b^2 =0.45$ and $b^2 =0.55$ respectively. Consequently, the \mbox{IBM-CM-2} can describe adequately the empirical $\be{4}{1}{2}{1}$ and $\be{4}{1}{2}{2}$ rates. However, this structure leads to other noticeable discrepancies.
In particular, the calculated values ${\be{2}{2}{0}{2}\!=\!47}$, ${\be{2}{3}{2}{1}\!=\!46}$ and ${\be{2}{2}{2}{1}\!=\!0.55}$ W.u. are at variance with the experimental values of $1.8^{+1.4}_{-0.6}$,  $7.6^{+6.5}_{-2.3}$ and $46^{+35}_{-14}$~W.u., respectively.
The EXVAM calculation seems to encounter a similar problem, while it produces two strong transitions from the $4^+_1$ state, it exhibits major discrepancies for $\be{2}{1}{0}{1}=42.5$~W.u., $\be{2}{1}{0}{2}=7.38$~W.u. and $\be{4}{2}{2}{1}=59$~W.u., which are measured to be $1.1^{+0.3}_{-0.2}$, between 11.5. and 71.3 and $0.6^{+0.17}_{-0.12}$~W.u, respectively.

Additional notable discrepancies of the MCSM with the experimental data are for the calculated values $\be{2}{2}{2}{1}=8.7$, $\be{4}{2}{2}{2}=76$ W.u. (for MCSM-1), and $\be{4}{2}{2}{1}=103$ W.u. (for MCSM-2), which are measured to be $46^{+35}_{-14}$, $4.6^{+1.7}_{-1.3}$, and $0.6^{+0.17}_{-0.12}$~W.u., respectively.
Another interesting aspect to compare between the different calculations is the transition $2^+_2 \to 0^+_3$, which has not been measured. In both IBM-CM calculations, this transition is weak, where in the MCSM and EXVAM calculations it is strong. The reason for the difference is that in both IBM-CM calculations the $2^+_2$ and $0^+_3$ states are part of the same $n_d\!\approx\!2$ multiplet (see \cref{sec:92-96zr-region} for more details), whereas in the MCSM and EXVAM calculations these states are part of the same deformed band.
\subsection{The $^{100}$Zr isotope}\label{sec:comp-100zr}
For $^{100}$Zr, a comparison between the present work, IBM-CM-2 \cite{GarciaRamos2019,GarciaRamos2020}, mean-field based IBM calculation (IBM-MF) \cite{Nomura2016c}, MCSM \cite{Togashi2016} and the experimental $B(E2)$ values is given in \cref{tab:100Zr-be2}.
One sees a considerable similarity between the present work and that of  \cite{GarciaRamos2019,GarciaRamos2020}, except for the values of $\be{4}{2}{2}{2},~\be{2}{3}{0}{3}$ and $\be{4}{3}{2}{3}$, which are strong in the present work but weak in \cite{GarciaRamos2019,GarciaRamos2020} and have no experimental data.
The IBM-MF calculation \cite{Nomura2016c} reproduces well the yrast band transitions; however, it does not reproduce the important $\be{0}{2}{2}{1}$ transition (see \cref{sec:98-102zr-region} for more details).
The MCSM calculation offers a more qualitative rather than quantitative agreement with the experimental data, where not many transitions were calculated.

The spherical state in $^{100}$Zr is identified in the present work and in \cite{GarciaRamos2019,GarciaRamos2020} as the $0^+_2$ state. However, the present work calculated the spherical $2^+$ state to be $2^+_5$, while it is $2^+_2$ in \cite{GarciaRamos2019,GarciaRamos2020}. The main source of the difference is the large value for the $\kappa^{(A)}$ parameter of the normal quadrupole operator, \cref{eq:ham_a}, that is used in \cite{GarciaRamos2019,GarciaRamos2020} ($-0.02326$~MeV) compared to the present work ($-0.006$~MeV). The MCSM has identified the spherical state as the $0^+_4$, in contrast to the experimental data that exhibits only three $0^+$ states. The rest of the calculated lower three $0^+$ states serve as band heads of prolate, oblate and another prolate deformed bands. The IBM-MF calculation \cite{Nomura2016c}, has identified only oblate and prolate configurations for the lowest $0^+$ states, without spherical states. 

The two IBM-CM calculations and MCSM all show a large jump in $B(E2;2^+_1\rightarrow0^+_1)$, between $^{98}$Zr and $^{100}$Zr, typical of a first-order QPT. This is in contrast with the IBM-MF and other mean-field based calculations~\cite{Delaroche2010, Nomura2016c, Mei2012}, which due to their character smooth out the phase transitional behavior, and show no such jump at the critical-point of the QPT (see Fig.~2 of~\cite{Singh2018}).

\subsection{Heavier isotopes}\label{sec:comp-heavier}
The observed peak in $B(E2;2^+_1\rightarrow0^+_1)$ for $^{104}$Zr (see \cref{fig:be2}), is reproduced by the present work and IBM-CM-2 \cite{GarciaRamos2019,GarciaRamos2020} but not by the MCSM \cite{Togashi2016} nor the IBM-MF \cite{Nomura2016c} calculations.
For the region of $^{106-110}$Zr, the IBM-CM-2 calculates a prolate-deformed band where in the current work it is $\gamma$-unstable deformed.
For $^{110}$Zr, the MCSM calculates a proton $6p$--$6h$ (approximately) intruder prolate-deformed ground-band and another proton $4p$--$4h$ (approximately) triaxial-deformed band.

\subsection{General remarks}\label{sec:general-remarks}
In general, the results of the present IBM-CM calculation resemble those obtained in the MCSM (which focuses on spectra and $E2$ rates) and the IBM-CM-2.
However, there are some noticeable differences. Specifically, the inclusion of more than two configurations in the MCSM in which their deformation evolves differently from the present work and the IBM-CM-2. 
The underlying physics in our and IBM-CM-2 study is similar to that of Refs.~\cite{Federman1979,Heyde1985,Heyde1987,Sieja2009}, with a shell-model interpretation of 0p-0h and 2p-2h proton excitation, which use a different formal language, where the lowering in energy and developed collectivity of the intruder configuration are governed by the relative magnitude of $V_{pn}$ (especially its monopole and quadrupole components) and the energy gaps between spherical shell-model states near shell and subshell closures. A more direct relation between the two approaches necessitates a proton-neutron version~of~the~IBM.

\section{Conclusions and outlook}\label{sec:conclu}
We have performed a quantum and classical analysis for the entire chain of $_{40}$Zr isotopes, from neutron number 52 to 70, within the framework of the IBM-CM. The quantum analysis examined the spectra and properties of individual isotopes as well as the evolution of energy levels and other observables (two-neutron separation energies, $E2$ and $E0$ transition rates, isotope shifts and magnetic moments) along the chain. Special attention has been devoted to changes in the configuration-content and symmetry-content of wave functions, and their impact on relevant order parameters. A classical analysis, based on coherent states, examined individual shapes and their evolution with neutron number. In general, the calculated results, obtained by a fitting procedure described in the Appendix, are found to be in excellent agreement with the empirical data.

The results of the comprehensive analysis suggest a complex phase structure in these isotopes, involving two configurations. The normal $A$ configuration remains spherical in all isotopes considered. The intruder $B$ configuration undergoes first a spherical to axially deformed U(5)-SU(3) QPT, with a critical point near $A\approx100$, and then an axially deformed to $\gamma$-unstable SU(3)-SO(6) crossover.
In parallel to the gradual shape evolution within configuration $B$, the two configurations cross near neutron number 60, and the ground state changes from configuration $A$ to configuration $B$. The two configurations are weakly mixed and retain their purity before and after the crossing, which are the defining ingredients of intertwined QPTs (IQPTs).

There are several further observables that  would be worthwhile to measure. Specifically, measuring in $^{98}$Zr the $E2$ transition rates for the $2^+_1 \to 0^+_2$, will shed light on the deviations between experiment and theory, discussed in \cref{sec:comp-98zr}. Measuring the $2^+_2 \to 0^+_3$ transition is also of interest, in order to determine the structure of these states, either as part of an $n_d\!\approx\!2$ triplet or a deformed band. It would also be insightful to employ a more microscopic IBM calculation, such as IBM-2, to further determine the structure of the enigmatic $2^+_1$, $2^+_2$ and $4^+_1$ states.
For $^{100}$Zr, it would be interesting to measure $E2$ transitions from different $2^+$ states to the $0^+_2$ state in order to identify the spherical $2^+$ state. For $^{102-104}$Zr, measuring the $0^+_2\to0^+_1$ and $0^+_3\to0^+_1$ $E0$ transition rates would help verify the evolution of deformation and choice of parameters for the $E0$ transition operator, \cref{eq:e0}.

The present work on the Zr isotopes provides evidence of intertwined quantum phase transitions (IQPTs) in nuclei. It sets the path for new investigations of IQPTs in other nuclei and other physical systems. In particular, our method of calculation could also be applied to the $_{38}$Sr isotopes, which show similar features~\cite{Mach1989}, as opposed to $_{42}$Mo isotopes, where IQPTs appear to be less pronounced.

\section*{Acknowledgments}
This work is supported by the US-Israel Binational Science Foundation Grant No. 2016032 and is based on part of a Ph.D. thesis by N.G. carried out at the Hebrew University, Jerusalem, Israel. N.G. acknowledges support by the Israel Academy of Sciences of a Postdoctoral Fellowship Program in Nuclear Physics. We thank R.F. Casten and J.E. Garc\'ia-Ramos for enlightening discussions.

\appendix

\section{Fitting procedure}\label{app:fitting}
\begin{table*}[t]
\begin{center}
\caption{\label{tab:parameters}
\small
Parameters of the IBM-CM Hamiltonian, \cref{eq:ham-cm}, are in MeV and $\chi$ is dimensionless. The first row of the Table lists the number of neutrons, and  particle-bosons $(N,N+2)$ or hole-bosons $(\bar N,\bar N+2)$ in the $(A,B)$ configurations.}
\begin{tabular}{lcccccccccc}
\hline
          & $52(1,3)$ & $54(2,4)$  & $56(3,5)$ &  $58(4,6)$ & $60(5,7)$ & 
$62(6,8)$ & $64(7,9)$ & $66(8,10)$ & $68(\bar7,\bar9)$	& $70(\bar6,\bar8)$ 
\\[2pt]
\hline
$\epsilon^{(A)}_d$   & 0.9 & 0.8  & 	1.82   & 1.75   & 1.2 &    
1.2 & 1.2 & 1.2 & 1.2 & 1.2 \\
$\kappa^{(A)}$ & $-0.005$ & $-0.005$ & $-0.005$ & $-0.007$ & $-0.006$ & $-0.006$ & $-0.006$ & $-0.006$ & $-0.006$ & $-0.006$ \\
$\epsilon^{(B)}_d$  & 0.35 & 0.37 & 0.6 & 0.45 & 0.3 & 0.15 & 0 &  0 & 0 & 0.15\\
$\kappa^{(B)}$ & $-0.02$ & $-0.02$ & $-0.015$ & $-0.02$ & $-0.02$ & $-0.025$ & $-0.0275$ & $-0.03$ & $-0.0275$ & $-0.025$ \\
$\kappa^{\prime(B)}$ & 0.01 & 0.01 & 0.01 & 0.01 & 0.0075 & 0.01 & 0.0125 & 0.0125 & 0.0125 & 0.01 \\
$ \chi $ & $-0.6$ & $-0.6$ & $-0.6$ & $-0.6$ & $-1.0$ & $-1.0$ & $-0.75$ & $-0.25$ & $-0.25$ & 0 \\
$\Delta^{(B)}_p$ & 1.6 & 1.6 & 1.84 & 1.43 & 0.8 & 0.8 & 0.8 & 0.8 & 0.8 & 0.8 \\
$\omega$ & 0.1 & 0.1 & 0.02 & 0.02 & 0.02 & 0.02 & 0.02 & 0.02 & 0.02 & 0.02 \\
\hline 
\end{tabular}
\end{center}
\end{table*}
\begin{table}[t]
\begin{center}
\caption{\label{tab:92-98Zr-levels}
\small
Experimental levels of $^{92-110}$Zr that are assigned to configuration-$B$ and used to fit the parameters of $\hat H_B$ \eqref{eq:ham_b}.
For $^{92-98}$Zr, the indicated levels correspond to calculated states dominated by U(5) components with $n_d\approx0,~1,~2,~3$ within the $B$ configuration part of the wave function $\ket{\Psi_B;[N+2],L}$, \cref{eq:wf} (see \cref{sec:results-evo} for more details).}
\begin{tabular}{lc}
\hline
$^{92}$Zr & $0^+_2,~2^+_2,~(4^+_2,~2^+_3,~0^+_3),~(6^+_1,~4^+_3,~3^+_1,~2^+_5)$ \\
$^{94}$Zr & $0^+_2,~2^+_2,~(4^+_2,~2^+_3),~(6^+_1,~4^+_3,~3^+_1,~2^+_5)$ \\
$^{96}$Zr& $0^+_2,~2^+_2,~(4^+_1,~2^+_3,~0^+_3),~(6^+_4,~4^+_3,~2^+_4,~0^+_4)$ \\
$^{98}$Zr& $0^+_2,~2^+_1,~(0^+_3,~2^+_2,~4^+_1),~(6^+_1,~4^+_3,~3^+_1,~2^+_4,~0^+_4)$\\
$^{100}$Zr& $0^+_1,~2^+_1,~4^+_1,~0^+_3,~2^+_2,~6^+_1,~2^+_3$\\
$^{102}$Zr& $0^+_1,~2^+_1,~4^+_1,~0^+_2,~6^+_1,~2^+_2,~2^+_3,~3^+_1$\\
$^{104}$Zr& $0^+_1,~2^+_1,~4^+_1,~6^+_1$\\
$^{106}$Zr& $0^+_1,~2^+_1,~4^+_1,~2^+_2,~6^+_1$\\
$^{108}$Zr& $0^+_1,~2^+_1,~4^+_1,~6^+_1$\\
$^{110}$Zr& $0^+_1,~2^+_1,~4^+_1,~2^+_2$\\
\hline 
\end{tabular}
\end{center}
\end{table}
The parameters of the Hamiltonian \cref{eq:ham_ab,eq:mixing} and $E2$ transition operator \eqref{eq:te2} are determined from a combined fit to the data on spectra and $E2$ transitions. Typically, in each nucleus there are about 10 known energy levels and between 2 and 15 $E2$ transitions. For those nuclei where there are fewer levels and $E2$ transitions known, the parameters have been extrapolated using continuity criteria and results from other IBM calculations such as those of Sambataro and Molnar \cite{Sambataro1982} for the Mo isotopes ($Z\!=\!42$).
We allow a gradual change between adjacent isotopes, but take into account the proposed shell-model interpretation for the structure evolution in this region \cite{Federman1979, Heyde1985, Heyde1987}. The derived Hamiltonian parameters, given in \cref{tab:parameters} and \cref{fig:params}, are consistent with those of previous calculations in this mass region \cite{Sambataro1982, Duval1983, PadillaRodal2006}.

For configuration $A$, the states are associated with seniority-like neutron single-particle excitations~\cite{Federman1979}. They comprise the experimental $0^+_1,~2^+_1,~4^+_1$ states of $^{92,94}$Zr, the $0^+_1,~2^+_1,~3^+_1,~4^+_2$ states of $^{96}$Zr and the $0^+_1,~2^+_3,~4^+_2$ states of $^{98}$Zr. Due to the fact that the IBM-CM describes collective low-lying states rather than single-particle excitations, we only include in the fit the corresponding $0^+$ and $2^+$ states and exclude the others. These $0^+$ and $2^+$ states are generated by the configuration~$A$ Hamiltonian~\eqref{eq:ham_a}, $\hat H_{A}$.
It is possible to introduce an additional term, $\hat n_d(\hat n_d-1)$, to $\hat H_A$ to raise the other configuration~$A$ states higher in energy, while keeping the $0^+,2^+$ states at the same energy. We choose, however, not to do so for simplicity.

In Table~\ref{tab:92-98Zr-levels} we give the states of each isotope that were used to fit the parameters of the configuration~$B$ Hamiltonian~\eqref{eq:ham_b}.
For $^{92-96}$Zr, the values of $\epsilon_d^{(B)}$ and $\kappa^{(B)}$ follow the trend of the lowest configuration~$B$ $0^+,~2^+$ and $4^+$ states, which is approximately constant for neutron numbers 52 and 54, i.e., $E(0^+_\text{2})\!=\!1.38,~1.30$ MeV,  $E(2^+_\text{2})\!=\!1.85,~1.67$ MeV and $E(4^+_\text{2})\!=\!2.40,~2.33$ MeV, respectively. Then, at neutron number 56, a large jump occurs due to the closure of the neutron $2d_{5/2}$ subshell~\cite{Auerbach1965}. The $\kappa^{\prime(B)}$ parameter is fitted to reproduce the energy difference between the $2^+_3,~4^+_2$ states in $^{92-94}$Zr and $2^+_3,~4^+_1$ states in $^{96}$Zr.
For $^{98-106}$Zr, we expand the $B$ configuration parameters as a function of the boson number $N$ \cite{IachelloArimaBook}:
\begin{align}\label{eq:parametrization}
\epsilon_d^{(B)}(N) & = \epsilon_d^{(B)}(N_0) \nonumber\\
					& + \frac{\partial \epsilon_d^{(B)}}{\partial N}\Big|_{N=N_0} (N-N_0) + \ldots \approx \epsilon_0 - \theta N~, \nonumber \\
\kappa^{(B)}(N) & = \kappa^{(B)}(N_0) \nonumber\\
				& + \frac{\partial \kappa^{(B)}}{\partial N}\Big|_{N=N_0} (N-N_0) + \ldots \approx \kappa_0~, \nonumber \\
\kappa^{\prime(B)}(N) & = \kappa^{\prime(B)}(N_0) \nonumber\\
					  & + \frac{\partial \kappa^{\prime(B)}}{\partial N}\Big|_{N=N_0} (N-N_0) + \ldots \approx \kappa^\prime_0~.
\end{align}
As valence neutrons are added to the higher shell orbitals, deformation is increased \cite{Federman1979}. This is taken care of by the reduction of the value of $\epsilon_d^{(B)}$, while $\kappa^{(B)}$ and $\kappa^{\prime(B)}$ are kept approximately constant.
For $^{104}$Zr, in the vicinity of mid-shell, deformation is maximal and we set $\epsilon_d^{(B)}\!=\!0$. Consequently, we fit the rest of the $B$ configuration parameters to $(\epsilon_d^{(B)},\kappa^{(B)},\kappa^{\prime(B)})\!=\!(0,-0.0275,0.0125)$ MeV to reproduce the experimental $0^+_1,~2^+_1,~4^+_1,~6^+_1,~8^+_1,10^+_1,~12^+_1$ states, which are assumed to be part of configuration~$B$. To obtain a gradual increase in deformation (and decrease in $\epsilon_d^{(B)}$), from neutron number 56 to 66, we determine in \cref{eq:parametrization} $\epsilon_0\!=\!1.35$ and $\theta\!=\!0.15$ MeV.
For $^{108-110}$Zr, we use neutron holes and impose a symmetry about mid-shell on all parameters (except $\chi$), in accord with microscopic aspects of the IBM \cite{IachelloTalmi1987}. That is, we use the same parameters for $^{108}$Zr and $^{104}$Zr and for $^{110}$Zr and $^{102}$Zr.

For $^{92-100}$Zr the parameter $\kappa^{(A)}$ of configuration~$A$ was determined from the relation $\kappa^{(B)}\approx3\kappa^{(A)}$, reflecting the fact that configuration~$A$ is more spherical. The parameter $\epsilon_d^{(A)}$ was fitted accordingly to approximately reproduce the experimental energy difference between the first $2^+$ and $0^+$ states in configuration $A$. The parameter $\Delta_p$ is determined so as to reproduce approximately the offset energy between the two configurations. The parameter of the mixing term in \cref{eq:mixing}, $\omega$, is determined from transitions between states from different configurations. This parameter is kept constant, except for $^{92,94}$Zr, where the $A$ configuration space is small ($N\!=\!1,2$, respectively). For $^{102-110}$Zr there are not enough data to determine configuration~$A$ states and therefore $\epsilon_d^{(A)}$, $\kappa^{(A)}$, $\Delta_p$ and $\omega$ are set to have the same values as for $^{100}$Zr.

The parameter $\chi$ of \cref{eq:q-op} is taken, for simplicity, to be the same for both configurations $(A)$ and $(B)$ and constant for $^{92-98}$Zr, where deformation is weaker. It was determined for $^{100-102}$Zr from the energy of the first excited $0^+$ state in configuration~$B$. For $^{106,110}$Zr, it was determined from the energies of the $2^+_2$ and $4^+_1$, which are close in energy.
The boson $E2$ effective charges were determined to be $e^{(A)}\!=\!0.9$ and $e^{(B)}\!=\!2.24$~(W.u.)$^{1/2}$ for the entire chain of isotopes from the $2^+\to0^+$ transition within each configuration.
Fine tuning the parameters for individual isotopes can improve the fit; however the main conclusions of the analysis are not changed. 

Apart from some fluctuations due to the subshell closure at neutron number 56, filling the $2d_{5/2}$ orbital, the values of the parameters are a smooth function of neutron number and, in some cases, a constant, as can be seen in \cref{fig:params}. A notable exception is the sharp decrease by 1~MeV of the energy off-set parameter $\Delta_p$ beyond neutron number 56. Such a behavior was observed for the Mo and Ge chains~\cite{Sambataro1982,Duval1983,PadillaRodal2006} and, as noted in~\cite{Sambataro1982}, it reflects the effects of the isoscalar residual interaction, $V_{pn}$, between protons and neutrons occupying the partner orbitals $1g_{9/2}$ and $1g_{7/2}$, which is the established mechanism for descending cross shell-gap excitations and onset of deformation in this region~\cite{Federman1979,Heyde1985}. This trend in $\Delta_p$ agrees with shell-model estimates for the monopole correction of $V_{pn}$~\cite{Heyde1987}. 
It is interesting though that $\Delta_p$ retains a positive value for the entire chain, as opposed to previous works~\cite{Sambataro1982, Duval1983, PadillaRodal2006}. This suggests that the change in the ground state configuration near the Type~II critical point, $A\!\approx\!100$, is driven less from the change in $\Delta_p$ and more from the increase in deformation within the $B$ configuration.

\bibliography{refs}
\end{document}